\documentclass[manuscript, screen, nonacm]{jair}

\ifdefined\pdfsuppressptexinfo\pdfsuppressptexinfo=2\relax\fi

\makeatletter\let\hyxmp@parse@acmart\relax\makeatother

\newif\ifwithchecklist
\withchecklistfalse

\setcopyright{none}

\acmDOI{}
\acmArticle{}
\makeatletter
\def\@formatdoi#1{\edef\@tempa{#1}\ifx\@tempa\@empty\else
  \textsc{doi:} \href{https://doi.org/#1}{#1}\fi}
\AtBeginDocument{\fancyfoot{}\fancypagestyle{firstpagestyle}{\fancyhf{}%
  \fancyfoot[L]{\footnotesize Preprint.}}}
\makeatother

\RequirePackage[
  datamodel=acmdatamodel,
  style=acmauthoryear,
  backend=biber,
  giveninits=true,
  uniquename=init
  ]{biblatex}

\DeclareDelimFormat{multicitedelim}{\addcomma\space}
\usepackage{mathtools}
\usepackage{array}
\usepackage[capitalize,noabbrev]{cleveref}
\usepackage{placeins}
\usepackage{longtable}
\usepackage{needspace}
\usepackage{flafter}
\usepackage{tikz}
\usetikzlibrary{arrows.meta,calc}

\definecolor{optblue}{HTML}{0072B2}
\definecolor{satred}{HTML}{D55E00}
\definecolor{falsegray}{HTML}{6E6E6E}

\theoremstyle{acmplain}
\newtheorem{theorem}{Theorem}[section]
\newtheorem{proposition}{Proposition}[section]
\newtheorem{lemma}{Lemma}[section]
\newtheorem{corollary}{Corollary}[section]
\newtheorem*{mainresult}{Main Result}
\theoremstyle{acmdefinition}
\newtheorem{definition}{Definition}[section]
\newtheorem{example}{Example}[section]
\newtheorem{remark}{Remark}[section]
\theoremstyle{acmplain}
\makeatletter
\let\c@proposition\c@theorem
\let\c@lemma\c@theorem
\let\c@corollary\c@theorem
\let\c@definition\c@theorem
\let\c@example\c@theorem
\let\c@remark\c@theorem
\makeatother

\newcommand{\R}{\mathcal{R}}                 
\newcommand{\hstar}{h^{\star}}               
\newcommand{\pre}{\mathrm{pre}}
\newcommand{\add}{\mathrm{add}}
\newcommand{\del}{\mathrm{del}}
\newcommand{\act}[3]{\langle #1,\,#2,\,#3\rangle} 
\DeclareMathOperator{\Opt}{Opt}
\DeclareMathOperator{\Tok}{Tok}    
\newcommand{\IW}{\mathrm{IW}}

\newcommand{\ws}{w_{\mathrm{set}}}
\newcommand{\wc}{w_{\mathrm{chain}}}
\newcommand{\wg}{w_{\mathrm{goal}}}
\newcommand{\W}{\mathcal{W}}

\newcommand{\slack}{\rho}                    
\newcommand{\rhoPi}{\rho(\Pi)}
\newcommand{\HS}{\tau}
\newcommand{\Hy}{\mathcal{H}}
\newcommand{\Pone}{\mathcal{P}_{1}}
\newcommand{\Pzero}{\mathcal{P}_{0}}

\newcommand{\Pclass}{\mathrm{P}}
\newcommand{\NPclass}{\mathrm{NP}}

\newcommand{\SigmaTwo}{\Sigma_2^{\mathrm{P}}}

\newcommand{\Wtwo}{\mathrm{W}[2]}

\newcommand{\Pphi}{\Pi_{\varphi,m}}
\newcommand{\PC}{\Pi_{C}}
\newcommand{\ctl}[1]{\mathsf{#1}}            
\newcommand{\sgoal}{s_{\star}}               

\newcommand{\VerAInstances}{384 tasks from 96 formulas (16 adversarial and 80 seeded random, 10 of them unsatisfiable), $m\in\{2,4,8,16\}$, up to 2{,}950 reachable states. Definitional widths of all three notions at the thresholds $w-1$ and $w$ for 152 tasks with $m\le4$}
\newcommand{\VerAOutcome}{sizes, promises, depth-$D$ states and profiles as predicted, widths $1$ or $m+1$}
\newcommand{\VerBInstances}{4{,}000 seeded random solvable tasks with $\hstar\ge2$, 4 to 8 atoms and up to 119 reachable states, exact definitional widths. Of these, 2{,}935 have a unique optimal state sequence (259 of them several optimal action sequences), 629 of these a single goal state and 258 zero slack. Fast and reference oracles compared on 3{,}000 tasks}
\newcommand{\VerBOutcome}{no counterexample}
\newcommand{\VerDInstances}{508 hypergraphs (8 adversarial, 400 random with at most 6 elements, 100 random with 7 to 12 elements), definitional widths for all 508}
\newcommand{\VerDOutcome}{single path, zero slack, $w=\HS(\Hy)$ for all three notions}
\newcommand{\VerEInstances}{400 random circuits with up to 4 inputs, 6 gates and 5 outputs, definitional widths for 370. 111 $\exists\forall$-3-DNF sentences (11 fixed and 100 random, 56 of them true), circuit compared with \cref{eq:qbf-edge} on every input. Up to 23{,}497 reachable states}
\newcommand{\VerEOutcome}{promises hold, $w=1+\HS(\Hy_C)$, and $\HS(\Hy_\Psi)=n$ exactly for true sentences}
\newcommand{\VerMInstances}{5 deliberate errors in the three constructions and one violated hypothesis of \cref{thm:compiler}}
\newcommand{\VerMOutcome}{6 of 6 detected}
\newcommand{\VerSeconds}{80}
\newcommand{\VerPeakMemory}{93}
\newcommand{\VerSolverCNF}{92}
\newcommand{\VerSolverQBF}{111}

\newcommand{\VerProfileTasks}{384}

\newcommand{\ExtXTasks}{3{,}000}
\newcommand{\ExtXMulti}{1{,}686}

\newcommand{\ExtXNested}{461}

\newcommand{\ExtXMaxStates}{60}
\newcommand{\ExtXGreedyAbove}{21}
\newcommand{\ExtYTasks}{3{,}840}

\newcommand{\ExtZTight}{34}
\newcommand{\ExtZNested}{20}
\newcommand{\ExtHGraphs}{295}
\newcommand{\ExtHMaxPlan}{114}
\newcommand{\ExtSeconds}{25}

\crefname{equation}{Equation}{Equations}
\Crefname{equation}{Equation}{Equations}

\begin{document}

\title[Planning Width Is Hard to Approximate]{Planning Width Is Hard to Approximate Even Given the Unique Optimal Plan}

\author{Viet Bac Nguyen}
\orcid{0009-0000-5930-682X}
\email{bacngv@vnu.edu.vn}
\affiliation{%
  \department{Institute for Artificial Intelligence}
  \institution{VNU-University of Engineering and Technology, Hanoi}
  \country{Vietnam}
}

\renewcommand{\shortauthors}{Nguyen}
\authorsaddresses{}

\begin{abstract}
{\bf Background:}
For each fixed $k$, $\IW(k)$ finds an optimal plan in polynomial time on tasks of tuple-chain width at most $k$. For goal-admissible width, deciding whether the width is at most a given value is PSPACE-complete in general and coNP-complete for each fixed threshold $k\ge1$. How closely width can be estimated, even when the unique optimal plan is known, has remained open.

{\bf Objectives:}
We identify which states constrain a width certificate and determine, up to constant factors, how well width can be approximated as the number $N$ of atoms grows.

{\bf Methods:}
When all optimal plans visit the same states, we express set and chain width as hitting numbers. We then construct a STRIPS task in which a Boolean assignment determines which states compete with the supplied plan at a single depth.

{\bf Results:}
Even when the unique optimal plan is supplied and every reachable state lies on a goal-reaching execution at most one action longer than optimal, the width can be $1$ or $\Theta(N)$, and telling the two cases apart is NP-hard. This holds for set, chain and goal-admissible width, with binary branching, constant-size actions and a single reachable goal state. Unless $\Pclass=\NPclass$, no polynomial-time algorithm achieves a one-sided factor $o(N)$ or a symmetric factor $o(\sqrt{N})$. Elementary estimates attain both orders. For positive-precondition STRIPS tasks supplied with a unique optimal plan of length at least two, testing set or chain width against a threshold $k\ge1$ is coNP-complete for each fixed $k$ and $\SigmaTwo$-complete when $k$ is part of the input. Both results also hold for goal-admissible width if only one goal state is reachable. On an explicit state graph, a greedy algorithm approximates set width within a logarithmic factor, even when optimal plans branch.

{\bf Conclusions:}
A supplied plan certifies an upper bound on width. Along a unique optimal state sequence, a smaller set or chain certificate must distinguish each optimal state from every competing state reached within the same number of actions. Goal-admissible width imposes a separate condition at the final step.

\end{abstract}


\maketitle
\hypersetup{pdfcreator={LaTeX with hyperref}}

\section{Introduction}\label{sec:intro}

Iterative width search, $\IW(k)$, keeps a state in breadth-first search if it is the first to make some tuple of at most $k$ atoms true. For each fixed $k$, this rule finds an optimal plan in polynomial time on tasks of tuple-chain width at most $k$ \citep{LipovetzkyGeffner2012}. Many classical benchmark domains have small width for single-atom goals. Width also enters the design of general policies \citep{BonetGeffner2021,BonetGeffner2024} and of sketches that decompose a task into subproblems of bounded width \citep{DrexlerSeippGeffner2024}.

Consider a solved planning instance. Its optimal plan provides one way to certify width. The states on the plan, each regarded as a tuple of all its atoms, form a tuple chain (\cref{prop:states}). Could much smaller tuples certify the same plan? A chain starts with a tuple true in the initial state. Every shortest plan that achieves one tuple must extend by one action to a shortest plan for the next tuple. Every shortest plan for the last tuple must solve the task optimally. Thus even a tuple taken from the supplied plan imposes a condition on executions outside it. An admissible tuple set lets the next tuple depend on the state reached. A goal-admissible sequence uses a different condition on the last extension. We give the three definitions in \cref{sec:prelim}.

\citet{Song2026} classified the threshold problem for goal-admissible width. Testing whether this width is at most $k$ is PSPACE-complete when $k$ belongs to the input and coNP-complete for each fixed $k\ge1$. The PSPACE lower bound uses plan existence. In the fixed-threshold reduction, the unique optimal plan can already be written down without solving the source formula. At $k=1$ the construction separates width $1$ from width at least $2$, and the competing executions can lead to dead ends. Song et al.\ left open how well width can be approximated.

We ask how large the gap can be when those competing executions can also reach the goal. Write $\hstar$ for the optimal plan length. The \emph{slack} of a reachable state is the length of a shortest goal-reaching execution through it minus $\hstar$, with infinite slack if the goal is unreachable from that state. A single extra action suffices for a linear gap.

\begin{mainresult}[Informal]
Given a 3-CNF formula $\varphi$ and a power of two $m\ge2$, one can compute in time polynomial in $\|\varphi\|+m$ a STRIPS task and its unique optimal plan such that every reachable state has slack at most one, exactly one goal state is reachable, at most two actions apply in any reachable state, and every action has at most two preconditions and three effects. For set, chain and goal-admissible width, the value is $1$ if $\varphi$ is unsatisfiable and $m+1$ if it is satisfiable.
\end{mainresult}

Choosing $m$ larger than the formula size makes $m+1$ proportional to the number $N$ of atoms. For a width $w$, a \emph{one-sided} estimate with factor $\alpha$ satisfies $w\le\widehat w\le\alpha w$. A \emph{symmetric} estimate satisfies $w/\alpha\le\widehat w\le\alpha w$. Unless $\Pclass=\NPclass$, no polynomial-time algorithm achieves a one-sided factor $o(N)$ or a symmetric factor $o(\sqrt N)$. The interval $1\le w\le N$ gives a one-sided factor of $N$ and a symmetric factor of $\lceil\sqrt N\,\rceil$ (\cref{cor:one-sided,cor:symmetric}).

The proof identifies which states can rule out a candidate tuple. Suppose all optimal plans visit the same states $s_0,\ldots,s_{\hstar}$. A set or chain certificate contains a tuple of each cost $i$, where the cost of a tuple is the fewest actions needed to reach a state containing it. For $i\ge1$, any such tuple $t$ is contained in $s_i$ and must distinguish $s_i$ from every other state $u$ reachable within $i$ actions. We call these states $u$ its \emph{competitors}. To distinguish $s_i$ from $u$, the tuple must contain an atom of $s_i\setminus u$, so a smallest such tuple is a smallest set meeting every difference $s_i\setminus u$. Its size is the \emph{hitting number} at depth $i$. Taking the largest hitting number along the plan gives exactly set and chain width (\cref{thm:hitting}). Goal-admissible width has a separate terminal condition (\cref{prop:terminal}).

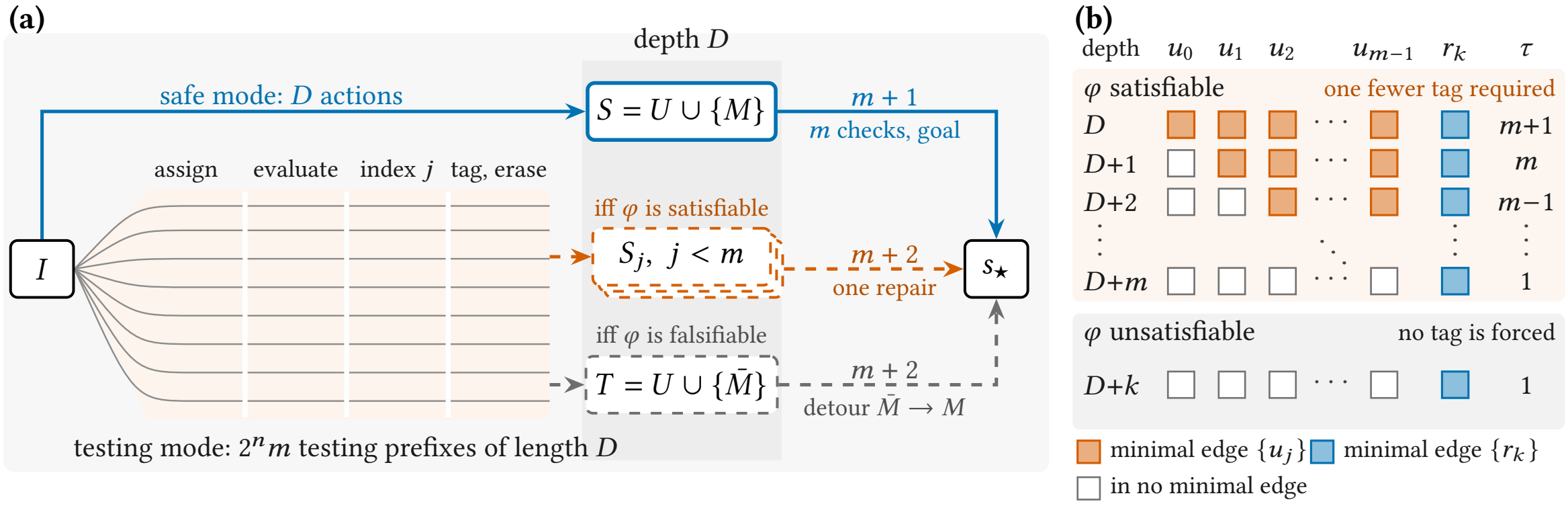
\begin{figure}[!htb]
\centering
\begin{tikzpicture}[x=1cm,y=1cm,
  state/.style={draw, rounded corners=2pt, minimum height=0.56cm, inner sep=2.6pt,
                font=\small, fill=white, line width=0.75pt},
  route/.style={-{Stealth[length=2.0mm,width=1.6mm]}, line width=0.95pt},
  lab/.style={font=\footnotesize, inner sep=1.5pt, text=black!90},
  tiny/.style={font=\scriptsize, inner sep=1pt, text=black!90},
  edge/.style={line width=0.6pt},
]
\colorlet{falsetext}{falsegray!70!black}
\colorlet{sattext}{satred!85!black}
\node[font=\small\bfseries, anchor=west] at (-0.15,4.45) {(a)};

\fill[black!3, rounded corners=3pt] (-0.08,0.0) rectangle (10.22,4.32);
\fill[black!7] (5.62,0.12) rectangle (7.58,4.05);
\node[lab, anchor=south] at (6.6,4.05) {depth $D$};

\fill[satred!7] (0.62,2.0) -- (1.35,2.78) -- (5.3,2.78) -- (5.3,0.52) -- (1.35,0.52) -- cycle;
\foreach \y in {0.70,0.98,1.26,1.54,1.82,2.10,2.38,2.62}
  \draw[black!45, line width=0.45pt] (0.62,2.0) .. controls (1.2,\y) .. (1.8,\y) -- (5.3,\y);
\foreach \x in {2.3,3.3,4.3}
  \draw[white, line width=1.4pt] (\x,0.54) -- (\x,2.76);
\node[tiny, anchor=base] at (1.72,2.91) {assign};
\node[tiny, anchor=base] at (2.8,2.91) {evaluate};
\node[tiny, anchor=base] at (3.8,2.91) {index $j$};
\node[tiny, anchor=base] at (4.8,2.91) {tag, erase};
\node[lab, anchor=west] at (0.55,0.22) {testing mode: $2^n m$ testing prefixes of length $D$};

\node[state, minimum width=0.62cm] (I) at (0.3,2.0) {$I$};
\node[state, draw=optblue, line width=1.05pt, minimum width=1.75cm] (S) at (6.6,3.55) {$S=U\cup\{M\}$};
\node[state, draw=satred, dashed, minimum width=1.75cm] (Sj3) at (6.72,2.0) {\phantom{$S_j$}};
\node[state, draw=satred, dashed, minimum width=1.75cm] (Sj2) at (6.66,2.06) {\phantom{$S_j$}};
\node[state, draw=satred, dashed, minimum width=1.75cm] (Sj) at (6.6,2.12) {$S_j,\ j<m$};
\node[state, draw=falsegray, dashed, minimum width=1.75cm] (T) at (6.6,0.86) {$T=U\cup\{\bar M\}$};
\node[state, minimum width=0.62cm] (G) at (9.70,2.0) {$\sgoal$};

\node[tiny, text=sattext, anchor=south] at (6.6,2.42) {iff $\varphi$ is satisfiable};
\node[tiny, text=falsetext, anchor=south] at (6.6,1.17) {iff $\varphi$ is falsifiable};

\draw[route, optblue] (I.north) |- (S.west)
  node[lab, pos=0.72, above, text=optblue] {safe mode: $D$ actions};
\draw[route, optblue] (S.east) -| (G.north);

\draw[route, satred, dashed] (5.3,2.12) -- (Sj.west);
\draw[route, falsegray, dashed] (5.3,0.86) -- (T.west);
\draw[route, satred, dashed] (Sj3.east) -- (G.west);
\draw[route, falsegray, dashed] (T.east) -| (G.south);

\foreach \yy/\col/\len/\why in {3.55/optblue/m+1/{$m$ checks, goal},
                                2.00/sattext/m+2/{one repair},
                                0.86/falsetext/m+2/{detour $\bar M\to M$}}{
  \node[lab, text=\col, anchor=south] at (8.60,\yy) {$\len$};
  \node[tiny, text=\col, anchor=north] at (8.60,\yy-0.05) {\why};
}

\begin{scope}[xshift=10.50cm,
  oncell/.style={line width=0.6pt},
  offcell/.style={draw=black!55, fill=white, line width=0.5pt}]
\node[font=\small\bfseries, anchor=west] at (-0.15,4.45) {(b)};

\fill[satred!6, rounded corners=2pt] (-0.05,1.68) rectangle (4.80,3.97);
\fill[black!5, rounded corners=2pt] (-0.05,0.42) rectangle (4.80,1.56);

\node[tiny, anchor=base west] at (0.0,4.10) {depth};
\foreach \x/\name in {1.02/u_0, 1.52/u_1, 2.02/u_2, 3.02/u_{m-1}, 3.72/r_k}
  \node[lab, anchor=base] at (\x,4.10) {$\name$};
\node[lab, anchor=base] at (4.42,4.10) {$\HS$};

\node[lab, anchor=base west] at (0.0,3.70) {$\varphi$ satisfiable};
\node[tiny, text=sattext, anchor=base east] at (4.75,3.70) {one fewer tag required};
\foreach \yy/\dep/\ks/\tv in {3.42/D/0/m{+}1, 3.04/D{+}1/1/m, 2.66/D{+}2/2/m{-}1, 1.88/D{+}m/4/1}{
  \node[lab, anchor=west] at (0.0,\yy) {$\dep$};
  \foreach \c/\x in {0/1.02, 1/1.52, 2/2.02, 3/3.02}{
    \ifnum\c<\ks
      \draw[offcell] (\x-0.13,\yy-0.13) rectangle ++(0.26,0.26);
    \else
      \draw[oncell, satred, fill=satred!50] (\x-0.13,\yy-0.13) rectangle ++(0.26,0.26);
    \fi
  }
  \node[lab, text=black!75] at (2.52,\yy) {$\cdots$};
  \draw[oncell, optblue, fill=optblue!45] (3.59,\yy-0.13) rectangle ++(0.26,0.26);
  \node[lab] at (4.42,\yy) {$\tv$};
}
\foreach \x in {0.22, 3.72, 4.42}
  \node[lab, text=black!75] at (\x,2.39) {$\vdots$};
\node[lab, text=black!75] at (2.52,2.27) {$\ddots$};

\node[lab, anchor=base west] at (0.0,1.30) {$\varphi$ unsatisfiable};
\node[tiny, anchor=base east] at (4.75,1.30) {no tag is forced};
\node[lab, anchor=west] at (0.0,0.86) {$D{+}k$};
\foreach \x in {1.02, 1.52, 2.02, 3.02}
  \draw[offcell] (\x-0.13,0.73) rectangle ++(0.26,0.26);
\node[lab, text=black!75] at (2.52,0.86) {$\cdots$};
\draw[oncell, optblue, fill=optblue!45] (3.59,0.73) rectangle ++(0.26,0.26);
\node[lab] at (4.42,0.86) {$1$};

\draw[oncell, satred, fill=satred!50] (0.0,0.09) rectangle ++(0.22,0.22);
\node[tiny, anchor=base west] at (0.28,0.15) {minimal edge $\{u_j\}$};
\draw[oncell, optblue, fill=optblue!45] (2.30,0.09) rectangle ++(0.22,0.22);
\node[tiny, anchor=base west] at (2.58,0.15) {minimal edge $\{r_k\}$};
\draw[offcell] (0.0,-0.27) rectangle ++(0.22,0.22);
\node[tiny, anchor=base west] at (0.28,-0.21) {in no minimal edge};
\end{scope}
\end{tikzpicture}%
\caption{The width gap for a formula with $n$ variables and $m$ tags (\cref{thm:main}). \textbf{(a)} The safe plan reaches $S$ at depth $D$ and then needs $m+1$ actions to reach the goal. Each of the $2^n m$ complete testing executions is in $S_j$ or $T$ at the same depth and then needs $m+2$ more actions. The dashed boxes indicate which states exist for the given formula. \textbf{(b)} A filled cell marks an atom that every separating tuple at that depth must contain. Some competitor lacks that atom and contains every other atom of the safe state. In hypergraph terms, the atom forms a singleton minimal edge. At depth $D+k$, the checker is at $r_k$, with $r_0=M$. Its predecessor forces $r_k$. If the formula is satisfiable, each as-yet unchecked tag $u_j$ with $j\ge k$ is also forced. The number of filled cells in a row is the hitting number $\HS$ at that depth (\cref{lem:profile}).}
\Description{The safe branch and the branches for satisfying and falsifying assignments reach their checkpoints at the same depth. The testing branches take one more action than the safe branch to reach the common goal. The adjacent grid gives the hitting numbers at successive checker depths. For a satisfiable formula they decrease from m+1 to 1. For an unsatisfiable formula they remain 1.}

\label{fig:construction}
\end{figure}

\Cref{fig:construction} shows how to force $m+1$ atoms into a single tuple. Both branches start with the same set of $m$ tags, $U=\{u_0,\ldots,u_{m-1}\}$. The \emph{safe} branch keeps every tag and reaches the checkpoint $S=U\cup\{M\}$ after $D$ actions. The \emph{testing} branch chooses a truth assignment and an index $j<m$. At depth $D$, a satisfying assignment gives $S_j=(U\setminus\{u_j\})\cup\{M,e_j\}$, where $e_j$ records the missing tag. A falsifying assignment gives $T=U\cup\{\bar M\}$, where $\bar M$ is a separate atom and not the negation of $M$. From $S$, the checker inspects each tag and proceeds to the goal. From $S_j$ or $T$ it spends one extra action, restoring $u_j$ or replacing $\bar M$ by $M$. Every complete testing execution thus finishes one step after the safe plan.

The safe state immediately before $S$ contains every tag and lacks $M$. It forces $M$ into any tuple that separates $S$ from its competitors, even if every assignment satisfies the formula. If the formula is satisfiable, each $S_j$ is reachable and $S\setminus S_j=\{u_j\}$. These $m$ competitors force all $m$ tags. If the formula is unsatisfiable, no testing execution has reached $M$ by depth $D$, so $M$ alone separates $S$.

With zero slack, the supplied unique optimal plan lists every competitor. With one unit of slack, a polynomial-size task can encode exponentially many competitors (\cref{tab:summary}). In the first case, choosing a smallest separating tuple can still be NP-hard, but greedy hitting set gives a logarithmic approximation. The same approximation order holds for set width whenever the reachable state graph is given explicitly, even with branching optimal plans (\cref{cor:explicit}).

In the second case, such a task can represent every circuit-defined hitting-set instance whose edges are nonempty (\cref{thm:compiler}). On these tasks, choosing tuples and testing them against all competitors gives a $\SigmaTwo$-complete decision problem for an input threshold, where $\SigmaTwo$ is the second level of the polynomial hierarchy. Even a polynomial-time algorithm with an NP oracle cannot guarantee a one-sided factor $N^{1-\varepsilon}$ for any fixed $0<\varepsilon<1$, unless $\SigmaTwo\subseteq\Pclass^{\NPclass}$ (\cref{cor:oracle-approx}).

\begin{table}[!htb]
\caption{Certifying width with a supplied unique optimal plan of length $\hstar\ge2$, where $N$ is the number of atoms and $\rhoPi$ the largest slack of a reachable state. Zero slack is a promise on the input, and the zero-slack lower bounds also assume a single goal atom. For goal-admissible width, the coNP-complete and $\SigmaTwo$-complete entries assume a single reachable goal state. In ``$a$ vs.\ $b$'', factor $a$ is attained in polynomial time and factor $b$ is not unless $\Pclass=\NPclass$, with $\gamma N$ for every fixed $0<\gamma<1/7$. The $\Theta$ entries are tight under the same assumption. ETH is the exponential time hypothesis, and para-coNP-hard means coNP-hard for a fixed parameter value. The slack-one lower bounds and the zero-slack decision lower bounds hold for actions with at most two preconditions and three effects, and the zero-slack approximation lower bounds use larger actions.}

\label{tab:summary}
\small
\begin{tabular}{@{}lll@{}}
\toprule
 & Zero slack, $\rhoPi=0$ & Slack at most one, $\rhoPi\le1$\\
\midrule
Competing states & earlier states of $\pi$ (explicit) & near-optimal executions (succinct)\\
Test $w\le k$, fixed $k\ge1$ & $N^{O(k)}\operatorname{poly}(\hstar)$ (\cref{cor:zero}) & coNP-complete (\cref{cor:fixed})\\
\quad parameterized by $k$ & $\Wtwo$-hard, no $f(k)N^{o(k)}$ under ETH (\cref{cor:zero-lower}) & para-coNP-hard (\cref{cor:fixed})\\
Test $w\le k$, $k$ in the input & NP-complete (\cref{cor:zero-lower}) & $\SigmaTwo$-complete (\cref{thm:sigma2})\\
One-sided factor & $1+\ln\hstar$ vs.\ $(1-\varepsilon)\ln(\hstar-1)$ (\cref{cor:zero,cor:zero-lower}) & $N$ vs.\ $\gamma N$, $\gamma<1/7$ (\cref{cor:one-sided})\\
Symmetric factor & $\Theta(\sqrt{\log\hstar})$ (\cref{cor:zero-lower}) & $\Theta(\sqrt N)$ (\cref{cor:symmetric})\\
\bottomrule
\end{tabular}
\end{table}

Optimal plans can also branch. \citet{BonetGeffner2024} observed that tuple sets can then have smaller width than chains. The difference is at most $\xi-1$, where $\xi$ is the largest number of optimal states at a common depth that are pairwise incomparable under inclusion. This bound is attained for every set width and every $\xi$, already at zero slack (\cref{thm:branching}). \Cref{thm:recursion} gives an exact recursion for set width in this case.

\Cref{sec:hitting} proves the structural results after the definitions in \cref{sec:prelim}. \Cref{sec:reduction,sec:approx} give the gap construction and its approximation bounds. \Cref{sec:map} treats exact certification and explicit state graphs, and \cref{sec:verification} reports exhaustive checks on small instances.

\section{Background}\label{sec:prelim}

\subsection{STRIPS Tasks}

We consider deterministic STRIPS tasks with unit costs and positive preconditions. A task is a tuple $\Pi=\langle F,I,A,G\rangle$, where $F$ is a finite set of atoms, $I\subseteq F$ is the initial state, $G\subseteq F$ is the goal, and $A$ is a finite set of actions. A state is a subset of $F$, and a goal state is a state that contains $G$. Each action $a$ has a precondition $\pre(a)$, an add list $\add(a)$ and a delete list $\del(a)$, all subsets of $F$, with $\add(a)\cap\del(a)=\varnothing$. We write $a=\act{\pre(a)}{\add(a)}{\del(a)}$. Action $a$ is applicable in a state $s$ if $\pre(a)\subseteq s$, and applying it yields the successor $s[a]=(s\setminus\del(a))\cup\add(a)$. The number of atoms is $N=|F|$.

A \emph{plan}, or \emph{execution}, is a sequence of actions applicable in turn from $I$. It may end in any state. Distinct action sequences count as distinct plans even when they visit the same states. A \emph{one-action extension} of $\pi$ is a plan $\pi a$ obtained by appending one applicable action. An execution is \emph{complete} if no action is applicable in its last state.

Let $\R$ be the set of states reachable from $I$. For $s\in\R$, the depth $d(s)$ is the length of a shortest plan ending in $s$, and $d_G(s)$ is the length of a shortest action sequence from $s$ to a goal state. If no goal state is reachable from $s$, then $d_G(s)=\infty$ and $s$ is a \emph{dead end}. A \emph{goal plan} ends in a goal state. Its minimum possible length is $\hstar=d_G(I)$. An \emph{optimal goal plan}, or simply an \emph{optimal plan} of the task, is a goal plan of length $\hstar$. Throughout, the tasks are solvable and $\hstar>1$. This excludes the cases for which earlier work uses special width-zero conventions.

Positive preconditions give the following monotonicity property for every action $a$ and states $s,u$.
\begin{equation}\label{eq:monotone}
\pre(a)\subseteq s\subseteq u
\quad\Longrightarrow\quad
a\text{ is applicable in }u\text{ and }s[a]\subseteq u[a].
\end{equation}
Consequently, any action sequence that leads from $s$ to a goal state also leads to a goal state from every superset of $s$.

\subsection{Iterative Width and Novelty}\label{sec:iw}

A \emph{tuple} is a set $t\subseteq F$. Given a set $\mathcal C$ of states, the \emph{novelty} of a state $s$ with respect to $\mathcal C$ is the size of a smallest tuple $t\subseteq s$ that is contained in no state of $\mathcal C$ \citep[Definition~6]{LipovetzkyGeffner2012}. If every subset of $s$ is contained in some state of $\mathcal C$, the novelty exceeds every bound $k\le N$. The search $\IW(k)$ is breadth-first search that prunes a generated state whenever its novelty with respect to the states generated before it exceeds $k$. Every state that survives makes some tuple of at most $k$ atoms true for the first time, so $\IW(k)$ expands at most $O(N^k)$ states. \citet[Theorem~10]{LipovetzkyGeffner2012} showed that $\IW(k)$ finds an optimal plan whenever the width of the task, in the sense of \cref{def:chain} below, is at most $k$. Novelty depends on the order in which states are generated, whereas width quantifies over all optimal plans. \Cref{rem:novelty} relates the two exactly.

\subsection{Tuples and Their Optimal Achievers}

A tuple $t$ is reachable if some state in $\R$ contains it. A plan \emph{for} $t$ is a plan that ends in a state containing $t$, and an optimal plan for a reachable tuple $t$ is a shortest one. Write $c(t)$ for this minimum length, the \emph{cost} of $t$, and $\Opt(t)$ for the set of endpoints of its optimal plans. Equivalently,
\begin{equation*}
c(t)=\min\{d(s): s\in\R,\ t\subseteq s\},
\qquad
\Opt(t)=\{s\in\R: t\subseteq s,\ d(s)=c(t)\}.
\end{equation*}
We call the states in $\Opt(t)$ the \emph{optimal achievers} of $t$.

\subsection{Three Notions of Width}

The three types of \emph{certificate} are admissible tuple sets, uniform optimal-goal chains (or simply uniform chains), and goal-admissible sequences. The width of a certificate is the number of atoms in its largest tuple.

\begin{definition}[Admissible tuple set, \citeauthor{BonetGeffner2024} \citeyear{BonetGeffner2024}]\label{def:set}
A finite set $Q$ of reachable tuples is \emph{admissible} if some $t\in Q$ holds in $I$ and the following extension condition holds. For every $t\in Q$ and every optimal plan $\pi$ for $t$, either $\pi$ is an optimal goal plan or it has a one-action extension that is an optimal plan for some $t'\in Q$. The width $\ws(\Pi)$ is the minimum of $\max_{t\in Q}|t|$ over admissible sets $Q$.
\end{definition}

\begin{definition}[Uniform optimal-goal chain]\label{def:chain}
A sequence $(t_0,\dots,t_{\hstar})$ of tuples is a \emph{uniform optimal-goal chain} if $c(t_i)=i$ for all $i$, every state in $\Opt(t_i)$ has a successor in $\Opt(t_{i+1})$ for $i<\hstar$, and every state in $\Opt(t_{\hstar})$ is a goal state. The width $\wc(\Pi)$ is the minimum of $\max_i|t_i|$ over such chains.
\end{definition}
This is the tuple-graph width of \citet[Definitions~1--3]{LipovetzkyGeffner2012}. In the sequence formulation of \citet{BonetGeffner2024} and \citet{DrexlerSeippGeffner2024}, every optimal plan for the last tuple must solve the task optimally. The tuples of a uniform chain form an admissible set, so $\ws\le\wc$.

\begin{definition}[Goal-admissible sequence, \citeauthor{Song2026} \citeyear{Song2026}]\label{def:goal}
A sequence $(t_0,\dots,t_L)$ of tuples with $L\ge1$ is \emph{goal-admissible} if
(i)~$t_0\subseteq I$,
(ii)~for each $i<L$, every optimal plan for $t_i$ has a one-action extension that is an optimal plan for $t_{i+1}$, and
(iii)~every optimal plan for $t_{L-1}$ has a one-action extension that is an optimal plan for $t_L\cup G$.
The width $\wg(\Pi)$ is the minimum of $\max_i|t_i|$ over goal-admissible sequences.
\end{definition}
\citet{Song2026} call sequences satisfying (i) and (ii) admissible. Every uniform chain also satisfies (iii), since an action from each optimal achiever of $t_{\hstar-1}$ reaches a state containing $t_{\hstar}\cup G$ at depth $\hstar$. Hence $\wg\le\wc$.

The choice of the next tuple distinguishes a set from a chain. In an admissible set the choice may depend on the endpoint of the current plan. A uniform chain prescribes a single next tuple for every optimal achiever at a given depth. Goal-admissible sequences also prescribe a single next tuple, but their terminal condition concerns $t_L\cup G$. Some optimal achievers of $t_L$ can therefore be non-goal states, and $L$ can exceed $\hstar$. We write $\W=\{\ws,\wc,\wg\}$ and use $w$ for any of the three widths. \Cref{sec:hitting} identifies when the widths agree and gives examples in which they differ.

\subsection{Optimality Slack}

\begin{definition}[Optimality slack]\label{def:slack}
The slack of a state $s\in\R$ is $\slack(s)=d(s)+d_G(s)-\hstar$, and the slack of the task is $\rhoPi=\max_{s\in\R}\slack(s)$.
\end{definition}
A state has slack zero exactly when it lies on an optimal goal plan, and a dead end has infinite slack. Under the promise $\rhoPi\le1$, every reachable state lies on a goal-reaching execution of length at most $\hstar+1$.

\subsection{Hypergraphs, Hitting Sets and Succinct Representations}\label{sec:hs}

A \emph{hypergraph} over a finite set $U$ is a finite family $\Hy$ of subsets of $U$, called edges. A set $t\subseteq U$ \emph{hits} $\Hy$ if it meets every edge, and the \emph{hitting number} $\HS(\Hy)$ is the minimum size of such a set, also known as the transversal number. An edge is \emph{minimal} if no other edge of $\Hy$ is a proper subset of it. Every edge contains a minimal edge, so a set hits $\Hy$ exactly when it meets every minimal edge. We set $\HS(\Hy)=\infty$ when $\varnothing\in\Hy$. Deciding $\HS(\Hy)\le k$ is NP-complete \citep{Karp1972}, since hitting set is set cover in dual form. If $\Hy$ is nonempty and has no empty edge, the greedy algorithm repeatedly picks an element that hits the largest number of remaining edges. It returns a hitting set of size at most $(1+\ln|\Hy|)\,\HS(\Hy)$ \citep{Chvatal1979}.

A hypergraph can also be given by a circuit. We use Boolean circuits with fan-in-two AND and OR gates whose inputs and outputs may be negated. Let $C$ be such a circuit with $n_C$ inputs and one output for each element of a set $U$. For $v\in\{0,1\}^{n_C}$, the edge $E_v$ is the set of elements whose outputs evaluate to $1$ on input $v$, and $\Hy_C=\{E_v: v\in\{0,1\}^{n_C}\}$ is the hypergraph \emph{represented succinctly} by $C$. It can have exponentially many edges in the size of $C$.

A language $\mathcal{L}$ is in $\SigmaTwo$ if and only if there are a polynomial $p$ and a polynomial-time predicate $\mathrm{Ver}$ such that $x\in\mathcal{L}$ exactly when some $y$ with $|y|\le p(|x|)$ satisfies $\mathrm{Ver}(x,y,z)$ for all $z$ with $|z|\le p(|x|)$. Deciding $\HS(\Hy_C)\le k$ is in $\SigmaTwo$, with a hitting set as $y$ and an input of $C$ as $z$. Succinct set cover, in which the sets are given by DNF formulas, is $\SigmaTwo$-complete \citep{Umans1999,SchaeferUmans2002} and reduces to this problem by duality. \Cref{lem:qbf} gives a self-contained proof of hardness for the instances we need. Truth of sentences $\exists X\,\forall Y\,\Phi(X,Y)$ with $\Phi$ in 3-DNF is $\SigmaTwo$-complete \citep{Stockmeyer1976,Wrathall1976}. For parameterized complexity we use the class $\Wtwo$, for which hitting set parameterized by $k$ is complete \citep{DowneyFellows2013}, and the exponential time hypothesis (ETH), which states that 3-SAT on $n$ variables cannot be solved in time $2^{o(n)}$.

\section{Certificates and Competing States}\label{sec:hitting}

The states on any optimal goal plan give a certificate. When all optimal plans visit the same states, the size of a smallest certificate is determined by the other states reachable at each depth.

\subsection{Certificates Along an Optimal Plan}\label{sec:free}

\begin{lemma}[Optimal-prefix containment]\label{lem:containment}
Let $Q$ be admissible. Every optimal plan for a tuple in $Q$ is a prefix of an optimal goal plan, and $Q$ contains a tuple of each cost $0,1,\dots,\hstar$. If all optimal goal plans visit the same state sequence $s_0,\dots,s_{\hstar}$, then every $t\in Q$ with $c(t)=i$ has $\Opt(t)=\{s_i\}$.
\end{lemma}
\begin{proof}
Let $\pi$ be an optimal plan for some $t\in Q$, so that $|\pi|=c(t)$. If $\pi$ is not an optimal goal plan, admissibility gives an action $a$ such that $\pi a$ is an optimal plan for some $t'\in Q$, and then $c(t')=|\pi a|=c(t)+1$. Repeating this step produces optimal plans $\pi,\pi a,\dots$ for tuples of $Q$ whose lengths increase by one at each step. The lengths belong to the finite set $\{c(u):u\in Q\}$, so this process terminates. Admissibility implies that the last plan is an optimal goal plan. Hence $\pi$ is a prefix of an optimal goal plan. Some tuple of $Q$ holds in $I$, so the same argument can start with the empty plan. The resulting sequence realizes each cost $0,1,\dots,\hstar$ in $Q$. Finally, assume that all optimal goal plans visit $s_0,\dots,s_{\hstar}$, and let $t\in Q$ with $c(t)=i$. Every state of $\Opt(t)$ is the endpoint of an optimal plan for $t$, which is a length-$i$ prefix of an optimal goal plan and therefore ends in $s_i$. Hence $\Opt(t)=\{s_i\}$.
\end{proof}

\begin{proposition}[The plan certifies itself]\label{prop:states}
Let $s_0,\dots,s_{\hstar}$ be the states visited by an optimal goal plan. Then $(s_0,\dots,s_{\hstar})$ is a uniform optimal-goal chain, and for every $w\in\W$
\begin{equation}\label{eq:trivial}
1\le w(\Pi)\le\max_{0\le i\le\hstar}|s_i|\le N.
\end{equation}
\end{proposition}
\begin{proof}
Let $a_1,\dots,a_{\hstar}$ be the plan. If $s_i\subseteq u$ with $d(u)<i$, then by \cref{eq:monotone} the actions $a_{i+1},\dots,a_{\hstar}$ lead from $u$ to a superset of $s_{\hstar}$, which is a goal state. Together with a shortest plan to $u$ this gives a goal plan of length $d(u)+\hstar-i<\hstar$, a contradiction. Hence $c(s_i)=i$, and $\Opt(s_i)$ consists of the supersets of $s_i$ at depth $i$. For such a state $u$ and $i<\hstar$, the successor $u[a_{i+1}]\supseteq s_{i+1}$ has depth at most $i+1$, hence exactly $c(s_{i+1})=i+1$, so it lies in $\Opt(s_{i+1})$. States in $\Opt(s_{\hstar})$ contain $s_{\hstar}\supseteq G$. This proves the chain property, and the upper bounds follow from $\ws\le\wc$ and $\wg\le\wc$.

For the lower bound, it suffices that every certificate contains a tuple of cost $1$, because such a tuple is nonempty (the empty tuple has cost $0$). An admissible set contains one by \cref{lem:containment}, and a uniform chain contains $t_1$. In a goal-admissible sequence $(t_0,\dots,t_L)$, conditions~(i) and~(ii) of \cref{def:goal} give $c(t_0)=0$ and $c(t_{i+1})=c(t_i)+1$ for $i<L$, so $c(t_i)=i$, and condition~(iii) gives $c(t_L\cup G)=L$. Every state containing $G$ is a goal state and has depth at least $\hstar$, so $L\ge\hstar\ge2$, and $t_1$ has cost $1$.
\end{proof}

\citet[footnote~2]{Song2026} observe that the width of a solvable task never exceeds its number of atoms. \Cref{prop:states} makes this bound constructive for all three notions. To improve on this bound, a certificate must replace states by smaller tuples.

\subsection{A Hitting-Set Characterization}\label{sec:characterization}

Assume that all optimal goal plans visit the same states $s_0,\dots,s_{\hstar}$. By \cref{lem:containment}, a certificate tuple of cost $i$ in an admissible set has $s_i$ as its only optimal achiever. Such a tuple must distinguish $s_i$ from every other state reachable in at most $i$ steps. The next theorem shows that these distinctions suffice, and that a tuple can be chosen independently at each depth.

\begin{theorem}[Competing-state characterization]\label{thm:hitting}
Suppose that all optimal goal plans of $\Pi$ visit the same state sequence $s_0,\dots,s_{\hstar}$, with $\hstar>1$. For $1\le i\le\hstar$, let
\begin{equation}\label{eq:H}
\Hy_i=\{s_i\setminus u: u\in\R,\ d(u)\le i,\ u\ne s_i\}.
\end{equation}
Every $\Hy_i$ is a nonempty family of nonempty sets, and
\begin{equation}\label{eq:hitting}
\ws(\Pi)=\wc(\Pi)=\max_{1\le i\le\hstar}\HS(\Hy_i).
\end{equation}
\end{theorem}
We call the states $u$ in \cref{eq:H} the \emph{competitors} of $s_i$, and $\Hy_i$ the \emph{competitor hypergraph} at depth $i$. A tuple $t\subseteq s_i$ is contained in a competitor $u$ exactly when it misses the edge $s_i\setminus u$, so $t$ hits $\Hy_i$ exactly when $s_i$ is the only state of depth at most $i$ that contains $t$. As $d(s_i)=i$, the hitting sets of $\Hy_i$ contained in $s_i$ are exactly the tuples $t\subseteq s_i$ with $c(t)=i$ and $\Opt(t)=\{s_i\}$.
\begin{proof}
Let $a_1,\dots,a_{\hstar}$ be an optimal goal plan, so that $s_i=s_{i-1}[a_i]$. Each $\Hy_i$ contains $s_i\setminus s_0$ and is therefore nonempty. Suppose that some edge is empty, that is, $s_i\subseteq u\ne s_i$ with $d(u)\le i$. By \cref{eq:monotone}, $a_{i+1},\dots,a_{\hstar}$ lead from $u$ to a goal state, and a shortest plan to $u$ followed by these actions is a goal plan of length $d(u)+\hstar-i$. If $d(u)<i$, this plan is shorter than optimal. If $d(u)=i$, it is an optimal goal plan whose $i$-th state is $u\ne s_i$. Both are impossible, so every edge is nonempty.

For the lower bound, let $Q$ be admissible and $1\le i\le\hstar$. By \cref{lem:containment}, $Q$ contains a tuple $t_i$ with $c(t_i)=i$ and $\Opt(t_i)=\{s_i\}$. If $t_i\subseteq u$ for some $u\ne s_i$ with $d(u)\le i$, then $c(t_i)<i$ or $u\in\Opt(t_i)$, and both are impossible. Hence $t_i$ meets every edge of $\Hy_i$, and $\max_{t\in Q}|t|\ge\HS(\Hy_i)$. Since $Q$ and $i$ are arbitrary, $\ws\ge\max_i\HS(\Hy_i)$. Uniqueness enters this lower bound through \cref{lem:containment}. When several optimal states are possible at depth $i$, the same tuple may serve more than one of them (\cref{rem:unique}).

For the upper bound, let $t_0=\varnothing$ and, for $i\ge1$, let $t_i$ be a minimum hitting set of $\Hy_i$. Every edge of $\Hy_i$ is a subset of $s_i$, so we may take $t_i\subseteq s_i$. No state of depth at most $i$ other than $s_i$ contains $t_i$, so $c(t_i)=i$ and $\Opt(t_i)=\{s_i\}$. Also $\Opt(t_0)=\{s_0\}$. Since $a_{i+1}$ maps $s_i$ to $s_{i+1}$ and $s_{\hstar}$ is a goal state, $(t_0,\dots,t_{\hstar})$ is a uniform chain of width $\max_i\HS(\Hy_i)$. So $\wc\le\max_i\HS(\Hy_i)$, and $\ws\le\wc$ gives \cref{eq:hitting}.
\end{proof}

Positive preconditions are used only to show that edges are nonempty. With negative preconditions the same proof gives \cref{eq:hitting} under the convention $\HS(\Hy)=\infty$ when $\varnothing\in\Hy$, in which case no admissible set or uniform chain exists.

\begin{remark}[Uniqueness of the optimal states]\label{rem:unique}
Uniqueness of the optimal state sequence cannot be dropped from \cref{eq:hitting}. If actions lead from $I=\{s\}$ to each of $\{x,y\}$, $\{x,z\}$ and $\{y,z\}$, and each of these reaches $G=\{g\}$ with one more action, then $(\{s\},\{x\},\{g\})$ is a uniform chain and $\ws=\wc=1$, whereas $\HS(\Hy_1)=2$ along each of the three optimal state sequences. Evaluating \cref{eq:hitting} along an arbitrarily chosen optimal trajectory can therefore overestimate the width, because a single tuple may serve several optimal states. The hypothesis allows several optimal action sequences, provided that they visit the same states. \Cref{sec:branching} treats tasks whose optimal plans branch.
\end{remark}

\begin{remark}[Novelty]\label{rem:novelty}
At depth $i$, $\HS(\Hy_i)$ is the novelty of $s_i$ relative to all other states of depth at most $i$ (\cref{sec:iw}). Thus \cref{thm:hitting} gives set and chain width as the largest such novelty along the optimal state sequence. The novelty that a planner actually observes also depends on which states it prunes and on their generation order.
\end{remark}

\subsection{Goal-Admissible Width}\label{sec:goal}

The terminal condition for $\wg$ allows the last extension to reach any goal state. If only one goal state is reachable, the last extension must reach it at its minimum depth. The following lemma fixes the length of the sequence in this case.

\begin{lemma}[One reachable goal state]\label{lem:weak}
If a task with $\hstar>1$ has exactly one reachable goal state, then every goal-admissible sequence $(t_0,\dots,t_L)$ has $L=\hstar$, and for $i<L$ every optimal plan for $t_i$ is a prefix of an optimal goal plan.
\end{lemma}
\begin{proof}
Condition~(ii) of \cref{def:goal} gives $c(t_{i+1})=c(t_i)+1$ for $i<L$, and condition~(i) gives $c(t_0)=0$, so $c(t_i)=i$ for all $i\le L$. Condition~(iii) gives $c(t_L\cup G)=L$. The only reachable state containing $G$ is the goal state, whose depth is $\hstar$, so $L=\hstar$. For $i<L$, an optimal plan for $t_i$ extends by condition~(ii) to an optimal plan for $t_{L-1}$ and then by condition~(iii) to a plan of length $\hstar$ that ends in a goal state. Hence it is a prefix of an optimal goal plan.
\end{proof}

\begin{proposition}[Terminal rule]\label{prop:terminal}
Under the hypotheses of \cref{thm:hitting}, suppose that $s_{\hstar}$ is the only reachable goal state, and let $\Hy^{<}_{\hstar}=\{s_{\hstar}\setminus u: u\in\R,\ d(u)<\hstar\}$. Then
\begin{equation}\label{eq:terminal}
\wg(\Pi)=\max\Bigl(\max_{1\le i<\hstar}\HS(\Hy_i),\ \HS(\Hy^{<}_{\hstar})\Bigr)\le\ws(\Pi)=\wc(\Pi).
\end{equation}
In particular, $\wg=\ws=\wc$ if $|G|=1$ or $\rhoPi=0$.
\end{proposition}
\begin{proof}
Let $(t_0,\dots,t_L)$ be goal-admissible. By \cref{lem:weak}, $L=\hstar$, and for $i<\hstar$ every optimal plan for $t_i$ is a prefix of an optimal goal plan. Hence $c(t_i)=i$ and $\Opt(t_i)=\{s_i\}$, and as in the proof of \cref{thm:hitting}, $t_i$ meets every edge of $\Hy_i$. Condition~(iii) extends an optimal plan for $t_{\hstar-1}$ to an optimal plan for $t_{\hstar}\cup G$ of length $\hstar$, which ends in a goal state of depth $\hstar$, that is, in $s_{\hstar}$. Hence $t_{\hstar}\subseteq s_{\hstar}$. Condition~(ii) gives $c(t_{\hstar})=\hstar$, so $t_{\hstar}$ is contained in no $u$ with $d(u)<\hstar$ and meets every edge of $\Hy^{<}_{\hstar}$. Taking the maximum of these lower bounds gives $\wg\ge\max\{\max_{1\le i<\hstar}\HS(\Hy_i),\HS(\Hy^{<}_{\hstar})\}$.

Conversely, take minimum hitting sets $t_i\subseteq s_i$ of $\Hy_i$ for $1\le i<\hstar$, a minimum hitting set $t_{\hstar}\subseteq s_{\hstar}$ of $\Hy^{<}_{\hstar}$, whose edges are nonempty because no state of depth less than $\hstar$ contains $G$, and $t_0=\varnothing$. For $i<\hstar$, $\Opt(t_i)=\{s_i\}$ as in \cref{thm:hitting}, and $c(t_{\hstar})=\hstar$. Fix an optimal goal plan $a_1,\dots,a_{\hstar}$. Every optimal plan for $t_i$ with $i<\hstar$ ends in $s_i$, and appending $a_{i+1}$ reaches $s_{i+1}\supseteq t_{i+1}$ at depth $i+1=c(t_{i+1})$. For $i=\hstar-1$, the state $s_{\hstar}$ also contains $t_{\hstar}\cup G$, and $c(t_{\hstar}\cup G)=\hstar$. So the sequence is goal-admissible, which proves the reverse inequality. The comparison with $\ws=\wc$ follows from $\Hy^{<}_{\hstar}\subseteq\Hy_{\hstar}$ and \cref{thm:hitting}.

If $G=\{g\}$, then $\{g\}$ meets every edge of $\Hy_{\hstar}$, because $s_{\hstar}$ is the only reachable state that contains $g$. Hence $\HS(\Hy^{<}_{\hstar})\le\HS(\Hy_{\hstar})\le1\le\HS(\Hy_1)$, the last inequality because $s_1\setminus s_0\in\Hy_1$. Both sides of \cref{eq:terminal} then equal $\max_{1\le i<\hstar}\HS(\Hy_i)=\max_{1\le i\le\hstar}\HS(\Hy_i)$.

Finally, suppose $\rhoPi=0$ and fix $u\in\R$. A shortest plan to $u$ followed by a shortest sequence from $u$ to a goal state has length $d(u)+d_G(u)=\hstar+\slack(u)=\hstar$. It is therefore an optimal goal plan, and $u=s_{d(u)}$ by hypothesis. Hence $\R=\{s_0,\dots,s_{\hstar}\}$ with $d(s_j)=j$. The states other than $s_{\hstar}$ of depth at most $\hstar$ then all have depth less than $\hstar$, so $\Hy^{<}_{\hstar}=\Hy_{\hstar}$.
\end{proof}

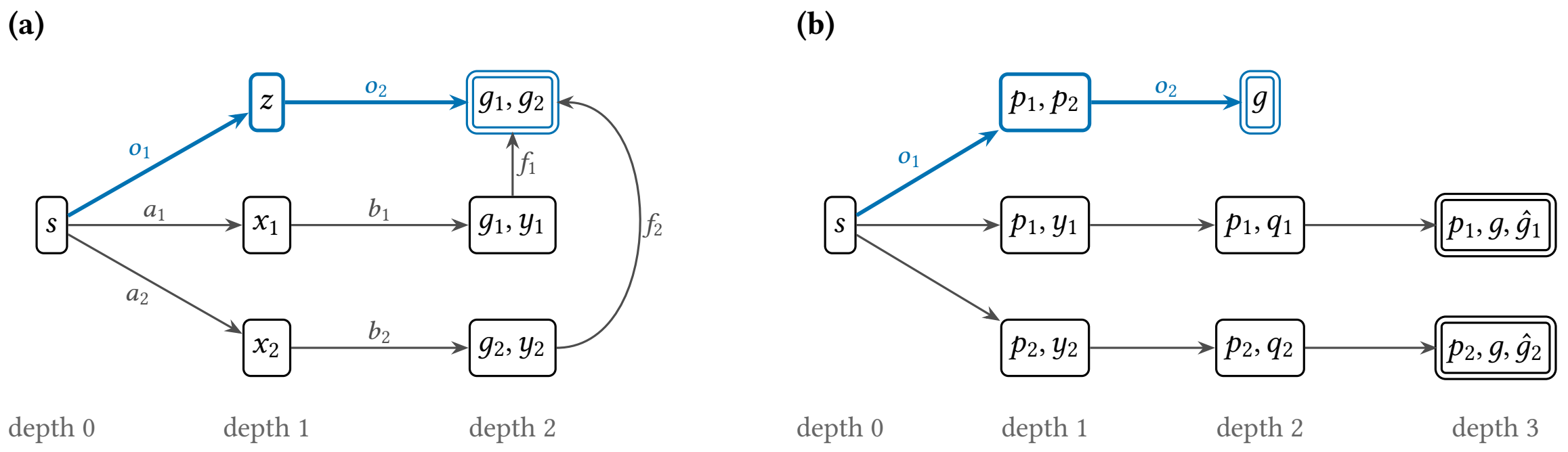
\begin{figure}[!htb]
\centering
\begin{tikzpicture}[x=1cm,y=1cm,
  st/.style={draw, rounded corners=2pt, minimum height=0.55cm, inner sep=2.5pt,
             font=\small, fill=white, line width=0.6pt},
  goal/.style={double, double distance=1.1pt},
  opt/.style={-{Stealth[length=1.9mm,width=1.5mm]}, line width=1.1pt, optblue},
  alt/.style={-{Stealth[length=1.7mm,width=1.3mm]}, line width=0.6pt, black!70},
  lab/.style={font=\scriptsize, inner sep=1pt},
  dep/.style={font=\footnotesize, text=black!60},
]
\node[font=\small\bfseries, anchor=west] at (-0.55,1.95) {(a)};
\node[st] (s)   at (0,0)      {$s$};
\node[st, draw=optblue, line width=1pt] (b) at (2.1,1.2) {$z$};
\node[st] (c1)  at (2.1,0)    {$x_1$};
\node[st] (c2)  at (2.1,-1.2) {$x_2$};
\node[st, goal, draw=optblue] (G) at (4.5,1.2) {$g_1,g_2$};
\node[st] (gy1) at (4.5,0)    {$g_1,y_1$};
\node[st] (gy2) at (4.5,-1.2) {$g_2,y_2$};
\draw[opt] (s) -- (b) node[lab, midway, above left] {$o_1$};
\draw[opt] (b) -- (G) node[lab, midway, above] {$o_2$};
\draw[alt] (s) -- (c1) node[lab, midway, above] {$a_1$};
\draw[alt] (s) -- (c2) node[lab, midway, below left] {$a_2$};
\draw[alt] (c1) -- (gy1) node[lab, midway, above] {$b_1$};
\draw[alt] (c2) -- (gy2) node[lab, midway, above] {$b_2$};
\draw[alt] (gy1) -- (G) node[lab, midway, right] {$f_1$};
\draw[alt] (gy2.east) .. controls (6.0,-1.2) and (6.0,1.2) .. (G.east)
  node[lab, pos=0.5, right] {$f_2$};
\foreach \x/\d in {0/0, 2.1/1, 4.5/2}
  \node[dep] at (\x,-2.0) {depth $\d$};

\begin{scope}[xshift=7.7cm]
\node[font=\small\bfseries, anchor=west] at (-0.55,1.95) {(b)};
\node[st] (s)    at (0,0)      {$s$};
\node[st, draw=optblue, line width=1pt] (p12) at (2.0,1.2) {$p_1,p_2$};
\node[st] (py1)  at (2.0,0)    {$p_1,y_1$};
\node[st] (py2)  at (2.0,-1.2) {$p_2,y_2$};
\node[st, goal, draw=optblue] (g) at (4.1,1.2) {$g$};
\node[st] (pq1)  at (4.1,0)    {$p_1,q_1$};
\node[st] (pq2)  at (4.1,-1.2) {$p_2,q_2$};
\node[st, goal] (h1) at (6.4,0)    {$p_1,g,\hat g_1$};
\node[st, goal] (h2) at (6.4,-1.2) {$p_2,g,\hat g_2$};
\draw[opt] (s) -- (p12) node[lab, midway, above left] {$o_1$};
\draw[opt] (p12) -- (g) node[lab, midway, above] {$o_2$};
\draw[alt] (s) -- (py1);
\draw[alt] (s) -- (py2);
\draw[alt] (py1) -- (pq1);
\draw[alt] (py2) -- (pq2);
\draw[alt] (pq1) -- (h1);
\draw[alt] (pq2) -- (h2);
\foreach \x/\d in {0/0, 2.0/1, 4.1/2, 6.4/3}
  \node[dep] at (\x,-2.0) {depth $\d$};
\end{scope}
\end{tikzpicture}
\caption{Reachable state graphs of \cref{ex:two-goal} (a) and \cref{ex:goal-states} (b), with states drawn in the column of their depth. The unique optimal plan is drawn with thick arrows, and goal states have a double border. In (a) the alternative executions reach the optimal goal state one action late. In (b) they end in two further goal states.}
\Description{Two small directed graphs drawn left to right. Panel a: the state s branches into z, x_1 and x_2. Thick arrows o_1 and o_2 lead from s to z and from z to the goal state containing g_1 and g_2. The thin path through x_1 reaches the state with g_1 and y_1, and the thin path through x_2 reaches the state with g_2 and y_2. Arrows f_1 and f_2 then lead from these two states back to the goal state with g_1 and g_2. Panel b: the state s branches into the state with p_1 and p_2, the state with p_1 and y_1, and the state with p_2 and y_2. Thick arrows o_1 and o_2 lead through the state with p_1 and p_2 to the goal state containing only g. Each of the two other branches continues through the state with p_k and q_k to a separate goal state containing p_k, g and hat g_k, at depth 3.}
\label{fig:examples}
\end{figure}

Two tasks with a unique optimal plan of length $2$ illustrate the terminal condition (\cref{fig:examples}). Both have slack one and satisfy $\wg<\ws=\wc$. In \cref{ex:two-goal}, only one goal state is reachable, but the goal has two atoms. A singleton terminal tuple also occurs in a non-goal state at depth $2$. In \cref{ex:goal-states}, the goal consists of one atom and several goal states are reachable. A goal-admissible sequence can end at one of the later goal states.

\begin{example}[Two-atom goal]\label{ex:two-goal}
Let $I=\{s\}$, $G=\{g_1,g_2\}$, $o_1=\act{\{s\}}{\{z\}}{\{s\}}$, $o_2=\act{\{z\}}{\{g_1,g_2\}}{\{z\}}$ and, for $j\in\{1,2\}$, $a_j=\act{\{s\}}{\{x_j\}}{\{s\}}$, $b_j=\act{\{x_j\}}{\{g_j,y_j\}}{\{x_j\}}$ and $f_j=\act{\{y_j\}}{\{g_{3-j}\}}{\{y_j\}}$. The unique optimal plan is $o_1o_2$ with $\hstar=2$. The other two executions reach $\{g_1,g_2\}$ after three actions, so there is one reachable goal state, no dead end, and $\rhoPi=1$. The depth-$2$ states $\{g_1,y_1\}$ and $\{g_2,y_2\}$ contribute the singleton edges $\{g_2\}$ and $\{g_1\}$ to $\Hy_2$, so $\ws=\wc=2$. Yet $\HS(\Hy_1)=\HS(\Hy^{<}_2)=1$ and $\wg=1$, with witness $(\{s\},\{z\},\{g_1\})$: the plan $o_1$ is the only optimal plan for $\{z\}$, and its extension $o_1o_2$ is an optimal plan both for $\{g_1\}$ and for $\{g_1\}\cup G$. The other optimal plan for $\{g_1\}$, which ends in $\{g_1,y_1\}$, is not constrained.
\end{example}

\begin{example}[Several goal states]\label{ex:goal-states}
Let $I=\{s\}$, $G=\{g\}$, $o_1=\act{\{s\}}{\{p_1,p_2\}}{\{s\}}$ and $o_2=\act{\{p_1,p_2\}}{\{g\}}{\{p_1,p_2\}}$. For $j\in\{1,2\}$, add the actions $\act{\{s\}}{\{p_j,y_j\}}{\{s\}}$, $\act{\{y_j\}}{\{q_j\}}{\{y_j\}}$ and $\act{\{q_j\}}{\{g,\hat g_j\}}{\{q_j\}}$. The unique optimal plan is $o_1o_2$ with $\hstar=2$, there are no dead ends, and $\rhoPi=1$, but $\{p_1,g,\hat g_1\}$ and $\{p_2,g,\hat g_2\}$ are further goal states. The depth-$1$ states $\{p_1,y_1\}$ and $\{p_2,y_2\}$ contribute the singleton edges $\{p_2\}$ and $\{p_1\}$ to $\Hy_1$, so $\ws=\wc=2$. However, $(\{s\},\{y_1\},\{q_1\},\{\hat g_1\})$ is goal-admissible with $L=3>\hstar$: its last three tuples have the respective unique optimal achievers $\{p_1,y_1\}$, $\{p_1,q_1\}$ and $\{p_1,g,\hat g_1\}$, the last of which contains $\{\hat g_1\}\cup G$. Hence $\wg=1<\HS(\Hy_1)$.
\end{example}

No goal-admissible sequence of width $1$ in \cref{ex:goal-states} has length $L=\hstar=2$. Its tuple $t_1$ would have cost $1$ and a single atom, so it would be $\{p_1\}$, $\{p_2\}$, $\{y_1\}$ or $\{y_2\}$. Each of these tuples has an optimal plan that ends in $\{p_j,y_j\}$ for some $j$, and no action leads from that state to a goal state. Algorithm~1 of \citet{Song2026} computes the optimal plan length $\ell$ in line~2 and, in line~6, guesses only sequences $(t_0,\dots,t_\ell)$ of that length. On this task, such sequences miss every certificate of width $1$, although the width defined in their Definitions~2 and~4 is $1$. If only one goal state is reachable, every goal-admissible sequence has $L=\hstar$, so the restriction matters only when several goal states are reachable. The example does not by itself bear on their PSPACE-completeness result, and we do not examine their separate proof for fixed thresholds.

\citet[p.~305]{Song2026} suggest a correspondence between the conjunctions in a goal-admissible sequence and admissible tuple sets. In \cref{ex:two-goal}, the sequence $(\{s\},\{z\},\{g_1\})$ is goal-admissible, yet its tuples fail the extension condition for an admissible set. An optimal plan for $\{g_1\}$ can end at $\{g_1,y_1\}$ at depth $2$. This state is not a goal state, and the set contains no tuple of cost $3$ to continue the plan. The complexity results of \citet{Song2026} concern the goal-admissible width defined in their paper.

The distinction also matters for search. The search $\IW(k)$ finds an optimal plan whenever $\ws\le k$ \citep[Theorem~6]{BonetGeffner2024}. In \cref{ex:goal-states}, $\wg=1$ while $\ws=\wc=2$, so $\wg\le k$ alone does not give this guarantee. Indeed, suppose that $\IW(1)$ generates $\{p_1,y_1\}$ and $\{p_2,y_2\}$ before $\{p_1,p_2\}$. Both atoms of $\{p_1,p_2\}$ have then appeared, so $\IW(1)$ prunes this state and returns a plan of length $3$.

\subsection{Several Optimal Trajectories}\label{sec:branching}

\Cref{thm:hitting} assumes that all optimal goal plans visit the same states, and \cref{rem:unique} shows that \cref{eq:hitting} can fail otherwise. To bound the widths without this assumption, we group the optimal states by depth. For $0\le i\le\hstar$, the \emph{optimal layer} $\mathcal O_i$ is the set of states reached after $i$ actions by some optimal goal plan. Every $s\in \mathcal O_i$ has $d(s)=i$, since a shorter plan to $s$ followed by the rest of the optimal goal plan would reach the goal in fewer than $\hstar$ actions. Two states are \emph{incomparable} if neither is a subset of the other, and an \emph{antichain} is a set of pairwise incomparable states. The \emph{optimal antichain number} $\xi(\Pi)$ is the largest size of an antichain contained in a single optimal layer. Thus $1\le\xi(\Pi)\le\max_i|\mathcal O_i|$, and $\xi(\Pi)=1$ exactly when every optimal layer is linearly ordered by inclusion, in particular when all optimal goal plans visit the same states.

\begin{theorem}[Several optimal trajectories]\label{thm:branching}
\begin{enumerate}\renewcommand{\labelenumi}{(\roman{enumi})}
\item Every task satisfies $\ws(\Pi)\le\wc(\Pi)\le\ws(\Pi)+\xi(\Pi)-1$.
\item For all integers $k,n\ge1$, there is a task $\Pi$ with zero slack, a single reachable goal state, optimal plan length $k+2$ and $\xi(\Pi)=n$, in which $\ws(\Pi)=k$ and $\wc(\Pi)=\wg(\Pi)=k+n-1$.
\end{enumerate}
\end{theorem}

The upper bound in (i) turns an admissible set into a uniform chain. For each optimal depth $i$, we pick a tuple $t_i$ of the set and an inclusion-minimal optimal achiever $m_i$ of $t_i$, its \emph{representative}, such that some action applicable in $m_i$ leads to a superset of $m_{i+1}$. Adding at most $\xi(\Pi)-1$ atoms to $t_i$ ensures that all its remaining optimal achievers contain $m_i$. Monotonicity then allows the same action to continue from each of them.

\begin{proof}[Proof of (i)]
The first inequality holds because the tuples of any uniform chain form an admissible set. For the second, let $Q$ be an admissible set with $\max_{t\in Q}|t|=\ws(\Pi)=k$. By \cref{lem:containment}, every optimal plan for a tuple $t\in Q$ is a prefix of an optimal goal plan, so every optimal achiever of $t$ lies in the optimal layer $\mathcal O_{c(t)}$.

\emph{Representatives.} We choose tuples $t_i\in Q$ with $c(t_i)=i$ and states $m_i\in\Opt(t_i)$ that are inclusion-minimal in $\Opt(t_i)$, for $i=0,\dots,\hstar$, together with actions $a_{i+1}$ for $i<\hstar$ such that $a_{i+1}$ is applicable in $m_i$ and $m_i[a_{i+1}]\supseteq m_{i+1}$. Let $t_0\in Q$ hold in $I$ and $m_0=I$, the only state of depth $0$. Given $t_i$ and $m_i$ with $i<\hstar$, a shortest plan to $m_i$ is an optimal plan for $t_i$. It has $i<\hstar$ actions and is therefore not an optimal goal plan, so by \cref{def:set} it has a one-action extension, by some action $a_{i+1}$, that is an optimal plan for some $t_{i+1}\in Q$. Hence $s'=m_i[a_{i+1}]\in\Opt(t_{i+1})$ and $c(t_{i+1})=i+1$. Let $m_{i+1}$ be an inclusion-minimal state of $\Opt(t_{i+1})$ contained in $s'$. It is inclusion-minimal in all of $\Opt(t_{i+1})$, because a state of $\Opt(t_{i+1})$ strictly inside $m_{i+1}$ would also lie inside $s'$. The state $m_{i+1}$ may be a proper subset of $s'$, so consecutive representatives need not be consecutive states of one execution. The inclusion $m_i[a_{i+1}]\supseteq m_{i+1}$ suffices for the construction below.

\emph{Refined tuples.} Fix $i$, let $\mathcal X_i$ be the set of states of $\Opt(t_i)$ that do not contain $m_i$, and let $\mathcal X_i^{\max}$ consist of the inclusion-maximal members of $\mathcal X_i$. The members of $\mathcal X_i^{\max}$ are pairwise incomparable. Each $u\in\mathcal X_i^{\max}$ is also incomparable with $m_i$: $m_i\not\subseteq u$ by the definition of $\mathcal X_i$, and $u\subsetneq m_i$ would contradict the minimality of $m_i$ in $\Opt(t_i)$. Hence $\mathcal X_i^{\max}\cup\{m_i\}$ is an antichain in $\mathcal O_i$, and $|\mathcal X_i^{\max}|\le\xi(\Pi)-1$. For each $u\in\mathcal X_i^{\max}$ choose an atom of $m_i\setminus u$, and let $t'_i$ consist of $t_i$ and these atoms. Then $t_i\subseteq t'_i\subseteq m_i$ and $|t'_i|\le k+\xi(\Pi)-1$. We claim that
\begin{equation}\label{eq:refined}
c(t'_i)=i
\qquad\text{and}\qquad
\Opt(t'_i)=\{s\in\Opt(t_i):m_i\subseteq s\}.
\end{equation}
Indeed, $t'_i\subseteq m_i$ and $d(m_i)=i$ give $c(t'_i)\le i$, and $t_i\subseteq t'_i$ gives $c(t'_i)\ge c(t_i)=i$. The optimal achievers of $t'_i$ are therefore the states of $\Opt(t_i)$ that contain $t'_i$. Every state that contains $m_i$ contains $t'_i$. Conversely, a state $s\in\Opt(t_i)$ that does not contain $m_i$ lies in $\mathcal X_i$, hence in some $u\in\mathcal X_i^{\max}$, and $t'_i$ contains an atom of $m_i\setminus u$, which is missing from $s$.

\emph{The chain.} We show that $(t'_0,\dots,t'_{\hstar})$ is a uniform chain. By \cref{eq:refined}, $c(t'_i)=i$. Let $i<\hstar$ and $s\in\Opt(t'_i)$, so that $m_i\subseteq s$ by \cref{eq:refined}. The action $a_{i+1}$ is applicable in $m_i$, so by \cref{eq:monotone} it is applicable in $s$, and
\begin{equation*}
s[a_{i+1}]\supseteq m_i[a_{i+1}]\supseteq m_{i+1}\supseteq t'_{i+1}.
\end{equation*}
As $d(s[a_{i+1}])\le i+1=c(t'_{i+1})$, the successor $s[a_{i+1}]$ is an optimal achiever of $t'_{i+1}$. Finally, $\Opt(t'_{\hstar})\subseteq\Opt(t_{\hstar})\subseteq \mathcal O_{\hstar}$, and the states of $\mathcal O_{\hstar}$ are goal states. The chain has width at most $k+\xi(\Pi)-1$.
\end{proof}

In the tasks of part~(ii), an admissible set can choose the next tuple according to the optimal achiever that a plan has reached, whereas a uniform chain must use one next tuple for all optimal achievers of its tuple of the same cost. At the branching depth, $k$ atoms distinguish the layer from the earlier states. A chain needs another $n-1$ atoms to distinguish one branch within that layer.

\begin{proof}[Proof of (ii)]
Let $P=\{p_1,\dots,p_k\}$ and $Y=\{y_1,\dots,y_n\}$, and let $q_1,\dots,q_k$, $x_1,\dots,x_n$ and $g$ be further atoms. Let
\begin{equation*}
J_j=Y\cup(P\setminus\{p_j\})\cup\{q_j\}\quad(1\le j\le k),\qquad
S_\ell=P\cup(Y\setminus\{y_\ell\})\quad(1\le\ell\le n).
\end{equation*}
The initial state is $J_1$ and the goal is $G=\{g\}$. For $1\le j<k$, the action $\act{\{q_j\}}{\{p_j,q_{j+1}\}}{\{p_{j+1},q_j\}}$ takes $J_j$ to $J_{j+1}$. For $1\le\ell\le n$, the actions $\act{\{q_k\}}{\{p_k\}}{\{q_k,y_\ell\}}$, $\act{S_\ell}{\{x_\ell\}}{S_\ell}$ and $\act{\{x_\ell\}}{\{g\}}{\{x_\ell\}}$ lead from $J_k$ to $S_\ell$, from $S_\ell$ to $\{x_\ell\}$ and from $\{x_\ell\}$ to $\{g\}$, respectively. A clock $q_j$ occurs only in $J_j$, the state $J_j$ lacks $p_j\in S_\ell$, and $S_d$ lacks $y_d\in S_\ell$ for $d\ne\ell$, so the action with precondition $S_\ell$ applies only in $S_\ell$. The reachable states are therefore $J_1,\dots,J_k$ at depths $0,\dots,k-1$, the states $S_\ell$ at depth $k$, the states $\{x_\ell\}$ at depth $k+1$ and the single goal state $\{g\}$ at depth $\hstar=k+2$. Every state lies on an optimal goal plan, so the slack is zero, and the optimal layers $\mathcal O_k=\{S_1,\dots,S_n\}$ and $\mathcal O_{k+1}=\{\{x_1\},\dots,\{x_n\}\}$ are antichains of size $n$, while the others are singletons. Hence $\xi(\Pi)=n$.

\emph{Set width.} By \cref{lem:containment}, an admissible set contains a tuple $t$ of cost $k$, and its optimal achievers lie in $\mathcal O_k$, so $t\subseteq S_\ell$ for some $\ell$. As $c(t)=k$, no $J_j$ contains $t$, and $S_\ell\setminus J_j=\{p_j\}$, so $P\subseteq t$ and $|t|\ge k$. Conversely, $\{\{q_1\},\dots,\{q_k\},P,\{x_1\},\dots,\{x_n\},\{g\}\}$ is admissible. The only optimal achiever of $\{q_j\}$ is $J_j$, whose successor contains $q_{j+1}$, or $P$ if $j=k$. The tuple $P$ has cost $k$ and the optimal achievers $S_1,\dots,S_n$, and each $S_\ell$ has the successor $\{x_\ell\}$, the only optimal achiever of the tuple $\{x_\ell\}$, whose successor is the goal state. Hence $\ws(\Pi)=k$.

\emph{Chain width.} Let $(t_0,\dots,t_{k+2})$ be a uniform chain. The tuple $t_{k+1}$ has cost $k+1$, so it is a nonempty subset of a state of depth $k+1$, that is, $t_{k+1}=\{x_\ell\}$ for some $\ell$, and $\Opt(t_{k+1})=\{\{x_\ell\}\}$. Every optimal achiever of $t_k$ has a successor in $\Opt(t_{k+1})$, and only $S_\ell$ reaches $\{x_\ell\}$, so $\Opt(t_k)=\{S_\ell\}$. The tuple $t_k\subseteq S_\ell$ is therefore contained in no $J_j$ and in no $S_d$ with $d\ne\ell$. The first condition forces $P\subseteq t_k$ as above, and the second forces $y_d\in t_k$ for every $d\ne\ell$, because $S_\ell\setminus S_d=\{y_d\}$. Hence $t_k=S_\ell$, which has $k+n-1$ atoms, and the chain $(\{q_1\},\dots,\{q_k\},S_1,\{x_1\},\{g\})$ attains this size. So $\wc(\Pi)=k+n-1$.

\emph{Goal-admissible width.} Let $(t_0,\dots,t_L)$ be goal-admissible. As $\{g\}$ is the only reachable goal state, \cref{lem:weak} gives $L=k+2$ and $c(t_i)=i$, so again $t_{k+1}=\{x_\ell\}$ for some $\ell$. Condition~(ii) of \cref{def:goal} requires every optimal plan for $t_k$ to have a one-action extension that ends in $\{x_\ell\}$, so $\Opt(t_k)=\{S_\ell\}$ and $|t_k|\ge k+n-1$ as before. With $\wg\le\wc$, this gives $\wg(\Pi)=k+n-1$.
\end{proof}

For $k=1$ the optimal plans have length $3$ for every $n$. The actions with precondition $S_\ell$ have $k+n-1$ preconditions, so the family does not have bounded action size. It has zero slack and a single reachable goal state, yet $\ws<\wg$ for $n\ge2$, so the last sentence of \cref{prop:terminal} depends on the hypothesis of \cref{thm:hitting}. \citet[Section~3]{BonetGeffner2024} observed that sets can yield smaller width than chains. \Cref{thm:branching} gives the largest possible difference when $\ws$ and $\xi$ are fixed.

\begin{corollary}[Nested optimal layers]\label{cor:nested}
If every optimal layer is linearly ordered by inclusion, then $\xi(\Pi)=1$ and $\ws(\Pi)=\wc(\Pi)$.
\end{corollary}

Such layers may contain many states, in which case \cref{thm:hitting} does not apply. For instance, if $I=\{s\}$, $G=\{g\}$, and the actions $\act{\{s\}}{\{a_1,\dots,a_j\}}{\{s\}}$ for $1\le j\le n$ are followed by the common action $\act{\{a_1\}}{\{g\}}{\{a_1,\dots,a_n\}}$, then $\mathcal O_1$ consists of $n$ nested states. \Cref{app:nested} gives the common value of $\ws$ and $\wc$ under the hypothesis of \cref{cor:nested} as a minimax of hitting numbers along sequences of optimal states. It reduces to \cref{eq:hitting} when every optimal layer is a single state, but not when a layer contains several nested states. For set width, \cref{app:recursion} gives an exact backward hitting-set recursion over the breadth-first layers that makes no assumption on the optimal layers.

\section{A Gap Reduction with a Known Optimum}\label{sec:reduction}

We construct a task in which satisfiability determines which competitors reach one state of the supplied optimal plan. Let $\varphi=C_1\wedge\dots\wedge C_\kappa$ be a CNF formula over $x_1,\dots,x_n$ with $n,\kappa\ge1$ and one to three literals per clause. Clauses and literals may repeat. Let $\ell=\sum_{i}|C_i|$ be the number of literal occurrences and $\|\varphi\|$ the binary encoding length of $\varphi$.

\begin{theorem}[Gap with a known unique optimum]\label{thm:main}
For every such $\varphi$ and every $m=2^b\ge2$, one can construct a STRIPS task $\Pphi$ and its unique optimal plan in time polynomial in $m+n+\ell$. The task has the following properties.
\begin{enumerate}\renewcommand{\labelenumi}{(\roman{enumi})}
\item $N=\Theta(m+n+\ell)$ and $|A|=O(m+n+\ell)$. The goal is a single atom. Every action has at most two preconditions and three effects, and at most two actions apply in any reachable state.
\item The reachable state graph is acyclic, has no dead ends and exactly one goal state, and $\rho(\Pphi)\le1$.
\item For every $w\in\W$, $w(\Pphi)=1$ if $\varphi$ is unsatisfiable and $w(\Pphi)=m+1$ if $\varphi$ is satisfiable.
\end{enumerate}
\end{theorem}
The running time is polynomial in the \emph{value} of $m$, since the task has one tag atom for each index $j<m$. All our applications choose $m$ polynomial in $\|\varphi\|$. The construction and its action schemas occupy \cref{sec:overview,sec:compilation}. After a worked instance (\cref{sec:small}), we prove the structural properties and compute the width in \cref{sec:checkpoint,sec:gap}.

\subsection{Overview of the Construction}\label{sec:overview}

The initial state is $I=U\cup\{\ctl{start}\}$, where $U=\{u_0,\dots,u_{m-1}\}$ is a set of \emph{tags}, and the goal is $G=\{g_\star\}$. The first action chooses one of two modes (\cref{fig:construction}a).

\paragraph{Safe mode.} The safe mode is a chain of $D$ actions through private control atoms $\ctl{safe}_1,\dots,\ctl{safe}_{D-1}$ that ends in the checkpoint state $S=U\cup\{M\}$. All $m$ tags remain present.

\paragraph{Testing mode.} A testing execution chooses an assignment and an index $j<m$. At the checkpoint, $u_j$ is missing exactly when the assignment satisfies $\varphi$. \Cref{tab:phases} gives the order and length of the phases. Each variable $x_i$ is stored as a one-hot pair $x_i^0,x_i^1$. Assigning the value $a$ makes exactly the atom $x_i^a$ present. The formula is evaluated by $\ell-1$ binary gates, namely left-associated disjunctions within each clause followed by left-associated conjunctions across clauses. A gate $z_t=f_t(y_t,y'_t)$ on one-hot pairs takes two guarded actions. The first copies the value $a$ of $y_t$ into a private pending atom, and the second reads the value $a'$ of $y'_t$ and writes $z_t^{f_t(a,a')}$. A negative literal uses its variable pair in reversed order, so exactly one gate action applies at each step even when inputs repeat or are complementary. The output pair $o^0,o^1$ is the pair of the last gate, or the possibly reversed variable pair of the only literal when $\ell=1$, and $o^1$ means that $\varphi$ is satisfied. A complete binary tree of height $b$ then selects $j$ through $b$ binary choices. At leaf $j$, a true output adds a marker $e_j$, and one further action deletes $u_j$ exactly when $e_j$ holds. Cleanup erases every one-hot pair except the output pair, and a final action replaces the output by $M$ if it is $o^1$ and by a separate atom $\bar M$ if it is $o^0$. After $D$ actions, the checkpoint state is $S_j=(U\setminus\{u_j\})\cup\{M,e_j\}$ for a satisfying assignment with index $j$, and $T=U\cup\{\bar M\}$ for a falsifying assignment.

\begin{table}[!htb]
\caption{The phases of a testing execution. The second column gives the number of actions performed before the phase starts. After the mode choice, only the assignment and the index tree branch. After $D=2n+3\ell+2b+2$ actions, every complete testing execution is in $S_j$ or $T$ (\cref{lem:merge}).}
\label{tab:phases}
\begin{tabular}{@{}lcccl@{}}
\toprule
Phase & Starts at depth & Actions & Binary choices & Effect\\
\midrule
Mode selection & $0$ & $1$ & $1$ & enter testing mode\\
Assignment & $1$ & $n$ & $n$ & write $x_i^{a_i}$ for all $i$\\
Circuit & $n+1$ & $2(\ell-1)$ & -- & write the output $o^{\varphi(a)}$\\
Index tree & $n+2\ell-1$ & $2b$ & $b$ & select $j<m$\\
Leaf & $n+2\ell+2b-1$ & $3$ & -- & add $e_j$ if and only if $o^1$ holds\\
Tag & $n+2\ell+2b+2$ & $1$ & -- & delete $u_j$ if and only if $e_j$ holds\\
Cleanup & $n+2\ell+2b+3$ & $n+\ell-2$ & -- & erase all pairs but $o$\\
Merge & $D-1$ & $1$ & -- & replace $o^1$ by $M$ and $o^0$ by $\bar M$\\
\midrule
Total & & $D$ & $n+b+1$ & \\
\bottomrule
\end{tabular}
\end{table}

\paragraph{Checker.} From $r_0=M$ the checker advances through $r_1,\dots,r_m$. At index $j$, a normal check replaces $r_j$ by $r_{j+1}$ and requires $u_j$. If $u_j$ is missing, its marker $e_j$ is present. The first repair action consumes $r_j$ and $e_j$ and adds a private atom $\ctl{rep}_j$. The second replaces $\ctl{rep}_j$ by $r_{j+1}$ and restores $u_j$. Repair therefore takes one action more than a normal check. The false detour turns $\bar M$ into $M$, also at the cost of one action, and the goal action replaces $r_m$ by $g_\star$.

The two modes reach the checkpoint at the same depth, so any testing state there competes with $S$. Each complete testing execution then takes one action more than the safe route to reach the goal. Thus the safe route is the unique optimal plan, and every reachable state has slack at most one.

For example, let $\varphi=x_1$ and $m=2$. At depth $D$ the safe mode reaches $S=\{u_0,u_1,M\}$. Setting $x_1$ to true gives $S_0=\{u_1,M,e_0\}$ or $S_1=\{u_0,M,e_1\}$ according to the chosen index. Setting $x_1$ to false gives $T=\{u_0,u_1,\bar M\}$ for either index. A tuple separating $S$ from these three states must contain both tags and $M$.

In the action schemas below, control tokens determine which step is possible, and data atoms determine which action of that step applies. A timing invariant makes this distinction precise.

\subsection{The Complete Action Set}\label{sec:compilation}

\paragraph{Indexing.} Enumerate the gates in topological order as $z_t=f_t(y_t,y'_t)$ for $1\le t<\ell$. Each ordered input pair is a variable pair, its reversal for a negative literal, or the output pair of an earlier gate. Gate outputs and pending atoms are private to their gate. Enumerate the variable and gate-output pairs other than the formula-output pair as $\chi_1,\dots,\chi_\omega$, where $\omega=n+\ell-2$. Cleanup erases the pairs in this order. When $\omega=0$ there is no erasure step.

A testing clock $q_k$ takes its index from the depth at which a testing action adds it. Let
\begin{equation}\label{eq:compiler-offsets}
\theta_t=1+n+2(t-1),\qquad
\psi_\delta=n+2\ell-1+2\delta,\qquad
K=n+2\ell+2b-1,\qquad
D=K+5+\omega.
\end{equation}

The atoms used by testing executions are of two kinds. \emph{Control tokens} record the step that an execution has reached. They are the clocks $q_k$, the gate-pending atoms $p_t^a$, the tree ticks $\eta_\delta$, and the leaf atoms $\hat\lambda_j$ and $\eta$. A pending atom $p_t^a$ stores the value $a$ of the first input of gate $t$, a tick $\eta_\delta$ marks a completed choice at level $\delta$ of the index tree, and $\hat\lambda_j$ and $\eta$ mark the two intermediate steps of the leaf phase. All other atoms of a testing state are \emph{data atoms}: the tags, the markers $e_j$, the tree nodes $\nu_\beta$, and the atoms of the one-hot pairs. The tree has a node $\nu_\beta$ for every binary string $\beta$ of length at most $b$, with root $\nu_{\mathrm{root}}$ for the empty string, and we identify the leaf $\lambda_j$ with $\nu_{\mathrm{bin}_b(j)}$.

\paragraph{Timing.} A testing action is \emph{scheduled at depth} $i$ if the schedule applies it after $i$ actions. The assignment of $x_i$ is scheduled at depth $i\le n$, the preparation and completion of gate $t$ at $\theta_t$ and $\theta_t+1$, the choice and advance of tree level $\delta$ at $\psi_{\delta}$ and $\psi_{\delta}+1$, the leaf preparation, output and advance at $K$, $K+1$ and $K+2$, tag deletion and retention at $K+3$, the erasure of $\chi_i$ at $K+3+i$, and the merge at $D-1$. Consecutive phases meet without gaps or overlaps, even when the circuit or cleanup phase is empty.
\begin{itemize}
\item The assignment ends at depth $n+1$. If $\ell\ge2$, then $n+1=\theta_1$, $\theta_t+2=\theta_{t+1}$ for $1\le t\le\ell-2$, and the last gate ends at $\theta_{\ell-1}+2=\psi_0$. If $\ell=1$, there are no gates, and the tree starts at $\psi_0=n+1$.
\item Since $m\ge2$, the tree has $b\ge1$ levels, $\psi_\delta+2=\psi_{\delta+1}$ for $0\le \delta\le b-2$, and the last level ends at $\psi_{b-1}+2=K$.
\item $K+4+\omega=D-1$. If $\omega=0$, which happens only for $n=\ell=1$, the tag step reaches $q_{D-1}$ directly.
\end{itemize}
By these identities, the scheduled depths partition $\{1,\dots,D-1\}$, so there is one scheduled testing step at each depth $1\le i<D$. In the rest of this subsection, $i$ denotes a depth and $\delta$ a tree level.

We set $\ctl{safe}_D=M$ and $r_0=M$, and we let $E_{j,0}=\varnothing$, $E_{j,1}=\{e_j\}$, $\mu_0=\bar M$ and $\mu_1=M$. \Cref{tab:compiler-schemas} lists all actions. In the table, $a,a'\in\{0,1\}$ and $0\le j<m$ wherever these indices occur, and all other names are fresh. The task has exactly the atoms that occur in $I$, $G$ or the table, and exactly the actions obtained by grounding the rows.

\begin{table}[!htb]
\caption{The grounded action schemas of $\Pphi$, grouped by phase. A schema yields one action for each combination of its indices, where $a,a'\in\{0,1\}$ and $0\le j<m$. The last column counts the resulting actions, and the total uses $D=2n+3\ell+2b+2$ and $\omega=n+\ell-2$. All preconditions are positive, and every action has at most two preconditions and at most three effects.}
\label{tab:compiler-schemas}
\small
\begin{tabular}{@{}l l l l l r@{}}
\toprule
Schema & $\pre$ & $\add$ & $\del$ & Indices & Count\\
\midrule
\multicolumn{6}{@{}l}{\emph{Mode selection and safe mode}}\\
\quad Select safe & $\ctl{start}$ & $\ctl{safe}_1$ & $\ctl{start}$ & & $1$\\
\quad Select test & $\ctl{start}$ & $q_1,\ \nu_{\mathrm{root}}$ & $\ctl{start}$ & & $1$\\
\quad Safe advance & $\ctl{safe}_i$ & $\ctl{safe}_{i+1}$ & $\ctl{safe}_i$ & $1\le i<D$ & $D-1$\\
\addlinespace
\multicolumn{6}{@{}l}{\emph{Assignment and circuit}}\\
\quad Assignment & $q_i$ & $q_{i+1},\ x_i^a$ & $q_i$ & $1\le i\le n$, $a$ & $2n$\\
\quad Gate preparation & $q_{\theta_t},\ y_t^a$ & $p_t^a$ & $q_{\theta_t}$ & $1\le t<\ell$, $a$ & $2(\ell-1)$\\
\quad Gate completion & $p_t^a,\ y_t'^{\,a'}$ & $q_{\theta_t+2},\ z_t^{f_t(a,a')}$ & $p_t^a$ & $1\le t<\ell$, $a,a'$ & $4(\ell-1)$\\
\addlinespace
\multicolumn{6}{@{}l}{\emph{Index tree}}\\
\quad Tree choice & $q_{\psi_\delta},\ \nu_\beta$ & $\eta_\delta,\ \nu_{\beta a}$ & $\nu_\beta$ & $0\le \delta<b$, $\beta\in\{0,1\}^\delta$, $a$ & $2(m-1)$\\
\quad Tree advance & $q_{\psi_\delta},\ \eta_\delta$ & $q_{\psi_\delta+2}$ & $q_{\psi_\delta},\ \eta_\delta$ & $0\le \delta<b$ & $b$\\
\addlinespace
\multicolumn{6}{@{}l}{\emph{Leaf and tag}}\\
\quad Leaf preparation & $q_K,\ \lambda_j$ & $\hat\lambda_j$ & $q_K,\ \lambda_j$ & $j$ & $m$\\
\quad Leaf output & $\hat\lambda_j,\ o^a$ & $\{\eta\}\cup E_{j,a}$ & $\hat\lambda_j$ & $j$, $a$ & $2m$\\
\quad Leaf advance & $\eta$ & $q_{K+3}$ & $\eta$ & & $1$\\
\quad Tag deletion & $q_{K+3},\ e_j$ & $q_{K+4}$ & $q_{K+3},\ u_j$ & $j$ & $m$\\
\quad Tag retention & $q_{K+3},\ o^0$ & $q_{K+4}$ & $q_{K+3}$ & & $1$\\
\addlinespace
\multicolumn{6}{@{}l}{\emph{Cleanup and merge}}\\
\quad Pair erasure & $q_{K+3+i},\ \chi_i^a$ & $q_{K+4+i}$ & $q_{K+3+i},\ \chi_i^a$ & $1\le i\le\omega$, $a$ & $2\omega$\\
\quad Output merge & $q_{D-1},\ o^a$ & $\mu_a$ & $q_{D-1},\ o^a$ & $a$ & $2$\\
\addlinespace
\multicolumn{6}{@{}l}{\emph{Checker}}\\
\quad False detour & $\bar M$ & $M$ & $\bar M$ & & $1$\\
\quad Normal check & $r_j,\ u_j$ & $r_{j+1}$ & $r_j$ & $j$ & $m$\\
\quad Repair entry & $r_j,\ e_j$ & $\ctl{rep}_j$ & $r_j,\ e_j$ & $j$ & $m$\\
\quad Repair exit & $\ctl{rep}_j$ & $r_{j+1},\ u_j$ & $\ctl{rep}_j$ & $j$ & $m$\\
\quad Goal & $r_m$ & $g_\star$ & $r_m$ & & $1$\\
\midrule
Total & \multicolumn{5}{r@{}}{$9m+6n+11\ell+3b-3$}\\
\bottomrule
\end{tabular}
\end{table}

\needspace{10\baselineskip}
Along a given testing execution, write $\Tok(i)$ for the control tokens prescribed after $i$ actions, where $1\le i\le D$. The schedule gives
\begin{equation*}
\Tok(i)=
\begin{cases}
\{p_t^a\} & \text{if } i=\theta_t+1 \text{ for a gate } t,\\
\{q_{\psi_{\delta}},\eta_{\delta}\} & \text{if } i=\psi_{\delta}+1 \text{ for a tree level } \delta,\\
\{\hat\lambda_j\} & \text{if } i=K+1,\\
\{\eta\} & \text{if } i=K+2,\\
\varnothing & \text{if } i=D,\\
\{q_i\} & \text{otherwise},
\end{cases}
\end{equation*}
where $a$ is the value copied by the preparation of gate $t$ and $j$ is the index of the leaf that the execution reached. The cases are mutually exclusive, because $\theta_t+1<\psi_0$ for every gate $t$, $\psi_{\delta}+1<K$ for every tree level $\delta$, and $K+2<D$. Every remaining depth $i<D$ is a depth at which an action of \cref{tab:compiler-schemas} adds the clock $q_i$, so $\Tok(i)$ is a well-defined set of atoms for every $1\le i\le D$. In the smallest instance of \cref{sec:overview}, with $\varphi=x_1$ and $m=2$, we have $\ell=b=1$ and $\omega=0$, so $\psi_0=2$, $K=4$ and $D=9$. The sets $\Tok(1),\dots,\Tok(9)$ are then $\{q_1\}$, $\{q_2\}$, $\{q_2,\eta_0\}$, $\{q_4\}$, $\{\hat\lambda_j\}$, $\{\eta\}$, $\{q_7\}$, $\{q_8\}$ and $\varnothing$.

Two properties justify the schedule. At each depth $1\le i<D$ of a testing execution, only actions of the scheduled step can apply (\emph{phase exclusion}). At least one such action applies (\emph{phase progress}). The timing invariant proves the first property from the control tokens. We prove the second in \cref{lem:merge} by following the data atoms.

\begin{quote}
\emph{Timing invariant.} For $1\le i\le D$, the state reached by a testing execution after $i$ actions contains exactly the control tokens in $\Tok(i)$ and, if $i\le K$, exactly one tree node $\nu_\beta$, where $|\beta|$ is the number of tree choices made so far. If $i<D$, every action applicable in this state is scheduled at depth $i$.
\end{quote}

\begin{proof}[Proof of the timing invariant]
We use induction on $i$. A testing execution starts with \emph{select test}, which produces $q_1$ and the root $\nu_{\mathrm{root}}$, and $\Tok(1)=\{q_1\}$. Now let $1\le i<D$, and assume that the claims about tokens and nodes hold at the depths $1,\dots,i$ and that the claim about applicable actions holds at the depths $1,\dots,i-1$. Then the actions after select test were scheduled at the depths $1,\dots,i-1$, in this order.

\emph{Applicable actions after $i$ actions.} Inspection of \cref{tab:compiler-schemas} shows that every testing action scheduled at a depth $i'$ requires a control token that the schedule assigns to depth $i'$, namely $q_{i'}$ in the last case of $\Tok$, and otherwise $p_t^a$, $\eta_{\delta}$ with $i'=\psi_{\delta}+1$, $\hat\lambda_j$ or $\eta$. Each control token is assigned to a single depth, except $q_{\psi_{\delta}}$, which is assigned to $\psi_{\delta}$ and $\psi_{\delta}+1$. An action scheduled at a depth $i'\ne i$ can therefore find its token in $\Tok(i)$ only if it is a tree choice of some level $\delta$ and $i=\psi_{\delta}+1$. The execution has then made $\delta+1$ tree choices, so the only node present has length $\delta+1$, whereas the tree choice requires a node of length $\delta$. Safe actions require $\ctl{start}$ or some $\ctl{safe}_i$, and checker actions require $\bar M$, some $\ctl{rep}_j$ or some $r_j$, including $r_0=M$. None of these atoms is present: select test deletes $\ctl{start}$, no testing action adds $\ctl{start}$, $\ctl{safe}_i$, $\ctl{rep}_j$ or $r_j$ with $j\ge1$, and testing actions add $M$ and $\bar M$ only through the merge, which is scheduled at $D-1$ and has not yet been applied. Hence every applicable action is scheduled at depth $i$.

\emph{Tokens and nodes after $i+1$ actions.} Every action that applies after $i$ actions deletes the tokens of $\Tok(i)$ and adds those of $\Tok(i+1)$, except a tree choice of level $\delta$, which keeps $q_{\psi_{\delta}}$ and adds $\eta_{\delta}$. Tree nodes are added only by select test and by tree choices, which replace $\nu_\beta$ by $\nu_{\beta a}$, and the only other action that deletes a node is the leaf preparation, which is scheduled at $K$ and deletes the leaf $\lambda_j$. Hence the state after $i+1$ actions satisfies the claims about tokens and nodes.
\end{proof}

On a testing execution, the invariant excludes safe and checker actions at depths $1,\dots,D-1$. At depth $D$, $\Tok(D)=\varnothing$ and $\ctl{start}$ is absent. Every testing action requires a control token or $\ctl{start}$, and checker actions add neither. A testing execution can therefore enter the checker only after $D$ actions and cannot return to testing.

\begin{lemma}[Size]\label{lem:size}
The task $\Pphi$ has
\begin{equation}\label{eq:size}
N=7m+6n+9\ell+4b+1
\qquad\text{and}\qquad
|A|=9m+6n+11\ell+3b-3.
\end{equation}
\end{lemma}
\begin{proof}
The atoms are the $m$ tags, the four atoms $\ctl{start}$, $g_\star$, $M$ and $\bar M$, the $D-1$ safe controls, the testing clocks, the $2n$ variable atoms, $2(\ell-1)$ gate-output atoms and $2(\ell-1)$ gate-pending atoms, the $2m-1$ tree nodes, the $b$ tree ticks and the leaf tick, and four families indexed by $j<m$, namely the leaf atoms $\hat\lambda_j$, the markers $e_j$, the checker atoms $r_{j+1}$ and the repair atoms $\ctl{rep}_j$. The testing clocks are $q_1,\dots,q_{n+1}$, one clock $q_{\theta_t+2}$ per gate, one clock $q_{\psi_\delta+2}$ per tree level, $q_{K+3}$, $q_{K+4}$, and $q_{K+4+i}$ for $1\le i\le\omega$, that is, $2n+2\ell+b$ clocks. Hence
\begin{equation*}
N=m+4+(D-1)+(2n+2\ell+b)+(2n+4\ell-4)+(2m+b)+4m=7m+6n+9\ell+4b+1,
\end{equation*}
using $D=2n+3\ell+2b+2$. Summing the last column of \cref{tab:compiler-schemas} and using $\omega=n+\ell-2$ gives $|A|$.
\end{proof}

\subsection{A Small Instance}\label{sec:small}

\begin{example}[The smallest instance]\label{ex:small}
For the task with $\varphi=x_1$ and $m=2$ in \cref{sec:overview}, $n=\kappa=\ell=b=1$. There are no gates or pair erasures ($\omega=0$). The pair $x_1^0,x_1^1$ holds the formula output. The checkpoint depth is $D=9$, the optimal plan has length $\hstar=D+m+1=12$, and \cref{lem:size} gives $34$ atoms and $35$ actions. The reachable graph has $43$ states. The four complete testing executions pass through $S_0$, $S_1$ or $T$ at the checkpoint. As seen above, distinguishing $S$ there requires all three atoms $u_0,u_1,M$, and \cref{lem:profile} shows that width $3$ suffices everywhere. For $\varphi=x_1\wedge\neg x_1$, all complete testing executions pass through $T$, and the singleton $\{M\}$ separates $S$ from every competitor at the checkpoint.
\end{example}

\subsection{The Checkpoint and the Promises}\label{sec:checkpoint}

\begin{lemma}[Checkpoint states]\label{lem:merge}
The safe mode reaches $S=U\cup\{M\}$ in exactly $D$ actions. On every complete execution that starts in testing mode, the state after $D$ actions is one of
\begin{equation*}
S_j=(U\setminus\{u_j\})\cup\{M,e_j\}\quad(j<m),\qquad T=U\cup\{\bar M\}.
\end{equation*}
Each $S_j$ is reachable if and only if $\varphi$ is satisfiable, and $T$ is reachable if and only if $\varphi$ has a falsifying assignment.
\end{lemma}
\begin{proof}
The safe mode applies \emph{select safe} and the $D-1$ safe advances, the last of which adds $\ctl{safe}_D=M$. All tags remain in the state. After the mode choice, safe and testing actions use disjoint control atoms, so the two routes cannot interleave before the checkpoint.

Consider an execution that starts in testing mode, and write $a\in\{0,1\}^n$ and $j<m$ for the assignment and the index it chooses. At each depth $1\le i<D$, the timing invariant of \cref{sec:compilation} restricts the available actions to the scheduled step. We check their preconditions to identify where the execution branches, what it carries to the next phase, and that the scheduled step always has an applicable action.
\begin{itemize}
\item \emph{Assignment.} At depth $i\le n$ the clock is $q_i$, and the two assignment actions for $x_i$ apply. After $n$ steps the state is $U\cup\{q_{n+1},\nu_{\mathrm{root}}\}\cup\{x_i^{a_i}:i\le n\}$.
\item \emph{Circuit.} Assume that before gate $t$ every written pair is one-hot and holds the value of its literal or gate under $a$. At clock $q_{\theta_t}$ exactly one preparation applies, namely the one that matches the present atom of the pair $y_t$, and then exactly one completion applies. Gate $t$ thus writes the atom of the pair $z_t$ that holds $f_t$ of the values of its two inputs, and the clock moves to $q_{\theta_t+2}$, which is $q_{\theta_{t+1}}$ if $t<\ell-1$ and $q_{\psi_0}$ if $t=\ell-1$. A reversed pair encodes the value of a negative literal, so by induction the output pair holds $o^{\varphi(a)}$. If $\ell=1$, there is no gate, and the output pair is the possibly reversed variable pair of the only literal, which holds $o^{\varphi(a)}$ after the assignment.
\item \emph{Index tree.} At level $\delta$ the clock is $q_{\psi_\delta}$ and the node is $\nu_\beta$ with $|\beta|=\delta$. The two tree choices replace $\nu_\beta$ by $\nu_{\beta0}$ or $\nu_{\beta1}$ and add $\eta_\delta$, after which only the tree advance applies, since no tree choice of level $\delta$ matches the deeper node. After $b$ levels the node is $\lambda_j$ and the clock is $q_K$, and every $j$ can be reached.
\item \emph{Leaf and tag.} Leaf preparation replaces $q_K$ and $\lambda_j$ by $\hat\lambda_j$. Exactly one leaf output applies, because the output pair is one-hot, and it adds $\eta$ and, if $\varphi(a)=1$, the marker $e_j$. Leaf advance then produces $q_{K+3}$. If $\varphi(a)=1$, then $e_j$ is present and $o^0$ is absent, so only tag deletion applies and $u_j$ is deleted. If $\varphi(a)=0$, then no marker is present and $o^0$ is present, so only tag retention applies. In both cases the clock becomes $q_{K+4}$.
\item \emph{Cleanup and merge.} Each pair $\chi_i$ is one-hot, so exactly one erasure applies at $q_{K+3+i}$, and the merge replaces $o^{\varphi(a)}$ by $\mu_{\varphi(a)}$ at $q_{D-1}$.
\end{itemize}
Each phase thus offers an applicable action at every one of its depths, so no prefix with fewer than $D=1+n+2(\ell-1)+2b+3+1+\omega+1$ actions is complete. A complete execution therefore performs all scheduled steps up to the merge. After these $D$ actions, all clocks, pending atoms, ticks, tree nodes and pairs have been deleted, so the state is $S_j$ if $\varphi(a)=1$ and $T$ if $\varphi(a)=0$. Since every assignment and every index can be chosen, $S_j$ is reachable for every $j$ if and only if $\varphi$ is satisfiable, and $T$ is reachable if and only if some assignment falsifies $\varphi$.
\end{proof}

\begin{lemma}[Lengths and promises]\label{lem:structure}
The unique optimal plan of $\Pphi$ is the safe chain followed by the $m$ normal checks and the goal action, so $\hstar=D+m+1$. Every other complete execution has length $\hstar+1$ and ends in the unique reachable goal state $\sgoal=U\cup\{g_\star\}$. The reachable graph is acyclic, has no dead ends and branching at most two, and $\rho(\Pphi)\le1$.
\end{lemma}
\begin{proof}
At index $j$ of the checker, the normal check requires $u_j$ and the repair entry requires $e_j$, and no reachable state that contains $r_j$ contains both. Indeed, checker atoms appear only from depth $D$ on, where by \cref{lem:merge} $e_j$ is present exactly when $u_j$ is missing, and the repair restores $u_j$ only after consuming $e_j$. From $S$ all tags are present, so the checker performs $m$ normal checks and the goal action, $m+1$ actions in total. From $T$, the checker takes the detour $\bar M\to M$ and then the same $m+1$ actions. From $S_j$, the checks before index $j$ are normal, the repair replaces one check by two actions, and the later checks are normal. By \cref{lem:merge}, a complete execution that starts in testing mode is in $T$ or in some $S_j$ after $D$ actions. It therefore has $D+1+(m+1)$ actions in the first case and $D+j+2+(m-j-1)+1$ in the second, that is, $D+m+2=\hstar+1$ in both. The safe route is deterministic and every testing route is longer, so the safe sequence is the unique optimal plan.

After the final check all tags are present and every marker has been consumed. Only the goal action adds $g_\star$, and no action applies afterward. Thus $\sgoal=U\cup\{g_\star\}$ is the only reachable goal state. Control atoms advance through the phases, checker indices only increase, and repair atoms are private, so the reachable graph is acyclic. By \cref{lem:merge} and the first paragraph, only the mode choice, the assignment and the tree choices offer two actions. Every partial execution extends to one of these goal-reaching executions, so no reachable state is a dead end. Finally, if $s$ is reached after $i$ actions of a complete execution $\pi'$ with $|\pi'|\le\hstar+1$, then $d(s)\le i$ and $d_G(s)\le|\pi'|-i$, so $\slack(s)\le|\pi'|-\hstar\le1$.
\end{proof}

\subsection{The Width Gap}\label{sec:gap}

By \cref{lem:structure}, $\Pphi$ has a unique optimal plan, so \cref{thm:hitting} gives its set and chain width as the largest hitting number along this plan. The next lemma shows that, for a satisfiable formula, this number peaks at the checkpoint and each check then removes one forced tag (\cref{fig:construction}b).

\begin{lemma}[Hitting profile]\label{lem:profile}
Let $s_0,\dots,s_{\hstar}$ be the states of the optimal plan of $\Pphi$. Then $\HS(\Hy_i)=1$ for $1\le i<D$ and for $i=\hstar$. For $0\le k\le m$,
\begin{equation}\label{eq:profile}
\HS(\Hy_{D+k})=
\begin{cases}
1, & \varphi\text{ is unsatisfiable},\\
m+1-k, & \varphi\text{ is satisfiable}.
\end{cases}
\end{equation}
\end{lemma}
\begin{proof}
\emph{Depths before the checkpoint and the final depth.} For $1\le i<D$, $s_i=U\cup\{\ctl{safe}_i\}$, and no other reachable state contains the private atom $\ctl{safe}_i$. So $\ctl{safe}_i$ belongs to every edge of $\Hy_i$, the set $\{\ctl{safe}_i\}$ hits $\Hy_i$, and $\HS(\Hy_i)=1$ because $\Hy_i$ is nonempty (\cref{thm:hitting}). The same argument applies to $s_{\hstar}=U\cup\{g_\star\}$. By \cref{lem:structure}, every complete execution ends in $\sgoal$, and $g_\star$ is added only by the goal action, after which no action applies, so $\sgoal$ is the only reachable state that contains $g_\star$. Testing executions reach it after $\hstar+1$ actions, but its depth is $\hstar$.

\emph{Checker depths.} Let $0\le k\le m$ and $i=D+k$, so that $s_i=U\cup\{r_k\}$ with $r_0=M$. Every competitor that lacks $r_k$ contributes an edge containing $r_k$. It remains to list the competitors that contain $r_k$, that is, the reachable states $s'\ne s_i$ with $d(s')\le i$ and $r_k\in s'$. The number of actions after which an execution reaches a state can exceed the depth of that state, because testing executions that rejoin the safe route reach its states one action later than the safe plan.
\begin{itemize}
\item \emph{Safe mode.} By \cref{lem:merge,lem:structure}, the safe route is deterministic and contains $r_k$ only in $s_i$.
\item \emph{Testing mode before the merge.} These states contain neither $M$ nor any other checker atom, because checker atoms first appear at depth $D$.
\item \emph{Falsifying testing executions.} They reach $T$ after $D$ actions, and the detour turns $T$ into $U\cup\{M\}=S$. From then on they visit the states of the safe route, so the only state containing $r_k$ that they reach is $s_i$ itself, after $D+k+1$ actions.
\item \emph{Satisfying testing executions with index $j$.} They reach $S_j$ after $D$ actions and perform normal checks at the indices $0,\dots,j-1$. If $k\le j$, they reach the state $(U\setminus\{u_j\})\cup\{r_k,e_j\}$ after $D+k$ actions, so its depth is at most $i$. If $k>j$, the repair at index $j$ takes two actions and restores $u_j$, so after $D+j+2$ actions they are in $U\cup\{r_{j+1}\}=s_{D+j+1}$. From then on they visit the states of the safe route, and the only state containing $r_k$ that they reach is $s_i$.
\end{itemize}
For each $j$ with $k\le j<m$, the competitor $(U\setminus\{u_j\})\cup\{r_k,e_j\}$ exists if and only if $\varphi$ is satisfiable and contributes the edge $\{u_j\}$ when it exists. There are no other competitors containing $r_k$. Every other edge contains $r_k$. The preceding safe state gives the singleton edge $s_i\setminus s_{i-1}=\{r_k\}$. In particular, at $k=0$ the predecessor $s_{D-1}=U\cup\{\ctl{safe}_{D-1}\}$ forces $M$ even if $\varphi$ is a tautology and $T$ is unreachable. The minimal edges of $\Hy_i$ are therefore
\begin{equation*}
\{r_k\}\ \text{and}\ \{u_j\}\ \text{for}\ k\le j<m\quad\text{if $\varphi$ is satisfiable},\qquad
\{r_k\}\ \text{alone}\quad\text{otherwise}.
\end{equation*}
These minimal edges are pairwise disjoint singletons, so every hitting set contains all of their atoms, and the set of these atoms hits $\Hy_i$ because every edge contains a minimal edge. Hence $\HS(\Hy_i)$ is the number of minimal edges, which is $1+(m-k)$ in the satisfiable case and $1$ otherwise. This is \cref{eq:profile}.
\end{proof}

\begin{proof}[Proof of \cref{thm:main}]
Part~(ii) is \cref{lem:structure}. For part~(iii), \cref{lem:structure} shows that all optimal goal plans visit the same states, and \cref{thm:hitting,lem:profile} give $\ws=\wc=\max_i\HS(\Hy_i)$, which is $m+1$ if $\varphi$ is satisfiable and $1$ otherwise. The goal is the single atom $g_\star$ and $\sgoal$ is the only reachable goal state, so \cref{prop:terminal} gives $\wg=\ws$. For part~(i), the operator bounds can be read off \cref{tab:compiler-schemas}, the branching bound is part of \cref{lem:structure}, and \cref{lem:size} gives $N=\Theta(m+n+\ell)$ and $|A|=O(m+n+\ell)$. The indices in the action table determine the safe plan independently of the truth of $\varphi$. Each atom name takes $O(\log(m+n+\ell))$ bits, so the task and plan have polynomial encoding length and can be written down in polynomial time.
\end{proof}

\section{Approximation Consequences}\label{sec:approx}

\Cref{thm:main} gives a gap linear in the number $N$ of atoms. We use it to bound two forms of approximation. An algorithm receives a task together with its unique optimal plan and runs in time polynomial in their combined encoding length. It outputs a rational number in binary. A \emph{one-sided} $\alpha$-approximation returns $\widehat w$ with $w\le\widehat w\le\alpha(N)\,w$, and a \emph{symmetric} one returns $\widehat w$ with $w/\alpha(N)\le\widehat w\le\alpha(N)\,w$. Write $\Pone$ for the class of task--plan pairs in which the supplied plan is the unique optimum and the task has slack at most one, a single reachable goal state, a single goal atom, an acyclic reachable graph, branching at most two, and actions with at most two preconditions and three effects.

\begin{corollary}[One-sided order]\label{cor:one-sided}
Let $w\in\W$. Returning $\max_i|s_i|$ over the states of the supplied plan, or simply $N$, is a polynomial-time one-sided $N$-approximation on every solvable positive-precondition task with $\hstar>1$. Unless $\Pclass=\NPclass$, for every constant $0<\gamma<1/7$ no polynomial-time algorithm is a one-sided $\alpha$-approximation on $\Pone$ with $\alpha(N)\le\gamma N$ for all sufficiently large $N$. In particular, none achieves $\alpha(N)=o(N)$.
\end{corollary}
\begin{proof}
The upper bound is \cref{eq:trivial}. For the lower bound, take a 3-CNF formula $\varphi$ and rename its occurring variables to $x_1,\dots,x_n$, so that $n\le\ell\le\|\varphi\|$. Choose $m=2^{\lceil2\log_2(\|\varphi\|+2)\rceil}$, so that $(\|\varphi\|+2)^2\le m<2(\|\varphi\|+2)^2$. \Cref{thm:main} produces a pair in $\Pone$ in time polynomial in $\|\varphi\|$, and by \cref{eq:size}
\begin{equation*}
7m\le N\le 7m+15\sqrt m+4\log_2m+1,
\end{equation*}
because $n\le\ell\le\|\varphi\|<\sqrt m$ gives $6n+9\ell\le15\ell<15\sqrt m$, and $b=\log_2m$. Hence $N\le7m+o(m)$ and
\begin{equation*}
\gamma N\le7\gamma m+o(m)<m+1\qquad\text{for all sufficiently long }\varphi,
\end{equation*}
since $7\gamma<1$. As $N\ge7m$ grows with $\|\varphi\|$, the bound $\alpha(N)\le\gamma N$ also holds for all sufficiently long $\varphi$. If $\varphi$ is unsatisfiable, then $w=1$ and $\widehat w\le\alpha(N)\,w\le\gamma N<m+1$. Otherwise $\widehat w\ge w=m+1$. Comparing $\widehat w$ with $m+1$ therefore decides satisfiability in polynomial time for all sufficiently long formulas, and a finite table handles the rest.
\end{proof}

\begin{corollary}[Symmetric order]\label{cor:symmetric}
Let $w\in\W$. Returning the integer $\lceil\sqrt N\,\rceil$ is a symmetric approximation with factor $\lceil\sqrt N\,\rceil$ on every solvable positive-precondition task with $\hstar>1$. Unless $\Pclass=\NPclass$, for every constant $0<\gamma<1/\sqrt7$ no polynomial-time algorithm is a symmetric $\alpha$-approximation on $\Pone$ with $\alpha(N)\le\gamma\sqrt N$ for all sufficiently large $N$. In particular, none achieves $\alpha(N)=o(\sqrt N)$.
\end{corollary}
\begin{proof}
Let $q=\lceil\sqrt N\,\rceil$ and $\widehat w=q$. By \cref{eq:trivial}, $1\le w\le N\le q^2$. The first inequality gives $\widehat w=q\le qw$, and the last two give $w/q\le q=\widehat w$, so $\widehat w$ is a symmetric approximation with factor $q$. For the lower bound, use the instances from the proof of \cref{cor:one-sided}. Since $\gamma^2<1/7$, the argument given there yields $\alpha(N)^2\le\gamma^2N<m+1$ for all sufficiently long $\varphi$. If $\varphi$ is unsatisfiable, then $w=1$ and $\widehat w\le\alpha(N)<\sqrt{m+1}$. Otherwise $w=m+1$ and $\widehat w\ge(m+1)/\alpha(N)>(m+1)/\sqrt{m+1}=\sqrt{m+1}$. Comparing the rational number $\widehat w^{\,2}$ with $m+1$ therefore decides satisfiability.
\end{proof}

Both orders are tight, and only the constants remain open, between $1/7$ and $1$ for one-sided estimates and between $1/\sqrt7$ and $1$ for symmetric ones. Our reduction cannot narrow these ranges, because every state on its optimal plan has exactly $m+1$ atoms. The estimate $\max_i|s_i|$ is therefore exact on satisfiable instances and off by a factor $m+1\approx N/7$ on the others. The lower bounds also apply to algorithms that output certificates instead of numbers, since the largest tuple cardinality in a returned certificate is a one-sided estimate.

\section{Explicit versus Succinct Competitors}\label{sec:map}

\Cref{thm:hitting} reduces certification to hitting sets of competitor hypergraphs. In the reduction of \cref{sec:reduction}, the plan and the safe checkpoint state are known, but whether some competitor misses a given tag is equivalent to the satisfiability of $\varphi$, because the task represents its competitors implicitly. The two slack regimes of \cref{tab:summary} correspond to two representations of these hypergraphs, explicit and succinct, and in each regime deciding an input threshold is complete for the same class as the corresponding hitting-set problem.

\subsection{Zero Slack: Explicit Hitting Sets}\label{sec:zero}

If all optimal plans follow the same state sequence and every reachable state has slack zero, the competitors at each depth are exactly the earlier states on that sequence. The hitting sets in \cref{thm:hitting} can then be read off the supplied plan.

\begin{corollary}[Zero slack]\label{cor:zero}
Under the hypotheses of \cref{thm:hitting}, if $\rhoPi=0$, then for every $w\in\W$
\begin{equation}\label{eq:zero}
w(\Pi)=\max_{1\le i\le\hstar}\HS\bigl(\{s_i\setminus s_j: 0\le j<i\}\bigr).
\end{equation}
Given $\Pi$ and an optimal goal plan, a uniform chain of width at most $(1+\ln\hstar)\,w(\Pi)$ can be computed in polynomial time, and $w(\Pi)\le k$ can be decided in time $N^{O(k)}\operatorname{poly}(\hstar)$ for each fixed $k$.
\end{corollary}
\begin{proof}
As shown at the end of the proof of \cref{prop:terminal}, $\rhoPi=0$ implies $\R=\{s_0,\dots,s_{\hstar}\}$ with $d(s_j)=j$. Moreover, $s_{\hstar}$ is the only reachable goal state, since $G\subseteq s_j$ with $j<\hstar$ would contradict $d_G(I)=\hstar$. The competitors of $s_i$ are therefore $s_0,\dots,s_{i-1}$, so $\Hy_i$ is the displayed hypergraph, with at most $\hstar$ edges, and \cref{thm:hitting,prop:terminal} give \cref{eq:zero}. Simulating the plan lists all edges. With $H_j=\sum_{i=1}^{j}1/i$ denoting the $j$th harmonic number, greedy hitting set is within a factor $H_{\hstar}\le 1+\ln\hstar$ of optimal at each depth \citep{Chvatal1979}, and the resulting sets form a chain as in the proof of \cref{thm:hitting}. For fixed $k$, it suffices to test every tuple of at most $k$ atoms of each $s_i$.
\end{proof}
Under the hypotheses of \cref{thm:hitting} but with arbitrary slack, the same greedy argument gives a factor $1+\ln|\R|$ for $\ws$ and $\wc$ whenever the reachable graph is given explicitly.

Conversely, every explicit hitting-set instance can be realized at a single depth of a zero-slack task.

\begin{proposition}[Path embedding]\label{prop:path}
Let $\Hy=\{E_0,\dots,E_{m-1}\}$ be a hypergraph with $m\ge1$ nonempty edges over a finite set $U$, where edges may repeat. One can compute in polynomial time a STRIPS task $\Pi_{\Hy}$ with $|U|+m+1$ atoms and $m+1$ actions, together with its unique optimal plan, such that the reachable states form a single path of $m+2$ states, $\hstar=m+1$, the goal is a single atom, $\rho(\Pi_{\Hy})=0$, and $w(\Pi_{\Hy})=\HS(\Hy)$ for every $w\in\W$. Moreover, every tuple of cost $m$ in an admissible set, a uniform chain or a goal-admissible sequence is a hitting set of $\Hy$, and every hitting set $Z$ of $\Hy$ yields in polynomial time a uniform chain of width $|Z|$.
\end{proposition}
\begin{proof}[Proof outline]
The task walks through the complements of the edges. Its state at depth $j<m$ is $s_j=(U\setminus E_j)\cup\{q_j\}$, where $q_j$ is a clock atom. For $j<m-1$, action $a_j$ advances the clock and changes $U\setminus E_j$ to $U\setminus E_{j+1}$. Action $a_{m-1}$ adds $E_{m-1}$ and deletes the last clock, reaching $s_m=U$. A final action with precondition $U$ adds the goal atom. Every state lies on this single path, so $\rho(\Pi_\Hy)=0$, and by \cref{cor:zero} the competitors of $s_i$ are the states $s_j$ with $j<i$. At a depth $i<m$, the clock $q_i$ belongs to no other state, and at depth $m+1$ the goal atom belongs to no other state, so these depths have hitting number $1$. At depth $m$, the competitors $s_j$ contribute the edges
\begin{equation*}
s_m\setminus s_j=U\setminus\bigl((U\setminus E_j)\cup\{q_j\}\bigr)=E_j\qquad(0\le j<m),
\end{equation*}
so $\Hy_m=\Hy$ and $w(\Pi_\Hy)=\max(1,\HS(\Hy))=\HS(\Hy)$. A certificate tuple of cost $m$ is therefore exactly a hitting set of $\Hy$ contained in $U$. \Cref{app:path} lists the actions and verifies the remaining claims.
\end{proof}

\begin{corollary}[Zero-slack lower bounds]\label{cor:zero-lower}
Let $\Pzero$ be the set of task--plan pairs $(\Pi,\pi)$ such that $\pi$ is the unique optimal plan of $\Pi$, $\rhoPi=0$, $\hstar>1$ and the goal is a single atom, and let $w\in\W$.
\begin{enumerate}\renewcommand{\labelenumi}{(\roman{enumi})}
\item The promise problem of deciding $w(\Pi)\le k$, for $(\Pi,\pi)\in\Pzero$ and $k$ given in the input, is NP-complete.
\item Parameterized by $k$, this problem is $\Wtwo$-hard. Assuming ETH, no algorithm decides it in time $f(k)\,N^{o(k)}$ for a computable function $f$.
\item For every $0<\varepsilon<1$, unless $\Pclass=\NPclass$, no polynomial-time algorithm achieves a one-sided factor $(1-\varepsilon)\ln(\hstar-1)$ on $\Pzero$ for all sufficiently large $\hstar$, and none achieves a symmetric factor $\sqrt{(1-\varepsilon)\ln(\hstar-1)}$ for all sufficiently large $\hstar$. Both statements also hold for algorithms that output certificates.
\end{enumerate}
There are polynomial-time algorithms with one-sided factor $1+\ln\hstar$ and symmetric factor $\sqrt{1+\ln\hstar}$.
\end{corollary}
\begin{proof}
(i) By \cref{cor:zero}, $w(\Pi)\le k$ holds if and only if each $s_i$ contains a tuple of at most $k$ atoms that meets every set $s_i\setminus s_j$ with $j<i$. The $\hstar$ tuples form a certificate of polynomial size that is checked by simulating $\pi$, so the problem is in NP. Hardness follows from the NP-completeness of hitting set \citep{Karp1972} and \cref{prop:path}, which preserves the threshold.

(ii) \Cref{prop:path} preserves $k$ exactly and maps an instance with $|U|$ elements and $m$ edges to a task with $N=|U|+m+1$ atoms. Hitting set parameterized by $k$ is $\Wtwo$-hard \citep{DowneyFellows2013}. Dominating set on an $n$-vertex graph is the hitting-set instance whose edges are the $n$ closed neighbourhoods, which gives $N=2n+1$, and it admits no $f(k)\,n^{o(k)}$ algorithm under ETH \citep{ChenEtAl2006}.

(iii) \citet{DinurSteurer2014} showed, by a gap-producing reduction, that for every $\varepsilon>0$ it is NP-hard to approximate set cover on universes of $n$ elements within $(1-\varepsilon)\ln n$. In hitting-set form, the $n$ elements are the edges, so \cref{prop:path} produces tasks with $\hstar-1=m=n$ and the same optimum. An algorithm that achieves the one-sided factor $(1-\varepsilon)\ln(\hstar-1)$ for all sufficiently large $\hstar$ would therefore approximate set cover within $(1-\varepsilon)\ln n$ for all sufficiently large $n$. Instances with a bounded number $n$ of elements can be solved exactly in time polynomial in the input length, for example by dynamic programming over the $2^n$ subsets of the universe, even when sets are repeated. For a gap with integer bounds $w_0<w_1$, a symmetric estimate with factor $\alpha<\sqrt{w_1/w_0}$ distinguishes $w\le w_0$ from $w\ge w_1$ by comparing $\widehat w^{\,2}$ with $w_0w_1$. Applying this to the same set-cover gaps gives the symmetric lower bound. A certificate of width $k$ also yields a hitting set of size at most $k$ by \cref{prop:path}.

For the upper bound, write $h=\hstar$ and let $\bar w$ be the width of the greedy chain in \cref{cor:zero}. Then $w\le\bar w\le H_h w$. To obtain a rational symmetric estimate, choose a positive dyadic rational $\zeta$ with $H_h\le\zeta^2\le H_h+1/8$. Bisection with rational square comparisons finds such a $\zeta$ in polynomial time. Since $h\ge2$, integral comparison gives $1+\ln h-H_h\ge\ln2-1/2>1/8$. Thus
\[
    \frac{w}{\sqrt{1+\ln h}}\le\frac{w}{\zeta}\le\frac{\bar w}{\zeta}
    \le\frac{H_h}{\zeta}w\le\zeta w\le\sqrt{1+\ln h}\,w .
\]
\end{proof}

On these paths, the hardness lies entirely in choosing a smallest separating tuple. The path embedding uses a goal action with $|U|$ preconditions and actions with up to $|U|+2$ effects. Actions of constant size suffice for NP-hardness, at the price of one extra atom in every certificate and a longer plan.

\begin{proposition}[Path embedding with bounded actions]\label{prop:bounded}
Let $\Hy$ be a hypergraph with $m\ge1$ nonempty edges, possibly repeated, over a set $U$ of $n$ elements. One can compute in polynomial time a STRIPS task $\Pi^{\mathrm b}_{\Hy}$ together with its unique optimal plan with the following properties.
\begin{enumerate}\renewcommand{\labelenumi}{(\roman{enumi})}
\item The reachable states form a single path, every state other than the goal state has exactly one applicable action, the goal is a single atom, $\rho(\Pi^{\mathrm b}_{\Hy})=0$, and every action has at most two preconditions and at most three effects.
\item $w(\Pi^{\mathrm b}_{\Hy})=1+\HS(\Hy)$ for every $w\in\W$.
\end{enumerate}
On pairs $(\Pi,\pi)\in\Pzero$ with these action bounds, deciding $w(\Pi)\le k$ remains NP-complete and $\Wtwo$-hard parameterized by $k$. Under ETH it admits no $f(k)N^{o(k)}$-time algorithm for any computable $f$.
\end{proposition}
\begin{proof}[Proof outline]
For each edge $E_j$ in turn, the task prepares data in which every element $u_i$ is present if $u_i\notin E_j$ and is replaced by an absence marker $\bar u_i$ if $u_i\in E_j$. A scanner token then passes through $r_1,\dots,r_n$. It advances at $u_i$ and fails at the first marker $\bar u_i$, which exists because $E_j\ne\varnothing$, and an edge label $\ctl{lab}_j$ selects the next preparation. After the last edge the task restores $U$, the final scan succeeds and reaches the goal. Each preparation step changes one data atom and switches a private clock atom, which gives three effects. At the start $U\cup\{r_1\}$ of the final scan, the initial state contributes the edge $\{r_1\}$, and the entry state of the scan for $E_j$ contributes the edge $E_j$, so a certificate tuple of that cost needs $1+\HS(\Hy)$ atoms. Every other state is separated from the earlier ones by at most this many atoms. \Cref{app:bounded} lists the actions and verifies (i) and (ii). For the consequence, map an instance $(\Hy,k)$ of hitting set to $(\Pi^{\mathrm b}_{\Hy},k+1)$. The reduction proves NP-hardness and $\Wtwo$-hardness \citep{Karp1972,DowneyFellows2013}, and membership in NP follows from \cref{cor:zero-lower}(i). For the time bound, take the $n$ closed neighbourhoods of an $n$-vertex dominating-set instance as the edges. The bounded-action construction has $N=O(n^2)$ atoms (\cref{app:bounded}) and threshold $k+1$. An algorithm with running time $f(k')N^{o(k')}$ at threshold $k'$ would decide dominating set in time $f(k+1)n^{o(k)}+n^{O(1)}$, contrary to the ETH lower bound of \citet{ChenEtAl2006}.
\end{proof}

The approximation lower bound in \cref{cor:zero-lower}(iii) uses the shorter path of \cref{prop:path}, for which $\hstar-1$ equals the number of edges.

Explicit competitors also arise when the reachable state graph of a task is enumerated, as for small training instances, whether or not the slack is zero and whether or not the optimal plan is unique. For set width, the greedy factor carries over, and it depends only on the number of reachable states.

\begin{corollary}[Explicit state graphs]\label{cor:explicit}
Suppose that the reachable states of a task, their atoms and the transitions between them are given explicitly. An admissible set of width at most $H_{|\R|-1}\,\ws(\Pi)$ can be computed in time polynomial in $N$ and in the size of this graph. For every $0<\varepsilon<1$, unless $\Pclass=\NPclass$, no polynomial-time algorithm returns, on all such inputs with sufficiently large $|\R|$, an estimate $\widehat w$ with $\ws(\Pi)\le\widehat w\le(1-\varepsilon)\ln(|\R|-2)\,\ws(\Pi)$, even if the graph is a single path.
\end{corollary}
\begin{proof}
The algorithm runs a backward recursion over the breadth-first layers of the graph, characterized in \cref{app:recursion}, with Chvátal's greedy rule in place of exact hitting sets. \Cref{app:greedy} describes it and proves the bound. For the lower bound, a task in $\Pzero$ has the reachable states $s_0,\dots,s_{\hstar}$ (proof of \cref{prop:terminal}), so its graph, with $|\R|=\hstar+1$, can be computed from the supplied plan in polynomial time, and \cref{cor:zero-lower}(iii) applies with $\ln(\hstar-1)=\ln(|\R|-2)$.
\end{proof}

The recursion of \cref{app:recursion} maintains two sets at each depth $i$. The set $B_i$ contains the states that can continue a certificate, namely those with a successor in $V_{i+1}$ if $i<\hstar$ and the goal states if $i=\hstar$. A state of $B_i$ is kept in $V_i$ only if, in addition, a tuple within the budget separates it from every earlier state and from every state of depth $i$ outside $B_i$. A kept state contributes its tuple to the certificate, and admissibility requires only that the other optimal achievers of this tuple can continue, that is, lie in $B_i$. The greedy version keeps this distinction with greedy tuples, which lets a single harmonic factor cover all depths. The running time is polynomial in the explicit graph, whose size can be exponential in the number of STRIPS atoms.

\subsection{Slack One: Succinct Hitting Sets}\label{sec:one}

At slack one, the task can encode a family of competitors by a Boolean circuit. In \cref{sec:reduction}, testing a satisfying assignment with index $j$ contributes a singleton edge $\{u_j\}$. We now allow a testing execution to delete a whole set of tags. This construction realizes the hitting number of any circuit-defined family of nonempty edges. It gives the variable-threshold classification and an approximation lower bound for algorithms with access to an NP oracle.

\begin{theorem}[Succinct hitting sets as width]\label{thm:compiler}
Let $C$ be a circuit with $n_C\ge1$ inputs and outputs indexed by a nonempty set $U$, and suppose that $\Hy_C$ has no empty edge. One can compute, in time polynomial in the size of $C$, a STRIPS task $\PC$ and its unique optimal plan such that the pair belongs to $\Pone$ and
\begin{equation}\label{eq:compiler}
w(\PC)=1+\HS(\Hy_C)\qquad\text{for every }w\in\W.
\end{equation}
\end{theorem}
As in \cref{sec:reduction}, the safe route and every testing route reach their checkpoints after the same number of actions. A testing execution chooses a circuit input $v$, which selects the edge $E_v$ of $\Hy_C$, and deletes the tags in $E_v$. The tags present at the safe checkpoint $S$ and absent from this testing state are precisely those in $E_v$. Each such state therefore contributes the edge $E_v$ to the competitor hypergraph. Unlike in $\Pphi$, a testing execution may miss many tags, and the checker must still charge it only one extra action. \Cref{app:compiler} gives the grounded actions and proves the testing invariant. We describe the checker here.

\paragraph{Safe and testing mode.} Write $U=\{u_0,\dots,u_{p-1}\}$. The initial state is $U\cup\{\ctl{start}\}$, the goal is a single atom $g$, and a safe mode of $D$ actions through private atoms reaches $S=U\cup\{M\}$. Testing mode chooses an input $v\in\{0,1\}^{n_C}$ through $n_C$ binary choices and evaluates $C$ with the two-action gate gadget of \cref{sec:compilation}, one gadget per gate. For each element $u_\ell$ in turn it then spends two actions, which delete $u_\ell$ and add a marker $e_\ell$ if the output for $u_\ell$ is $1$, and change nothing otherwise. Finally it erases all one-hot pairs and adds $M$. As in \cref{sec:compilation}, a clock that advances with the depth synchronizes all testing executions.

\begin{lemma}[Testing executions]\label{lem:testing}
The safe mode reaches $S=U\cup\{M\}$ in exactly $D$ actions. Every complete execution that starts in testing mode chooses an input $v\in\{0,1\}^{n_C}$, and its state after $D$ actions is
\begin{equation}\label{eq:testing-state}
T_v=(U\setminus E_v)\cup\{M\}\cup\{e_\ell:u_\ell\in E_v\},
\qquad\text{so that}\qquad S\setminus T_v=E_v.
\end{equation}
Every $v\in\{0,1\}^{n_C}$ is chosen by some complete execution. Only the mode selection and the input choices offer two applicable actions, and every other state before depth $D$ offers exactly one.
\end{lemma}
The proof is a phase-by-phase verification like that of \cref{lem:merge} and is given in \cref{app:compiler}.

\paragraph{Checker.} The checker of $\Pphi$ repairs a missing tag with one extra action. Here $T_v$ may miss many tags, and charging one action per missing tag would raise the slack to $|E_v|$. The checker therefore has two copies, with checker atoms $r_\ell$ and $r'_\ell$. In the unprimed copy, which the safe route uses, every index $\ell$ takes two actions. A normal check requires the tag and passes through an intermediate atom,
\begin{equation*}
r_\ell,u_\ell\ \to\ \ctl{chk}_\ell\ \to\ r_{\ell+1}.
\end{equation*}
If the tag is missing, the only applicable action starts the \emph{first repair}, which takes three actions, restores $u_\ell$ and switches to the primed copy,
\begin{equation*}
r_\ell,e_\ell\ \to\ \ctl{rep}_\ell\ \to\ \ctl{rep}'_\ell,u_\ell\ \to\ r'_{\ell+1}.
\end{equation*}
In the primed copy, a present tag is checked and a missing tag is repaired in two actions each,
\begin{equation*}
r'_\ell,u_\ell\ \to\ \ctl{chk}'_\ell\ \to\ r'_{\ell+1}
\qquad\text{and}\qquad
r'_\ell,e_\ell\ \to\ \ctl{fix}_\ell\ \to\ r'_{\ell+1},u_\ell.
\end{equation*}
The goal action adds $g$ from $r_p$ or from $r'_p$, with $r_0=M$.

The primed copy records that the first missing tag has already cost the testing route an extra action. For example, with $p=3$ and $E_v=\{u_0,u_2\}$, the continuation from $T_v$ spends three actions repairing $u_0$, two checking $u_1$, two repairing $u_2$, and one reaching the goal. The safe route takes seven actions after depth $D$, and this testing route takes eight.

\begin{lemma}[Checker]\label{lem:checker}
The unique complete continuation from $S$ consists of the $p$ normal checks and the unprimed goal action, which are $2p+1$ actions. Let $\ell_0=\min\{\ell:u_\ell\in E_v\}$, which exists because $E_v\ne\varnothing$. The unique complete continuation from $T_v$ performs normal checks for $\ell<\ell_0$, the first repair at $\ell_0$, and for $\ell_0<\ell<p$ a primed check if $u_\ell\in T_v$ and a primed repair otherwise, followed by the primed goal action. It has
\begin{equation*}
2\ell_0+3+2(p-\ell_0-1)+1=2p+2
\end{equation*}
actions, ends in $U\cup\{g\}$, and offers exactly one applicable action in every state before the goal.
\end{lemma}
\begin{proof}
When the checker is at index $\ell$, exactly one of $u_\ell$ and $e_\ell$ is present. At depth $D$ this follows from \cref{eq:testing-state}, and it is preserved, because a check keeps $u_\ell$, a repair consumes $e_\ell$ and restores $u_\ell$ before the index advances, and no checker action touches the tags and markers of other indices. Hence at index $\ell$ exactly one of the normal check and the first repair applies in the unprimed copy, and exactly one of the primed check and the primed repair in the primed copy. Each intermediate atom $\ctl{chk}_\ell$, $\ctl{rep}_\ell$, $\ctl{rep}'_\ell$, $\ctl{chk}'_\ell$ or $\ctl{fix}_\ell$ enables exactly one action, the next step of its check or repair. From $S$, all tags are present and no marker is, so the checker performs $p$ normal checks of two actions each and then the goal action. From $T_v$, the tags $u_\ell$ with $\ell<\ell_0$ are present and are checked normally. At $\ell_0$ the tag is missing and $e_{\ell_0}$ is present, so the first repair applies. It takes three actions, restores $u_{\ell_0}$ and leads to $r'_{\ell_0+1}$. Every later index takes two actions, through a primed check or a primed repair, and the primed goal action follows. Counting gives $2\ell_0+3+2(p-\ell_0-1)+1=2p+2$ actions. Every marker has been consumed, every tag restored, and the goal action deletes the last checker atom, so the final state is $U\cup\{g\}$.
\end{proof}

\begin{proof}[Proof of \cref{thm:compiler}]
\emph{Promises.} By \cref{lem:testing,lem:checker}, the safe route has length $D+2p+1$, and every complete execution that starts in testing mode has length $D+2p+2$, because every $E_v$ is nonempty. Safe and testing control atoms are disjoint until $M$ is added at depth $D$, and $T_v\ne S$ because $T_v$ misses a tag. Hence $\hstar=D+2p+1$, and the safe plan is the unique optimal plan. All complete executions end in $U\cup\{g\}$, which is therefore the only reachable goal state, and no action applies there. Clocks, tag indices and checker indices only increase, and the control atoms of each step are private, so the reachable graph is acyclic. Branching is at most two by \cref{lem:testing,lem:checker}. Every partial execution can be completed, so there are no dead ends, and the argument at the end of the proof of \cref{lem:structure} gives $\rho(\PC)\le1$. The goal is the single atom $g$, every action has at most two preconditions and three effects (\cref{tab:compiler-sigma}), and the number of atoms and actions is $O(n_C+n_G+p)$ for $n_G$ gates (\cref{app:compiler}). Since all optimal goal plans visit the same states $s_0,\dots,s_{\hstar}$, \cref{thm:hitting,prop:terminal} give $w(\PC)=\max_i\HS(\Hy_i)$ for every $w\in\W$.

\emph{Depths before the checkpoint and the final depth.} For $1\le i<D$, $s_i$ contains the private atom $\ctl{safe}_i$, and $s_{\hstar}=U\cup\{g\}$ contains $g$, which no other state of depth at most $\hstar$ contains. As in \cref{lem:profile}, these depths have hitting number $1$.

\emph{The checkpoint.} The competitors of $S$ are the earlier safe states, the testing states before the merge, and the states $T_v$. Every earlier safe state contributes the edge $\{M\}$, every testing state before the merge misses $M$, and $T_v$ contributes $E_v$ by \cref{eq:testing-state}. Thus
\begin{equation*}
\Hy_D=\{\{M\}\}\ \cup\ \{\text{edges containing }M\}\ \cup\ \{E_v:v\in\{0,1\}^{n_C}\}.
\end{equation*}
Every hitting set of $\Hy_D$ contains $M$, and since $M\notin U\supseteq E_v$, its remaining atoms must hit every $E_v$. Conversely, $M$ together with a hitting set of $\Hy_C$ hits $\Hy_D$. Hence $\HS(\Hy_D)=1+\HS(\Hy_C)$.

\emph{After the checkpoint.} Let $D<i<\hstar$. The safe state $s_i$ is $U\cup\{r_\ell\}$ if $i=D+2\ell$ and $U\cup\{\ctl{chk}_\ell\}$ if $i=D+2\ell+1$. A competitor that does not contain this checker atom contributes an edge that contains it, so it remains to find the competitors that do contain it. By \cref{lem:checker}, with $\ell_0(v)=\min\{\ell:u_\ell\in E_v\}$, a testing execution that chose $v$ contains unprimed checker atoms only up to its first repair. It performs normal checks at the indices $\ell<\ell_0(v)$ and reaches $r_\ell$, for $\ell\le \ell_0(v)$, at depth exactly $D+2\ell$, as the safe route does. The competitors that contain $r_\ell$ are therefore the states $(U\setminus E_v)\cup\{r_\ell\}\cup\{e_{\ell'}:u_{\ell'}\in E_v\}$ with $\ell\le \ell_0(v)$, at depth $D+2\ell$. Those that contain $\ctl{chk}_\ell$ are the corresponding states with $\ctl{chk}_\ell$ in place of $r_\ell$, at depth $D+2\ell+1$, with $\ell<\ell_0(v)$. Each of them differs from $s_i$ by missing exactly the tags in $E_v$, so it contributes the edge $E_v$. Every edge of $\Hy_i$ thus contains the checker atom of $s_i$ or equals some $E_v$, and this atom together with a minimum hitting set of $\Hy_C$ hits $\Hy_i$. Hence $\HS(\Hy_i)\le1+\HS(\Hy_C)$ for $D<i<\hstar$. The maximum over all depths is therefore attained at $D$, which proves \cref{eq:compiler}.
\end{proof}

If $E_v=\varnothing$ for some input $v$, then $T_v=S$ by \cref{eq:testing-state}, so a testing execution reaches the checkpoint state of the safe route at the same depth, and the optimal plan is no longer unique. The reduction of \cref{sec:reduction} is a sharpened special case of \cref{thm:compiler}. Its testing executions miss at most one tag each, so a repair costing one extra action keeps the slack at one, and falsifying assignments are routed to a state that misses $M$ rather than a tag, which yields width $1$ instead of $2$ in the unsatisfiable case.

We first apply the checkpoint constructions to fixed and variable thresholds. \Cref{cor:oracle-approx} then uses the same circuit construction to transfer an approximation gap.

\begin{corollary}[Fixed thresholds at slack one]\label{cor:fixed}
Fix $k\ge1$. For $w\in\{\ws,\wc\}$, the set of pairs $(\Pi,\pi)$ such that $\Pi$ is a positive-precondition STRIPS task, $\pi$ is its unique optimal plan, $|\pi|>1$, and $w(\Pi)\le k$ is coNP-complete. For $\wg$, the same holds for the promise problem whose inputs additionally have a single reachable goal state. Without this promise, the set for $\wg$ is coNP-hard and belongs to $\Pclass^{\NPclass}$. Hardness holds already on pairs in $\Pone$.
\end{corollary}
\begin{proof}
For membership, a pair $(\Pi,\pi)$ is outside the set if and only if (a)~the input is malformed or $|\pi|\le1$, (b)~$\pi$ is not a plan that ends in a goal state, (c)~some plan different from $\pi$ of length at most $|\pi|$ ends in a goal state, or (d)~none of (a)--(c) holds and $w(\Pi)>k$. Conditions (a) and (b) are checked in polynomial time, and (c) has the other plan as an NP witness. For (d) and $w\in\{\ws,\wc\}$, \cref{thm:hitting} gives a depth $i$ with $\HS(\Hy_i)>k$. A witness lists $i$ and, for each of the at most $(N+1)^k$ tuples $t\subseteq s_i$ with $|t|\le k$, a plan of length at most $i\le|\pi|$ that reaches some $s'\ne s_i$ with $t\subseteq s'$. The list has polynomial size because $\pi$ is explicit and $k$ is fixed. When $\pi$ is the unique optimum, a complete list exists exactly when $\HS(\Hy_i)>k$. Inputs satisfying (a), (b) or (c) are already outside the language. For $\wg$ under the promise, \cref{prop:terminal} expresses $\wg(\Pi)$ through the hitting numbers of $\Hy_1,\dots,\Hy_{|\pi|-1}$ and $\Hy^{<}_{|\pi|}$, and the same witness works, with plans of length less than $|\pi|$ at depth $|\pi|$. Without the promise, let $(t_0,\dots,t_L)$ be a goal-admissible sequence of width at most $k$. Conditions~(i) and~(ii) of \cref{def:goal} give $c(t_i)=i$ for all $i$, so the tuples are pairwise different and $L<(N+1)^k$. A polynomial-time algorithm with an NP oracle first checks (a)--(c). For every tuple $t$ with $|t|\le k$, and for every set $t\cup G$ with such a tuple $t$, it then determines the cost or finds that the cost is at least $(N+1)^k$, by asking whether some plan of length at most $j$ reaches a state that contains the set. Given these costs, conditions~(ii) and~(iii) for a pair of tuples reduce to comparisons of costs and to queries whether some state of depth $c(t)$ that contains $t$ has no successor containing $t'$, or no successor containing $t'\cup G$. Each such query has an NP witness, namely a plan to that state. A goal-admissible sequence of width at most $k$ is then a path in a graph on at most $(N+1)^k$ tuples, so the set belongs to $\Pclass^{\NPclass}$.

For hardness, let $m=\max(2,2^{\lceil\log_2k\rceil})$, a constant with $m+1>k$. Mapping $\varphi$ to $\Pphi$ and its optimal plan takes polynomial time, and by \cref{thm:main} the image satisfies $w\le k$ if and only if $\varphi$ is unsatisfiable.
\end{proof}

When the threshold is part of the input, the choice of a small tuple is existential and its separation from all competitors is universal. We encode this alternation in a hitting-set instance. One atom from each of $n$ disjoint pairs represents an assignment to the existential variables. To test it against a 3-DNF formula and a fixed universal assignment, choose one literal from every term. The formula is false exactly when some such selection consists entirely of false literals. Let
\begin{equation*}
\Psi=\exists x_1\cdots x_n\,\forall y_1\cdots y_{n'}\ \Phi,\qquad \Phi=\Phi_1\vee\dots\vee\Phi_\kappa,
\end{equation*}
where $n\ge1$ and each term $\Phi_j$ is a conjunction of one to three literals over the $x_i$ and $y_i$. Let $U_\Psi=\{x_i^0,x_i^1:1\le i\le n\}\cup\{\ctl{z}\}$, and associate the atom $x_i^1$ with the literal $x_i$ and $x_i^0$ with $\neg x_i$. For $y\in\{0,1\}^{n'}$ and a selection $\sigma=(\sigma_1,\dots,\sigma_\kappa)$ of one literal $\sigma_j$ from each term $\Phi_j$, let
\begin{equation}\label{eq:qbf-edge}
E_{y,\sigma}=
\begin{cases}
U_\Psi, & \text{if some }\sigma_j\text{ is a }y\text{-literal true under }y,\\
\{\ctl{z}\}\cup\{\text{atoms of the }x\text{-literals among }\sigma_1,\dots,\sigma_\kappa\}, & \text{otherwise},
\end{cases}
\end{equation}
and let $\Hy_\Psi$ consist of the pairs $P_i=\{x_i^0,x_i^1\}$ for $1\le i\le n$ and the edges $E_{y,\sigma}$ for all $y$ and $\sigma$. The extra atom $\ctl{z}$ belongs to every edge $E_{y,\sigma}$ but to no pair. It makes every edge nonempty and yields a hitting set of size $n+1$ in all cases, whereas a hitting set of size $n$ must consist of exactly one atom from each pair.

\begin{lemma}[Quantified formulas as succinct hitting sets]\label{lem:qbf}
Every edge of $\Hy_\Psi$ is nonempty, $\Hy_\Psi=\Hy_{C_\Psi}$ for a circuit $C_\Psi$ computable in time polynomial in the length of $\Psi$, and
\begin{equation}\label{eq:qbf-tau}
\HS(\Hy_\Psi)=
\begin{cases}
n, & \Psi\text{ is true},\\
n+1, & \Psi\text{ is false}.
\end{cases}
\end{equation}
\end{lemma}
\begin{proof}
Pairs are nonempty, and every other edge contains $\ctl{z}$. For the circuit, an input consists of a mode bit, a pair index in $\lceil\log_2n\rceil$ bits, the assignment $y$, and, for each term, the position of $\sigma_j$ in $\lceil\log_2|\Phi_j|\rceil$ bits. Codes out of range are read as the first index, so they produce no further edges. In pair mode the circuit outputs $P_i$ for the decoded index $i$, and in formula mode it outputs $E_{y,\sigma}$. Whether an element of $U_\Psi$ belongs to the output depends only on the mode bit, on the decoded pair index, on the decoded positions $\sigma_j$ and on the values of the selected $y$-literals. Comparing each code with the at most $n+3\kappa$ indices that it can take gives a Boolean formula of polynomial size for each output, so $C_\Psi$ has polynomial size.

The $n$ pairs are disjoint, so $\HS(\Hy_\Psi)\ge n$. The set $\{\ctl{z}\}\cup\{x_i^0:1\le i\le n\}$ meets every pair and every other edge, so $\HS(\Hy_\Psi)\le n+1$. It remains to show that a hitting set of size $n$ exists if and only if $\Psi$ is true.

For an assignment $a\in\{0,1\}^n$ to the existential variables, let $X_a=\{x_i^{a_i}:1\le i\le n\}$. A set of size $n$ that meets all $n$ disjoint pairs contains exactly one atom of each pair and therefore equals $X_a$ for some $a$, and in particular it does not contain $\ctl{z}$. The atom of an $x$-literal belongs to $X_a$ exactly when the literal is true under $a$. Fix $a$, an assignment $y\in\{0,1\}^{n'}$ to the universal variables, and a selection $\sigma$. If some selected $y$-literal is true under $y$, then $E_{y,\sigma}=U_\Psi$ meets $X_a$. Otherwise, every selected $y$-literal is false, and $X_a$ meets $E_{y,\sigma}$ exactly when some selected $x$-literal is true under $a$. In both cases,
\begin{equation*}
X_a\cap E_{y,\sigma}\ne\varnothing
\quad\Longleftrightarrow\quad
\text{some }\sigma_j\text{ is true under }(a,y).
\end{equation*}
Quantifying over all selections,
\begin{align*}
X_a\text{ meets }E_{y,\sigma}\text{ for every }\sigma
&\iff\text{every selection }\sigma\text{ contains a literal that is true under }(a,y)\\
&\iff\text{there is no selection of one false literal from each term}\\
&\iff\text{some term }\Phi_j\text{ contains no false literal}\\
&\iff\Phi(a,y)\text{ is true}.
\end{align*}
Since $X_a$ meets every pair, $X_a$ hits $\Hy_\Psi$ if and only if $\Phi(a,y)$ holds for every $y\in\{0,1\}^{n'}$. Hence a hitting set of size $n$ exists if and only if some $a$ satisfies $\forall y\,\Phi(a,y)$, that is, if and only if $\Psi$ is true. Together with $n\le\HS(\Hy_\Psi)\le n+1$, this proves \cref{eq:qbf-tau}.
\end{proof}

\begin{theorem}[Variable thresholds]\label{thm:sigma2}
For $w\in\{\ws,\wc\}$, the set of triples $(\Pi,\pi,k)$ such that $\Pi$ is a positive-precondition STRIPS task, $\pi$ is its unique optimal plan, $|\pi|>1$ and $w(\Pi)\le k$ is $\SigmaTwo$-complete. It is $\SigmaTwo$-hard already on triples with $(\Pi,\pi)\in\Pone$. For $\wg$, the same holds for the promise problem whose inputs additionally have a single reachable goal state.
\end{theorem}
\begin{proof}
\emph{Membership.} Let $h=|\pi|$, and let $s_0,\dots,s_h$ be the states visited by $\pi$, which are computed in polynomial time. If $\pi$ is not a plan that ends in a goal state, or if $h\le1$, the triple is rejected. Otherwise we claim that the triple belongs to the set if and only if there are tuples $t_1,\dots,t_h$ with $t_i\subseteq s_i$ and $|t_i|\le k$ such that every action sequence $\pi'$ of length at most $h$ satisfies the following two conditions.
\begin{enumerate}\renewcommand{\labelenumi}{(\alph{enumi})}
\item If $\pi'$ is a plan that ends in a goal state, then $\pi'=\pi$.
\item For every $1\le i\le h$, if $\pi'$ is a plan of length at most $i$ that ends in a state $s'\ne s_i$, then $t_i\not\subseteq s'$.
\end{enumerate}
Condition~(a) for all $\pi'$ says exactly that $\pi$ is the unique optimal plan. Under it, all optimal goal plans visit $s_0,\dots,s_h$, and condition~(b) for all $\pi'$ says that, for each $i$, no state of depth at most $i$ other than $s_i$ contains $t_i$, that is, $t_i$ hits $\Hy_i$. By \cref{thm:hitting}, tuples $t_i$ with these properties exist if and only if $w(\Pi)\le k$, which proves the claim. The claim has the form $\exists(t_1,\dots,t_h)\ \forall\pi'\ \mathrm{Ver}$, where the tuples have polynomial size, $\pi'$ has polynomial length, and the predicate $\mathrm{Ver}$ that checks (a) and (b) runs in polynomial time. This is the form required by the definition of $\SigmaTwo$ in \cref{sec:hs}, so the set is in $\SigmaTwo$. A threshold $k\ge N$ only weakens the bound on $|t_i|$. For $\wg$ under the promise, \cref{prop:terminal} replaces the condition on $t_h$ by $t_h\not\subseteq s'$ for all $s'$ reached by plans of length less than $h$.

\emph{Hardness.} Given $\Psi$, compute $C_\Psi$ by \cref{lem:qbf} and the pair $(\Pi_{C_\Psi},\pi)\in\Pone$ by \cref{thm:compiler}. By \cref{eq:compiler,eq:qbf-tau}, $w(\Pi_{C_\Psi})=n+1$ if $\Psi$ is true and $w(\Pi_{C_\Psi})=n+2$ otherwise, for every $w\in\W$. Hence $\Psi$ is true if and only if $(\Pi_{C_\Psi},\pi,n+1)$ belongs to the set. Truth of $\exists\forall$-3-DNF sentences is $\SigmaTwo$-complete \citep{Stockmeyer1976,Wrathall1976}, and adding an unused existential variable ensures $n\ge1$.
\end{proof}

The distinction between fixed and variable thresholds persists even on $\Pone$. The existential choice of a small tuple in \cref{thm:sigma2} raises exact certification to the second level of the polynomial hierarchy. The compiler also gives a quantitative consequence for algorithms that use SAT as an oracle.

\begin{corollary}[Approximation with an NP oracle]\label{cor:oracle-approx}
Fix $0<\varepsilon<1$ and $w\in\W$. Unless $\SigmaTwo\subseteq\Pclass^{\NPclass}$, no deterministic polynomial-time algorithm with an NP oracle gives a one-sided $N^{1-\varepsilon}$-approximation for $w$ on $\Pone$.
\end{corollary}
\begin{proof}
Consider a succinct set-cover instance $\mathcal S$ of encoding length $\|\mathcal S\|$, given by 3-DNF formulas $\Delta_1,\ldots,\Delta_p$ whose satisfying assignments together cover $\{0,1\}^n$. For every fixed $0<\varepsilon'<1$, approximating the minimum cover size to within $\|\mathcal S\|^{1-\varepsilon'}$ is $\SigmaTwo$-hard \citep[Theorem~7.2]{TaShmaUmansZuckerman2007}. Define a circuit $C$ with output $j$ equal to $\Delta_j$, so its edge on assignment $v$ is $E_v=\{j:\Delta_j(v)=1\}$. Every $E_v$ is nonempty. A set of formula indices covers all assignments exactly when it meets every $E_v$, so the minimum cover size is $\tau_C=\HS(\Hy_C)\ge1$.

After discarding unused variables, the circuit has $O(\|\mathcal S\|)$ inputs, outputs and fan-in-two gates. Indeed, each term and each disjunction of terms use at most a constant number of gates per literal occurrence. A dummy input suffices if a constant is needed. By \cref{thm:compiler,app:compiler}, the constructed pair lies in $\Pone$, has $N=O(\|\mathcal S\|)$ atoms, and satisfies $w=1+\tau_C$. A one-sided estimate of the stated quality would give
\[
    \tau_C\le\widehat w\le N^{1-\varepsilon}(1+\tau_C)
      \le 2N^{1-\varepsilon}\tau_C\le\|\mathcal S\|^{1-\varepsilon/2}\tau_C
\]
for all sufficiently large $\|\mathcal S\|$. This estimate distinguishes the hard set-cover gaps for $\varepsilon'=\varepsilon/2$ with the same NP oracle. The finitely many shorter inputs can be solved directly. Hence such an algorithm would place $\SigmaTwo$ in $\Pclass^{\NPclass}$.
\end{proof}

\section{Mechanical Verification}\label{sec:verification}

We enumerated the reachable state graphs of small tasks and computed width in two ways. The first calculation starts from \cref{def:set,def:chain,def:goal}. For each tuple of at most $k$ atoms it records the minimum cost and all optimal achievers. It then tests $w\le k$ by a greatest fixed point for admissible sets, a backward induction over costs for uniform chains, or a forward search for goal-admissible sequences. The search for goal-admissible sequences allows lengths other than $\hstar$. Each test has both a direct reference implementation and a faster implementation. Neither requires a unique optimal plan or a single reachable goal state.

The second calculation builds the competitor hypergraphs of \cref{thm:hitting,prop:terminal} and computes their hitting numbers exactly. For the comparisons reported below, the definitional tests had a budget of 300,000 tuples per check. On larger instances of the reduction we computed the hitting numbers on the entire reachable graph.

\begin{table}[!htb]
\caption{Mechanical checks, organized by the claim they test. ``Definitional'' means that widths are decided by brute force directly from \cref{def:set,def:chain,def:goal}, within the tuple budget described in the text. All checks passed.}
\label{tab:verification}
\small
\begin{tabular}{@{}>{\raggedright\arraybackslash}p{3.5cm}>{\raggedright\arraybackslash}p{8.6cm}>{\raggedright\arraybackslash}p{2.9cm}@{}}
\toprule
Claim & Instances and method & Outcome\\
\midrule
Gap reduction (\cref{thm:main}, \cref{lem:size,lem:merge,lem:structure,lem:profile}) &
\VerAInstances &
\VerAOutcome\\
\addlinespace
General statements (\cref{lem:containment,prop:states}, \cref{thm:hitting}, \cref{prop:terminal}, \cref{cor:zero}) &
\VerBInstances &
\VerBOutcome\\
\addlinespace
\Cref{rem:unique}, \cref{ex:two-goal,ex:goal-states} &
the three hand-built tasks, definitional widths &
values as stated\\
\addlinespace
Path embedding (\cref{prop:path}) &
\VerDInstances &
\VerDOutcome\\
\addlinespace
Succinct compiler and $\SigmaTwo$ reduction (\cref{thm:compiler}, \cref{lem:qbf}) &
\VerEInstances &
\VerEOutcome\\
\addlinespace
Several optimal trajectories, recursion and greedy algorithm (\cref{thm:branching,cor:nested,prop:gamma,thm:recursion,cor:explicit}) &
Suite~X. \ExtXTasks\ random tasks with at most \ExtXMaxStates\ reachable states, \ExtXMulti\ of them with several optimal state sequences and \ExtXNested\ with $\xi(\Pi)=1$ and several states in some optimal layer. All \ExtYTasks\ eligible graph--goal pairs on four states. Families of \cref{thm:branching}(ii) for \ExtZTight\ pairs $(k,n)$ and \ExtZNested\ families with nested layers. Definitional widths &
bounds, recursion and family widths as stated, greedy sets admissible within $H_{|\R|-1}$, above $\ws$ in \ExtXGreedyAbove\ random tasks\\
\addlinespace
Bounded-action embedding (\cref{prop:bounded}) &
\ExtHGraphs\ hypergraphs on one to six elements, plans of up to \ExtHMaxPlan\ actions, definitional widths &
single path, bounded actions, $w=1+\HS(\Hy)$\\
\addlinespace
Sensitivity of the checks &
\VerMInstances &
\VerMOutcome\\
\bottomrule
\end{tabular}
\end{table}

Suite~A checks the gap construction against the action schemas in \cref{tab:compiler-schemas}. The formulas include unit clauses, repeated or complementary literals, tautological clauses, unused variables and repeated gate inputs. One unsatisfiable instance contains all eight clauses on three variables. For each of the \VerProfileTasks\ tasks we counted the atoms, actions and optimal action sequences, checked the graph properties in \cref{thm:main}, and recorded the states reached at the checkpoint. The hitting numbers at every depth matched \cref{eq:profile}. Of the 192 tasks with $m\le4$, 152 also fit the budget for a direct check of all three widths. \Cref{tab:sizes} gives the counts for the two examples in \cref{ex:small}.

Suite~B tests the characterizations on random solvable tasks. It compares all three widths with the bounds in \cref{prop:states} and applies \cref{thm:hitting,prop:terminal,cor:zero} whenever their hypotheses hold. Every retained task fit the tuple budget. On the 3,000 tasks with four to six atoms we also compared the fast and reference implementations. Suite~C checks \cref{rem:unique,ex:two-goal,ex:goal-states} directly, and Suite~D checks the path embedding.

Suite~E tests the succinct compiler on random circuits and checks the edges in \cref{eq:qbf-edge} on every input of the circuits compiled from quantified formulas. The direct width test fit 370 of the 400 random circuits. For tasks built from quantified formulas we used the hypergraph calculation. As a separate check, we used Glucose~4 through PySAT \citep{IgnatievMorgadoMarquesSilva2018} on the \VerSolverCNF\ distinct CNFs and on the \VerSolverQBF\ quantified sentences, using counterexample-guided search for the latter. All solver answers agreed with exhaustive evaluation. Suite~M tested sensitivity to deliberate changes in the construction. The checks detected all five changes, including omission of a tag restoration and omission of the extra action charged for a first repair. A sixth test allowed an empty circuit edge and found the resulting second optimal plan.

\begin{table}[!htb]
\caption{Sizes of the generated tasks for the satisfiable formula $x_1$ and the unsatisfiable formula $x_1\wedge\neg x_1$ of \cref{ex:small}. The columns give the number of reachable states, atoms and actions, the optimal plan length and the checkpoint depth. All values agree with \cref{lem:size,lem:structure}, and the maximal slack is $1$ in every task.}
\label{tab:sizes}
\begin{tabular}{@{}lrrrrrr@{}}
\toprule
Formula & $m$ & $|\R|$ & $N$ & $|A|$ & $\hstar$ & $D$\\
\midrule
$x_1$ & $2$ & $43$ & $34$ & $35$ & $12$ & $9$\\
 & $4$ & $82$ & $52$ & $56$ & $16$ & $11$\\
 & $8$ & $170$ & $84$ & $95$ & $22$ & $13$\\
 & $16$ & $392$ & $144$ & $170$ & $32$ & $15$\\
\addlinespace
$x_1\wedge\neg x_1$ & $2$ & $43$ & $43$ & $46$ & $15$ & $12$\\
 & $4$ & $67$ & $61$ & $67$ & $19$ & $14$\\
 & $8$ & $113$ & $93$ & $106$ & $25$ & $16$\\
 & $16$ & $203$ & $153$ & $181$ & $35$ & $18$\\
\bottomrule
\end{tabular}

\end{table}

Suite~X checks the results for branching optimal plans and the bounded-action path embedding. It compares the three widths on random tasks and on all eligible graphs built from four pairwise incomparable two-atom states, and tests \cref{thm:recursion} at every threshold. Set and chain width agreed on every random task and every four-state graph. Among the tasks tested, only the family in \cref{thm:branching}(ii) attained the upper bound with $\wc>\ws$. The greedy algorithm always returned an admissible set within the factor $H_{|\R|-1}$, including the \ExtXGreedyAbove\ random tasks on which it returned more than the minimum width. For the bounded-action embedding, we tested every simple family of nonempty edges on at most three elements and random families on four to six elements. Each task had a single reachable path and width $1+\HS(\Hy)$.

The checks are written in Python and use only the standard library and, for the solver cross-check, PySAT. The generators use the fixed seeds 20261117 (Suites A--M) and 20261118 (Suite~X). The final run used Python~3.13.7, with PySAT~1.9.dev22 for the solver cross-check. Suites A--M took about \VerSeconds~seconds with a peak memory of \VerPeakMemory~MB, and Suite~X took about \ExtSeconds~seconds. The code, raw results and logs are not part of this preprint.

\section{Related Work}\label{sec:related}

\subsection{Width and Its Certificates}
\citet{LipovetzkyGeffner2012} introduced tuple-graph width, denoted here by $\wc$, and proved the optimality guarantee for $\IW(k)$. \citet{BonetGeffner2024} related admissible tuple sets to optimal policies, building on \citet{BonetGeffner2021}. Theorem~28 of \citet{BonetGeffner2024} gives a necessary condition for set width at most $k$. Some optimal trajectory must consist entirely of states that optimally achieve tuples of size at most $k$. When the optimal state sequence is unique, \cref{thm:hitting} identifies a necessary and sufficient condition. At each depth, one such tuple must exclude every competitor. Goal-admissible width as defined by \citet{Song2026} constrains the last extension to $t_L\cup G$ (\cref{sec:goal}). \citet{DrexlerSeippGeffner2024} use sketches to express decompositions into subproblems of bounded width. The learning method of \citet{DrexlerSeippGeffner2022} uses the weaker notion of satisficing width, which permits suboptimal subplans.

\subsection{Complexity of Width}
For tasks with polynomially bounded plans, \citet{Song2025} study the function that maps a task to its width. They report $\mathrm{OptP}[O(\log\log\|\Pi\|)]$-hardness and an upper bound of $\mathrm{OptP}[O(\log\|\Pi\|)]$ with an oracle for the same class, where $\|\Pi\|$ is the encoding length of the task and the bracketed term bounds the output length in bits \citep{Krentel1988}. They define width through admissible sets. The proof of their Theorem~3 uses their Equation~(1) to identify this minimum with the minimum over admissible sequences. In \cref{ex:two-goal}, the sequence $(\{s\},\{z\},\{g_1\})$ meets their sequence definition with width~$1$, although the set width is~$2$. The upper bound for set width therefore needs a different justification.

\citet{Song2026} proved PSPACE-completeness for goal-admissible width with an input threshold and coNP-completeness for each fixed $k\ge1$. Their fixed-threshold construction already supplies a unique optimal plan and, at $k=1$, separates width $1$ from width at least $2$. It uses a counter with $2^k-1$ enumeration actions and permits dead ends. In \cref{thm:main}, the checkpoint competitors force a separate atom for each possible missing tag. The resulting gap is linear for all three widths, and the checker keeps every state within one action of optimality using only bounded-size actions.

\citet{Gadzhiev2026} computes tuple-chain width by compilation to SAT. Other inapproximability results for STRIPS concern optimal plan cost in monotone planning \citep{BackstromJonssonOrdyniak2021} and the maximum number of goals a plan can satisfy \citep{TanGrastien2026}.

\subsection{The Polynomial Hierarchy in Planning}
For propositional STRIPS, plan existence and plan existence with a binary-encoded length bound are PSPACE-complete \citep{Bylander1994}. For a polynomial length bound, a proposed plan can be checked by simulation, and binary search with logarithmically many NP-oracle queries then finds the optimal length whenever a plan within the bound exists. In conformant planning, uncertainty about states or effects gives rise to second-level complexity even for polynomial-length plans \citep{BaralKreinovichTrejo2000,Turner2002}. In \cref{thm:sigma2}, the task is deterministic and fully observable and its optimal plan is supplied. The existential choice is a tuple at each depth. The universal condition tests the chosen tuples against competing executions.

\subsection{Novelty and Other Measures of Hardness}
Best-first width search combines novelty with goal-directed estimates \citep{LipovetzkyGeffner2017}. Width-based methods also search simulators and pixel domains \citep{LipovetzkyRamirezGeffner2015,JunyentJonssonGomez2019}. \citet{Singh2021} approximate novelty through sampling and Bloom filters, and \citet{DoldHelmert2024} compare novelty with other planning hardness measures. The novelty of a state depends on which states have already been generated. \Cref{rem:novelty} identifies the comparison set that recovers width on a unique optimal state sequence.

\subsection{Hitting Sets and Covering}
Greedy hitting set \citep{Chvatal1979} gives the upper bounds in \cref{cor:zero,cor:explicit}. The path embedding transfers the decision and parameterized lower bounds from explicit hitting set \citep{Karp1972,DowneyFellows2013,ChenEtAl2006}. For approximation, \citet{Feige1998} proved the logarithmic threshold for set cover under the assumption $\NPclass\not\subseteq\mathrm{DTIME}(n^{O(\log\log n)})$. \citet{DinurSteurer2014} obtained the $(1-\varepsilon)\ln n$ lower bound under $\Pclass\ne\NPclass$. The embedding preserves the number of edges as $\hstar-1$ (\cref{cor:zero-lower}).

The implicit hitting-set algorithm of \citet{ChandrasekaranEtAl2011} alternates between choosing a hitting set for the known edges and asking an oracle for an unhit edge. For the hypergraphs in \cref{thm:hitting}, this oracle asks whether a competing execution can reach the proposed tuple within a given number of actions. Succinct set cover is complete for $\SigmaTwo$ \citep{Umans1999,SchaeferUmans2002}. \citet{Umans1999} proved polynomial-factor hardness for succinct set cover. The stronger exponent of \citet{TaShmaUmansZuckerman2007} transfers to width through \cref{thm:compiler,cor:oracle-approx}. The linear gap of \cref{thm:main} uses only 3-SAT. At its checkpoint, all minimal edges are singletons. Deciding which singletons occur already suffices for the gap.

\section{Conclusion}\label{sec:conclusion}

Along a unique optimal state sequence, the competitors at each positive depth determine the size of the smallest tuple there in a set or chain certificate. With zero slack, simulating the supplied plan lists all of them. Within one action of optimality, their existence can encode satisfiability, and the best polynomial-time approximation guarantees have orders $N$ and $\sqrt N$ for one-sided and symmetric estimates. The sharp change occurs even when every complete execution reaches the same goal and the supplied optimal plan is easy to construct.

For a learned decomposition evaluated on a small training instance, an enumerated state graph supports an exact test of set width and a polynomial-time logarithmic approximation. These procedures also apply when the optimal plans branch. A supplied optimal demonstration gives an upper bound through its full-state tuple chain, but whether much smaller tuples suffice depends on how early other executions can reach them.

\subsection{Limitations}
The lower bounds describe worst-case certification for unit-cost STRIPS tasks with positive preconditions. How often competing states force large certificates on planning benchmarks is an empirical question. With several optimal state sequences, \cref{thm:recursion} still determines set width. For chain width, \cref{prop:gamma} gives an upper bound $\Gamma(\Pi)$ that is exact when the optimal states at each depth are ordered by inclusion. The examples attaining $\wc-\ws=\xi-1$ use actions of unbounded size. At zero slack the decision and ETH lower bounds hold for bounded-size actions, while the sharp logarithmic approximation lower bound uses the unrestricted path embedding. The variable-threshold upper bound in \cref{thm:sigma2} uses the supplied plan to bound the length of a competing execution.

\subsection{Open Questions}
The optimal constants in the linear and square-root approximation bounds remain open. At zero slack, can actions of bounded size enforce the same $(1-\varepsilon)\ln(\hstar-1)$ lower bound as the unrestricted path embedding? For branching optimal plans, one can describe a chain by its set of optimal achievers at each depth and the transitions linking successive sets. There can be exponentially many candidate sets. An exact formula for chain width in terms of individual states, analogous to the recursion for set width, would avoid this enumeration. A further question is whether the features used in a learned sketch can identify families of subproblems whose competitors can be enumerated efficiently.


\printbibliography

\clearpage
\appendix
\crefalias{section}{appendix}
\crefalias{subsection}{appendix}
\section{Notation}\label{app:notation}

\begin{table}[H]
\caption{Notation used in the article. The left pair of columns covers planning, width and hitting sets. The right pair covers the reductions of \cref{sec:reduction,sec:map}.}
\label{tab:notation}
\small
\addtolength{\tabcolsep}{-1pt}
\begin{tabular}{@{}llll@{}}
\toprule
Symbol & Meaning & Symbol & Meaning\\
\midrule
$\Pi=\langle F,I,A,G\rangle$, $N=|F|$ & task and its number of atoms & $\varphi$, $\|\varphi\|$ & formula and its length\\
$\act{\pre(a)}{\add(a)}{\del(a)}$ & an action as a triple & $n$, $\kappa$, $\ell$ & variables, clauses, literals\\
$s[a]$ & successor of $s$ under $a$ & $m=2^b$ & number of tags\\
$\R$ & reachable states & $U=\{u_0,\dots,u_{m-1}\}$ & tags\\
$d(s)$, $d_G(s)$ & depth and distance to a goal state & $D$ & checkpoint depth\\
$\hstar=d_G(I)$ & optimal plan length & $S$, $S_j$, $T$ & checkpoint states (\cref{lem:merge})\\
$c(t)$, $\Opt(t)$ & cost and optimal achievers & $\sgoal=U\cup\{g_\star\}$ & unique goal state of $\Pphi$\\
$\ws$, $\wc$, $\wg$, $\W$ & the three widths and their set & $M$, $\bar M$ & true and false merge atoms\\
$\slack(s)$, $\rhoPi$ & slack of a state and of $\Pi$ & $e_j$, $r_j$, $\ctl{rep}_j$ & marker, checker, repair atoms\\
$\Hy_i$ & competitor hypergraph (\cref{eq:H}) & $\omega=n+\ell-2$ & number of erased pairs\\
competitor of $s_i$ & $u\in\R$, $u\ne s_i$, $d(u)\le i$ & $r'_\ell$ & primed checker atom\\
$\HS(\Hy)$ & hitting number & $\Pone$, $\Pzero$ & promise classes\\
$C$, $n_C$, $E_v$, $\Hy_C$ & circuit, its inputs, an edge, its hypergraph & $\Pi_{\Hy}$, $\PC$ & path task and compiled task\\
$\widehat w$, $\alpha(N)$ & estimate and factor & $\Psi$, $\Hy_\Psi$ & sentence and its hypergraph\\
$\mathcal O_i$, $\xi(\Pi)$ & optimal layer and antichain number & $q_k$ and $\theta_t$, $\psi_\delta$, $K$ & clock and phase start depths\\
$H_j=\sum_{i=1}^{j}1/i$ & harmonic number & $\Pi^{\mathrm b}_{\Hy}$ & bounded-action path task\\
$\Hy^{<}_{\hstar}$ & terminal hypergraph (\cref{prop:terminal}) & $\Tok(i)$ & control tokens at depth $i$\\
\bottomrule
\end{tabular}
\end{table}

\section{Deferred Constructions}\label{app:proofs}

\subsection{Proof of \texorpdfstring{\cref{prop:path}}{Proposition 6.2}}\label{app:path}

Let $q_0,\dots,q_{m-1}$ and $g$ be fresh atoms, and let $F=U\cup\{q_0,\dots,q_{m-1},g\}$, $I=(U\setminus E_0)\cup\{q_0\}$ and $G=\{g\}$. For $0\le j<m-1$ define
\begin{equation*}
a_j=\bigl\langle\{q_j\},\ \{q_{j+1}\}\cup(E_j\setminus E_{j+1}),\ \{q_j\}\cup(E_{j+1}\setminus E_j)\bigr\rangle.
\end{equation*}
Let $a_{m-1}=\act{\{q_{m-1}\}}{E_{m-1}}{\{q_{m-1}\}}$, fix an element $u_\ast\in U$, and let $a_m=\act{U}{\{g\}}{\{u_\ast\}}$. Add and delete lists are disjoint in every action. Define
\begin{equation*}
s_j=(U\setminus E_j)\cup\{q_j\}\quad(0\le j<m),\qquad s_m=U,\qquad s_{m+1}=(U\setminus\{u_\ast\})\cup\{g\}.
\end{equation*}
For $j<m-1$, an element of $U$ belongs to $s_j[a_j]$ if and only if it lies outside $E_j$ and outside $E_{j+1}\setminus E_j$, or inside $E_j\setminus E_{j+1}$, that is, if and only if it lies outside $E_{j+1}$. Hence $s_j[a_j]=s_{j+1}$. Also $s_{m-1}[a_{m-1}]=s_m$ and $s_m[a_m]=s_{m+1}$.

In $s_j$ with $j<m$, the only clock is $q_j$, so $a_j$ is the only clock action that applies, and $a_m$ does not apply because the nonempty set $E_j$ is missing from $s_j$. In $s_m$ no clock is present and $a_m$ applies. In $s_{m+1}$ no action applies, because $u_\ast$ is missing. Thus the reachable states form the path $s_0,\dots,s_{m+1}$ with $d(s_j)=j$. Its only goal state is $s_{m+1}$, and its unique goal plan $a_0\cdots a_m$ has length $m+1>1$. Every state lies on this plan, giving $\rho(\Pi_{\Hy})=0$. The task has $|U|+m+1$ atoms and $m+1$ actions.

By \cref{cor:zero}, every $w\in\W$ equals $\max_i\HS(\{s_i\setminus s_j:j<i\})$. For $1\le i<m$ the clock $q_i$ belongs to no other state, and $g$ belongs only to $s_{m+1}$, so these depths have hitting number $1$. For $i=m$, $s_m\setminus s_j=E_j$ for every $j<m$, because $q_j\notin s_m$ and $s_j\cap U=U\setminus E_j$. Hence the maximum is $\max(1,\HS(\Hy))=\HS(\Hy)$, using $\HS(\Hy)\ge1$. By \cref{lem:containment} and the proof of \cref{thm:hitting}, every tuple of cost $m$ in an admissible set or a uniform chain has $s_m$ as its only optimal achiever and hits $\Hy_m=\{E_j:j<m\}$, and \cref{lem:weak} gives the same for goal-admissible sequences. Conversely, if $Z$ hits $\Hy$, then $(\varnothing,\{q_1\},\dots,\{q_{m-1}\},Z,\{g\})$ is a uniform chain of width $\max(1,|Z|)=|Z|$.\qed

\subsection{The Compiler of \texorpdfstring{\cref{thm:compiler}}{Theorem 6.6}}\label{app:compiler}

\Cref{tab:compiler-sigma} gives the grounded actions for the circuit construction of \cref{sec:one}.

Let $C$ have $n_C$ inputs $1,\dots,n_C$, $n_G$ gates $1,\dots,n_G$ in topological order, and $p$ outputs, indexed by $U=\{u_0,\dots,u_{p-1}\}$. Each gate $t$ computes $f_t\in\{\wedge,\vee\}$ of two literals $y_t,y'_t$, where a literal is an input or an earlier gate, possibly negated. Each output is such a literal $o_\ell$. Every input $i$ has a one-hot pair $x_i^0,x_i^1$ and every gate $t$ a one-hot pair $z_t^0,z_t^1$. As in \cref{sec:compilation}, the atom of a literal $y$ for value $a$, written $y^a$, is the atom of its node for value $a$, or for value $1-a$ if the literal is negated. Let $\chi_1,\dots,\chi_{n_C+n_G}$ enumerate the input and gate pairs.

As in \cref{sec:compilation}, the atoms of testing executions are control tokens, which record the step that an execution has reached, and data atoms. The control tokens are the testing clocks $q_j$, indexed by depth, the pending atoms $p_t^a$, which store the value $a$ of the first input of gate $t$, and the atoms $\ctl{del}_\ell$ and $\ctl{keep}_\ell$, which mark the middle of the tag step for $u_\ell$. Each token identifies the step that can execute next. The data atoms are the tags, the markers $e_\ell$ and the atoms of the pairs. The phases start at the offsets
\begin{equation*}
\theta^{\mathrm{gate}}_t=n_C+1+2(t-1),\qquad
\theta^{\mathrm{tag}}_\ell=n_C+2n_G+1+2\ell,\qquad
\theta^{\mathrm{erase}}_i=n_C+2n_G+2p+i,
\end{equation*}
with $D=2n_C+3n_G+2p+2$. Here $\theta^{\mathrm{gate}}_t$ is the depth at which gate $t$ starts, $\theta^{\mathrm{tag}}_\ell$ the depth at which the two tag actions for $u_\ell$ start, and $\theta^{\mathrm{erase}}_i$ the depth at which the pair $\chi_i$ is erased. Input $i$ is read at depth $i$. The gate phase, possibly empty, ends at $\theta^{\mathrm{tag}}_0$, the last tag step ends at $\theta^{\mathrm{erase}}_1$, and the last erasure reaches $q_{D-1}$. The safe mode passes through private atoms $\ctl{safe}_1,\dots,\ctl{safe}_{D-1}$, and we set $\ctl{safe}_D=M$ and $r_0=M$. The checker atoms $r_\ell$, $r'_\ell$, $\ctl{chk}_\ell$, $\ctl{chk}'_\ell$, $\ctl{rep}_\ell$, $\ctl{rep}'_\ell$ and $\ctl{fix}_\ell$ are those of \cref{sec:one}. \Cref{tab:compiler-sigma} lists the action schemas.

{\small
\renewcommand{\arraystretch}{0.92}
\setlength{\LTcapwidth}{\textwidth}
\begin{longtable}{@{}l l l l l@{}}
\caption{The action schemas of $\PC$. Each row is a schema with one grounded action for every value of its parameters, listed in the last column, and of $a,a'\in\{0,1\}$. In each group of rows, $\theta$ is the offset named in the group heading, and names other than $U$, $M$ and $g$ are fresh.}\label{tab:compiler-sigma}\\
\toprule
Schema & $\pre$ & $\add$ & $\del$ & Indices\\
\midrule
\endfirsthead
\multicolumn{5}{@{}l}{\tablename~\thetable\ (continued)}\\
\toprule
Schema & $\pre$ & $\add$ & $\del$ & Indices\\
\midrule
\endhead
\bottomrule
\endfoot
\bottomrule
\endlastfoot
\multicolumn{5}{@{}l}{\emph{Mode selection and safe mode}}\\
\quad Select safe & $\ctl{start}$ & $\ctl{safe}_1$ & $\ctl{start}$ & \\
\quad Safe advance & $\ctl{safe}_j$ & $\ctl{safe}_{j+1}$ & $\ctl{safe}_j$ & $1\le j<D$\\
\quad Select test & $\ctl{start}$ & $q_1$ & $\ctl{start}$ & \\
\multicolumn{5}{@{}l}{\emph{Input choice and circuit}, with $\theta=\theta^{\mathrm{gate}}_t$ in the gate rows}\\
\quad Input & $q_i$ & $q_{i+1},\ x_i^a$ & $q_i$ & $1\le i\le n_C$, $a$\\
\quad Gate preparation & $q_\theta,\ y_t^a$ & $p_t^a$ & $q_\theta$ & $1\le t\le n_G$, $a$\\
\quad Gate completion & $p_t^a,\ y_t'^{\,a'}$ & $q_{\theta+2},\ z_t^{f_t(a,a')}$ & $p_t^a$ & $1\le t\le n_G$, $a,a'$\\
\multicolumn{5}{@{}l}{\emph{Tags}, with $\theta=\theta^{\mathrm{tag}}_\ell$}\\
\quad Deletion, first step & $q_\theta,\ o_\ell^1$ & $\ctl{del}_\ell$ & $q_\theta,\ u_\ell$ & $0\le \ell<p$\\
\quad Deletion, second step & $\ctl{del}_\ell$ & $q_{\theta+2},\ e_\ell$ & $\ctl{del}_\ell$ & $0\le \ell<p$\\
\quad Retention, first step & $q_\theta,\ o_\ell^0$ & $\ctl{keep}_\ell$ & $q_\theta$ & $0\le \ell<p$\\
\quad Retention, second step & $\ctl{keep}_\ell$ & $q_{\theta+2}$ & $\ctl{keep}_\ell$ & $0\le \ell<p$\\
\multicolumn{5}{@{}l}{\emph{Cleanup and merge}, with $\theta=\theta^{\mathrm{erase}}_i$ in the erasure rows}\\
\quad Pair erasure & $q_\theta,\ \chi_i^a$ & $q_{\theta+1}$ & $q_\theta,\ \chi_i^a$ & $1\le i\le n_C+n_G$, $a$\\
\quad Merge & $q_{D-1}$ & $M$ & $q_{D-1}$ & \\
\multicolumn{5}{@{}l}{\emph{Checker, unprimed}}\\
\quad Normal check, first step & $r_\ell,\ u_\ell$ & $\ctl{chk}_\ell$ & $r_\ell$ & $0\le \ell<p$\\
\quad Normal check, second step & $\ctl{chk}_\ell$ & $r_{\ell+1}$ & $\ctl{chk}_\ell$ & $0\le \ell<p$\\
\quad First repair, first step & $r_\ell,\ e_\ell$ & $\ctl{rep}_\ell$ & $r_\ell,\ e_\ell$ & $0\le \ell<p$\\
\quad First repair, second step & $\ctl{rep}_\ell$ & $\ctl{rep}'_\ell,\ u_\ell$ & $\ctl{rep}_\ell$ & $0\le \ell<p$\\
\quad First repair, third step & $\ctl{rep}'_\ell$ & $r'_{\ell+1}$ & $\ctl{rep}'_\ell$ & $0\le \ell<p$\\
\quad Goal & $r_p$ & $g$ & $r_p$ & \\
\multicolumn{5}{@{}l}{\emph{Checker, primed}}\\
\quad Primed check, first step & $r'_\ell,\ u_\ell$ & $\ctl{chk}'_\ell$ & $r'_\ell$ & $1\le \ell<p$\\
\quad Primed check, second step & $\ctl{chk}'_\ell$ & $r'_{\ell+1}$ & $\ctl{chk}'_\ell$ & $1\le \ell<p$\\
\quad Primed repair, first step & $r'_\ell,\ e_\ell$ & $\ctl{fix}_\ell$ & $r'_\ell,\ e_\ell$ & $1\le \ell<p$\\
\quad Primed repair, second step & $\ctl{fix}_\ell$ & $r'_{\ell+1},\ u_\ell$ & $\ctl{fix}_\ell$ & $1\le \ell<p$\\
\quad Primed goal & $r'_p$ & $g$ & $r'_p$ & \\
\end{longtable}
}

The initial state is $U\cup\{\ctl{start}\}$, the goal is $\{g\}$, and the supplied plan is the safe chain followed by the two actions of the normal check for $\ell=0,\dots,p-1$ and the unprimed goal action. Its length is $\hstar=D+2p+1$, and every action has at most two preconditions and three effects. There are $O(D)$ clocks and safe controls, $O(n_C+n_G)$ pair and pending atoms, and $O(p)$ tags, markers and checker atoms. The table also has $O(n_C+n_G+p)$ grounded actions.

\begin{proof}[Proof of \cref{lem:testing}]
The safe mode reaches $U\cup\{M\}$ as in \cref{lem:merge}. After the testing mode is selected, induction on the number $j$ of actions gives the following invariant up to the merge.
\begin{itemize}
\item The only control token is the clock $q_j$, a pending atom, or an atom $\ctl{del}_\ell$ or $\ctl{keep}_\ell$. Every testing action deletes the token of its step and adds that of the next step.
\item Each pair that has been written and has not yet been erased is one-hot.
\item For every tag index $\ell'$ below the current one, $u_{\ell'}$ is present exactly when $e_{\ell'}$ is absent, and all later tags are present.
\end{itemize}
Every testing precondition contains the control token of its step, so only the actions of the current step apply. Safe and checker actions do not apply, because select test deletes $\ctl{start}$, no testing action adds a safe atom or a checker atom other than $M$, and $M=r_0$ is added only by the merge. The data atoms then select the actions as follows.
\begin{itemize}
\item \emph{Input.} The two actions of input $i$ offer the choices $a\in\{0,1\}$.
\item \emph{Gates.} Exactly one preparation and then exactly one completion apply, because the input pairs are one-hot, and by induction on $t$ the pair of gate $t$ holds the value of gate $t$ on $v$.
\item \emph{Tags.} In tag step $\ell$, exactly one of $o_\ell^1$ and $o_\ell^0$ holds, so exactly one first action applies, followed by its second action. The deletion branch removes $u_\ell$ and adds $e_\ell$, and the retention branch changes no tag, so after the tag steps the tags and markers are as in \cref{eq:testing-state}.
\item \emph{Cleanup and merge.} The $n_C+n_G$ erasure steps remove the present atom of each pair, one pair per step, and the merge removes $q_{D-1}$ and adds $M$.
\end{itemize}
Every step other than an input therefore offers exactly one applicable action, and an input offers two. No prefix that ends before the merge is complete, so a complete execution that starts in testing mode performs all $1+n_C+2n_G+2p+(n_C+n_G)+1=D$ steps up to the merge. After the merge no control token or pair atom remains, so its state after $D$ actions is $T_v$ for the input $v$ it chose. Every input can be chosen, and by \cref{lem:checker} the resulting prefix has a complete continuation.
\end{proof}

\FloatBarrier

\subsection{Proof of \texorpdfstring{\cref{prop:bounded}}{Proposition 6.4}}\label{app:bounded}

Let $U=\{u_1,\dots,u_n\}$ and let $\Hy$ consist of the edges $E_1,\dots,E_m$. The task $\Pi^{\mathrm b}_{\Hy}$ uses the atoms of $U$, absence markers $\bar u_1,\dots,\bar u_n$, scanner tokens $r_1,\dots,r_n$, a failure token $\ctl{fail}$, a goal atom $g$, edge labels $\ctl{lab}_1,\dots,\ctl{lab}_m$, and clock atoms, each of which is used by a single action. For $1\le j\le m$ let
\begin{equation*}
U_j=(U\setminus E_j)\cup\{\bar u_i:u_i\in E_j\}\cup\{\ctl{lab}_j\},
\end{equation*}
let $U_{m+1}=U$, and let $U^-_1=U$ and $U^-_j=U_{j-1}\setminus\{\ctl{lab}_{j-1}\}$ for $2\le j\le m+1$. The data $U_j$ represent each element $u_i$ by $u_i$ itself if $u_i\notin E_j$ and by $\bar u_i$ if $u_i\in E_j$, and $U^-_j$ are the data present when the preparation of $U_j$ begins.

\emph{Preparation $j$}, for $1\le j\le m+1$, turns $U^-_j$ into $U_j$. List the atoms of $U^-_j\setminus U_j$ and of $U_j\setminus U^-_j$ in any fixed order as $x_1,\dots,x_{n_j}$, and use fresh clock atoms $q_{j,0},\dots,q_{j,n_j}$. For $1\le l\le n_j$, the $l$-th step is $\act{\{q_{j,l-1}\}}{\{q_{j,l}\}}{\{q_{j,l-1},x_l\}}$ if $x_l\in U^-_j\setminus U_j$ and $\act{\{q_{j,l-1}\}}{\{q_{j,l},x_l\}}{\{q_{j,l-1}\}}$ if $x_l\in U_j\setminus U^-_j$. The entry action $\act{\{q_{j,n_j}\}}{\{r_1\}}{\{q_{j,n_j}\}}$ starts a scan.

\emph{Scans.} For $1\le i\le n$, the actions $\act{\{r_i,u_i\}}{\{r_{i+1}\}}{\{r_i\}}$, where $r_{n+1}=g$, and $\act{\{r_i,\bar u_i\}}{\{\ctl{fail}\}}{\{r_i\}}$ advance the scanner or report failure. For $1\le j\le m$, the recovery action $\act{\{\ctl{fail},\ctl{lab}_j\}}{\{q_{j+1,0}\}}{\{\ctl{fail},\ctl{lab}_j\}}$ starts preparation $j+1$.

The initial state is $I=U\cup\{q_{1,0}\}$ and the goal is $\{g\}$. Every action has at most two preconditions and at most three effects. Each of the $m+1$ preparations uses at most $2n+2$ fresh clocks. The other atoms consist of $3n$ data, marker and scanner atoms, $m$ edge labels, $\ctl{fail}$ and $g$. Thus $N\le(m+1)(2n+2)+3n+m+2$, and the task and its plan have polynomial encoding length.

\emph{The path.} Starting from $I$, preparation $1$ passes through states that consist of data and one clock and ends in $U_1\cup\{r_1\}$. For $j\le m$, the scan from $U_j\cup\{r_1\}$ visits $U_j\cup\{r_i\}$ for $i\le i_j$, where $u_{i_j}$ is the first element of $E_j$, reaches $U_j\cup\{\ctl{fail}\}$, and the recovery leads to $U^-_{j+1}\cup\{q_{j+1,0}\}$. After preparation $m+1$, the final scan visits $U\cup\{r_i\}$ for $1\le i\le n$ and ends in $U\cup\{g\}$. At every state before the goal exactly one action applies. A state with a clock contains no scanner token and no $\ctl{fail}$, and its clock is the precondition of exactly one action. A scan state $U_j\cup\{r_i\}$ contains no clock and no $\ctl{fail}$, and exactly one of $u_i$ and $\bar u_i$ belongs to $U_j$. A failure state $U_j\cup\{\ctl{fail}\}$ contains the label $\ctl{lab}_j$ and no other label, scanner token or clock. The goal state $U\cup\{g\}$ contains no clock, scanner token or $\ctl{fail}$, so no action applies. No state repeats, because clocks are fresh, the scan and failure states of edge $j$ carry the label $\ctl{lab}_j$ and distinct tokens, and the final scan states carry no label. The reachable states therefore form a single path $s_0,\dots,s_{\hstar}$ to the unique goal state $U\cup\{g\}$. The path supplies the unique goal plan and contains every reachable state, giving $\rho(\Pi^{\mathrm b}_{\Hy})=0$. This proves~(i).

\emph{The widths.} By \cref{cor:zero}, every $w\in\W$ equals $\max_{1\le i\le\hstar}\HS(\{s_i\setminus s_j:j<i\})$. Let $s=U\cup\{r_1\}$ be the first state of the final scan. The initial state gives the edge $s\setminus I=\{r_1\}$, and the first state $U_j\cup\{r_1\}$ of the scan for $E_j$ gives the edge $U\setminus U_j=E_j$. Since $r_1\notin U$, every hitting set at the depth of $s$ contains $r_1$ and a hitting set of $\Hy$, so the maximum is at least $1+\HS(\Hy)$. For the upper bound, let $Z$ be a minimum hitting set of $\Hy$, and assign to each state $s_i$ with $i\ge1$ a tuple $t\subseteq s_i$ that no earlier state contains. A state with a clock gets the clock. A scan state $U_j\cup\{r_i\}$ with $j\le m$ gets $\{\ctl{lab}_j,r_i\}$: the earlier states that contain $\ctl{lab}_j$ are states of preparation $j$, which contain no scanner token, and scan states of edge $j$ with a smaller token index. A failure state $U_j\cup\{\ctl{fail}\}$ gets $\{\ctl{lab}_j,\ctl{fail}\}$, because earlier failure states carry other labels and the earlier states with $\ctl{lab}_j$ do not contain $\ctl{fail}$. A final scan state $U\cup\{r_i\}$ gets $\{r_i\}\cup Z$: the earlier states that contain $r_i$ are scan states $U_j\cup\{r_i\}$, and $U_j$ contains no element of $E_j$, which $Z$ meets. The goal state gets $\{g\}$. As $\HS(\Hy)\ge1$, every assigned tuple has at most $1+\HS(\Hy)$ atoms, and each of them hits the corresponding hypergraph. This proves~(ii).\qed

\FloatBarrier

\section{Several Optimal Trajectories and Explicit State Graphs}\label{app:branching}

This appendix supplements \cref{sec:branching} and proves \cref{cor:explicit}. We write $L_i=\{s\in\R:d(s)=i\}$ for the breadth-first layers and $\mathcal R_{<i}=\{s\in\R:d(s)<i\}$.

\subsection{Nested Optimal Layers}\label{app:nested}

For a state $s$ in the optimal layer $\mathcal O_i$, let
\begin{equation*}
\delta(s)=\HS\bigl(\{s\setminus u:u\in\R,\ d(u)\le i,\ s\not\subseteq u\}\bigr).
\end{equation*}
Write $s\Rightarrow s'$ if some action applicable in $s$ leads to a superset of $s'$. Let $\Gamma(\Pi)$ be the minimum of $\max_{1\le i\le\hstar}\delta(s_i)$ over all sequences $s_0,\dots,s_{\hstar}$ with $s_0=I$, $s_i\in \mathcal O_i$ and $s_i\Rightarrow s_{i+1}$ for $i<\hstar$. Such sequences exist, since the states visited by an optimal goal plan form one.

\begin{proposition}\label{prop:gamma}
Every task satisfies $\ws(\Pi)\le\wc(\Pi)\le\Gamma(\Pi)\le\ws(\Pi)+\xi(\Pi)-1$. In particular, $\ws(\Pi)=\wc(\Pi)=\Gamma(\Pi)$ if every optimal layer is linearly ordered by inclusion. If every optimal layer consists of a single state $s_i$, then $\Gamma(\Pi)=\max_{1\le i\le\hstar}\HS(\Hy_i)$, which is \cref{eq:hitting}.
\end{proposition}
\begin{proof}
Let $s\in \mathcal O_i$. No state $u$ with $d(u)<i$ contains $s$, since the actions that lead an optimal goal plan from $s$ to the goal would lead from $u$ to a goal state by \cref{eq:monotone}, giving a goal plan with $d(u)+\hstar-i<\hstar$ actions. Every edge in the definition of $\delta(s)$ is therefore a nonempty subset of $s$, so a minimum hitting set $t(s)\subseteq s$ exists. A state $u$ with $d(u)\le i$ contains $t(s)$ exactly when $s\subseteq u$, and then $d(u)=i$. Hence $c(t(s))=i$ and $\Opt(t(s))=\{u\in L_i:s\subseteq u\}$.

For $\wc\le\Gamma$, take a sequence $s_0,\dots,s_{\hstar}$ attaining $\Gamma(\Pi)$ and the tuples $t(s_i)$, with $t(s_0)=\varnothing$. Let $i<\hstar$ and $u\in\Opt(t(s_i))$, so that $s_i\subseteq u$. An action $a$ applicable in $s_i$ with $s_i[a]\supseteq s_{i+1}$ is applicable in $u$, and $u[a]\supseteq s_i[a]\supseteq s_{i+1}\supseteq t(s_{i+1})$ by \cref{eq:monotone}. As $d(u[a])\le i+1=c(t(s_{i+1}))$, the state $u[a]$ is an optimal achiever of $t(s_{i+1})$. The optimal achievers of $t(s_{\hstar})$ contain $s_{\hstar}\supseteq G$. So the tuples form a uniform chain of width $\Gamma(\Pi)$.

For $\Gamma\le\ws+\xi-1$, consider the representatives $m_i$ and the refined tuples $t'_i$ of the proof of \cref{thm:branching}(i). They satisfy $m_0=I$, $m_i\in \mathcal O_i$ and $m_i\Rightarrow m_{i+1}$. By \cref{eq:refined}, every state of depth at most $i$ that contains $t'_i$ has depth $i$ and contains $m_i$. So $t'_i\subseteq m_i$ meets every edge in the definition of $\delta(m_i)$, and $\delta(m_i)\le|t'_i|\le\ws(\Pi)+\xi(\Pi)-1$.

If every optimal layer is linearly ordered by inclusion, then $\xi(\Pi)=1$ and the three quantities coincide. If every optimal layer is a single state $s_i$, the only sequence in the definition of $\Gamma(\Pi)$ is $s_0,\dots,s_{\hstar}$. By \cref{thm:hitting}, no state $u\ne s_i$ with $d(u)\le i$ contains $s_i$, so the edges defining $\delta(s_i)$ are those of $\Hy_i$.
\end{proof}

When an optimal layer contains several nested states, \cref{eq:hitting} cannot be applied along an optimal trajectory through the smaller one: if $s,s'\in \mathcal O_i$ with $s\subsetneq s'$, then $s\setminus s'=\varnothing$ is an edge of $\Hy_i$ along a trajectory through $s$. The function $\delta$ excludes such supersets. No tuple inside $s$ separates $s$ from them, and they need not be separated, because by monotonicity they continue like $s$. When $\xi(\Pi)>1$, $\Gamma(\Pi)$ can exceed $\wc(\Pi)$. In \cref{rem:unique}, $\xi(\Pi)=3$, $\wc(\Pi)=1$ and $\Gamma(\Pi)=2$.

\subsection{A Recursion for Set Width}\label{app:recursion}

Fix an integer $k\ge0$ and define sets $B_i,V_i\subseteq L_i$ backward for $i=\hstar,\dots,0$. Let $B_{\hstar}$ consist of the goal states in $L_{\hstar}$, and for $i<\hstar$ let $B_i$ consist of the states of $L_i$ that have a successor in $V_{i+1}$. For $s\in B_i$, let
\begin{equation*}
\mathcal F_i(s)=\{s\setminus u:u\in \mathcal R_{<i}\cup(L_i\setminus B_i)\},
\end{equation*}
and let $V_i$ consist of the states $s\in B_i$ with $\HS(\mathcal F_i(s))\le k$. At depths below $\hstar$, membership in $B_i$ guarantees a continuation. A certificate tuple of cost $i$ inside $s$ must exclude all earlier states and every state of $L_i\setminus B_i$. These are the states represented in $\mathcal F_i(s)$.

\begin{theorem}\label{thm:recursion}
For every task, $\ws(\Pi)\le k$ if and only if $I\in V_0$.
\end{theorem}
\begin{proof}
For $s\in B_i$ and $t\subseteq s$, the tuple $t$ hits $\mathcal F_i(s)$ exactly when $c(t)=i$ and $\Opt(t)\subseteq B_i$. Indeed, $c(t)\le d(s)=i$, a state of $\mathcal R_{<i}$ contains $t$ exactly when $c(t)<i$, and when $c(t)=i$, a state of $L_i\setminus B_i$ contains $t$ exactly when it is an optimal achiever of $t$.

Suppose that $Q$ is admissible with $\max_{t\in Q}|t|\le k$. We show by backward induction on $i$ that every optimal achiever $s$ of a tuple $t\in Q$ with $c(t)=i$ lies in $V_i$. By \cref{lem:containment}, $i\le\hstar$, and an optimal plan for $t$ that ends in $s$ is a prefix of an optimal goal plan. If $i=\hstar$, this plan is an optimal goal plan, so $s\in B_{\hstar}$. If $i<\hstar$, the plan is not an optimal goal plan, so it has a one-action extension that is an optimal plan for some $t'\in Q$ with $c(t')=i+1$, and the successor of $s$ that it reaches lies in $V_{i+1}$ by induction. Hence $s\in B_i$. In both cases $\Opt(t)\subseteq B_i$, so $t$ hits $\mathcal F_i(s)$ and $s\in V_i$. Applied to a tuple of $Q$ that holds in $I$, this gives $I\in V_0$.

Conversely, suppose that $I\in V_0$. For every $s\in V_i$, choose a minimum hitting set $t_s\subseteq s$ of $\mathcal F_i(s)$, so that $|t_s|\le k$, $c(t_s)=i$ and $\Opt(t_s)\subseteq B_i$, and let $Q=\{t_s:s\in V_i,\ 0\le i\le\hstar\}$. The tuple $t_I$ holds in $I$. An optimal plan for $t_s$ with $s\in V_i$ ends in some $u\in\Opt(t_s)\subseteq B_i$. If $i=\hstar$, then $u$ is a goal state and the plan is an optimal goal plan. If $i<\hstar$, then $u$ has a successor $u'\in V_{i+1}$, and the plan extended by the action to $u'$ is an optimal plan for $t_{u'}\in Q$. Hence $Q$ is admissible and $\ws(\Pi)\le k$.
\end{proof}

In the converse part, every optimal achiever $u$ of a stored tuple $t_s$ lies in $V_i$, because $t_s\subseteq u$ hits $\mathcal F_i(u)$ by the first paragraph of the proof. The argument uses only $u\in B_i$, which matters for the greedy version in \cref{app:greedy}.

If all optimal goal plans visit the same states $s_0,\dots,s_{\hstar}$, then $B_i\subseteq\{s_i\}$ for every $i$. Indeed, $B_{\hstar}$ consists of goal states of depth $\hstar$, and a state of $L_i$ with a successor in $B_{i+1}$ lies on an optimal goal plan if that successor does. Let $1\le i\le\hstar$. If $B_i$ is nonempty, then $B_i=\{s_i\}$ and $\mathcal F_i(s_i)=\Hy_i$, the hypergraph of \cref{thm:hitting}, so $V_i=\{s_i\}$ if $\HS(\Hy_i)\le k$ and $V_i=\varnothing$ otherwise. At depth $0$, the layer $L_0=\{I\}$ has no earlier states, so if $B_0=\{I\}$, the family $\mathcal F_0(I)$ has no edges and $V_0=B_0$. An empty $V_i$ makes all earlier sets $B_j$ and $V_j$ empty. Hence $I\in V_0$ exactly when $\HS(\Hy_i)\le k$ for every $1\le i\le\hstar$, and \cref{thm:recursion} reduces to \cref{eq:hitting} for set width.

\subsection{Proof of \texorpdfstring{\cref{cor:explicit}}{Corollary 6.5}}\label{app:greedy}

\emph{Algorithm.} For a budget $\bar k\in\{1,\dots,N\}$, define sets $B^{\mathrm g}_i$, $V^{\mathrm g}_i$ and families $\mathcal F^{\mathrm g}_i(s)$ as in \cref{app:recursion}, with one change. For $s\in B^{\mathrm g}_i$, the state $s$ is put into $V^{\mathrm g}_i$ if $\mathcal F^{\mathrm g}_i(s)$ has no empty edge and the greedy algorithm of \cref{sec:hs}, which repeatedly selects an atom of $s$ contained in the largest number of edges not yet hit, with ties broken by a fixed order of the atoms, returns a hitting set $t^{\mathrm g}(s)$ with $|t^{\mathrm g}(s)|\le \bar k$. If $I\in V^{\mathrm g}_0$, the run succeeds and returns $Q^{\mathrm g}=\{t^{\mathrm g}(s):s\in V^{\mathrm g}_i,\ 0\le i\le\hstar\}$. The algorithm tries $\bar k=1,2,\dots,N$ and stops at the first success. Each run computes at most $|\R|$ families with at most $|\R|-1$ edges of at most $N$ atoms, so the algorithm is polynomial in $N$ and in the size of the explicit graph.

\emph{Soundness.} The converse part of the proof of \cref{thm:recursion} uses only that each stored tuple lies in its state, hits the corresponding family and has at most the budgeted size, and that its optimal achievers lie in the corresponding set $B$. It therefore shows that $Q^{\mathrm g}$ is admissible and has width at most $\bar k$. We call the optimal achievers of a stored tuple $t^{\mathrm g}(s)$ other than $s$ its \emph{co-achievers}. They lie in $B^{\mathrm g}_i$, so each has a successor in $V^{\mathrm g}_{i+1}$ if $i<\hstar$, and each is a goal state if $i=\hstar$, which is all that admissibility requires. In contrast to the exact recursion, a co-achiever $u$ need not belong to $V^{\mathrm g}_i$: the greedy algorithm may return for $\mathcal F^{\mathrm g}_i(u)$ a hitting set larger than $\bar k$, although $t^{\mathrm g}(s)$ hits this family.

\emph{The bound.} Let $k=\ws(\Pi)$, let $V_i$ and $B_i$ be the sets of \cref{app:recursion} for this $k$, and let $\bar k_{\max}=\min(N,\lfloor H_{|\R|-1}k\rfloor)$. We show $V_i\subseteq V^{\mathrm g}_i$ for the run with budget $\bar k_{\max}$, by backward induction on $i$. The sets $B_{\hstar}$ and $B^{\mathrm g}_{\hstar}$ coincide, and if $V_{i+1}\subseteq V^{\mathrm g}_{i+1}$, then $B_i\subseteq B^{\mathrm g}_i$. Let $s\in V_i$. Then $s\in B^{\mathrm g}_i$, and the states excluded by $\mathcal F^{\mathrm g}_i(s)$ form a subset of those excluded by $\mathcal F_i(s)$, so $\mathcal F^{\mathrm g}_i(s)\subseteq\mathcal F_i(s)$. A hitting set of $\mathcal F_i(s)$ with at most $k$ atoms therefore hits $\mathcal F^{\mathrm g}_i(s)$, which has no empty edge and at most $|\R|-1$ edges, one for each excluded state. The greedy algorithm returns at most $H_{|\R|-1}k$ atoms \citep{Chvatal1979}, and at most $N$, so $|t^{\mathrm g}(s)|\le \bar k_{\max}$ and $s\in V^{\mathrm g}_i$. By \cref{thm:recursion}, $I\in V_0\subseteq V^{\mathrm g}_0$, so the run with budget $\bar k_{\max}$ succeeds. The first successful run, with some budget $\bar k\le \bar k_{\max}$, returns an admissible set of width at most $\bar k\le H_{|\R|-1}\ws(\Pi)$. The runs for different budgets are independent, so no monotonicity in $\bar k$ is needed.\qed

\FloatBarrier

\ifwithchecklist
\input{checklist}
\fi

\end{document}